%% file: cosmicstringreview.tex
\documentclass[12pt]{article}

\usepackage{graphicx}
\usepackage{amsfonts,amssymb}
\input{definitions.tex}

\def \bk{\mbox{\boldmath$k$}}
\def \bxd{\mbox{\boldmath $\dot{X}$}}
\def \bx {\mbox{\boldmath $X$}}

\def \bu {\mbox{\boldmath $u$}}
\def \x {\mbox{\boldmath $x$}}
\def \l {\mbox{\boldmath $l$}}
\def \by {\mbox{\boldmath $Y$}}

\title{
\sc{ A Tutorial on Links between Cosmic String Theory and
Superstring Theory}\footnote{ Based on lectures given at the
Cosmology in the Laboratory Conference (COSLAB), Imperial College,
University of Leiden and Dhaka University in 2004-2005. }}
\author{ Mahbub Majumdar \\
${}$ \\
\textsf{Theoretical Physics}\\
\textsf{Imperial College} \\
\textsf{Blackett Laboratory} \\
\textsf{Prince Consort Road, London SW7 2BZ, UK}\\
\textsf{Email:} {\tt m.majumdar@ic.ac.uk}  }

\begin{document}

\setlength{\baselineskip}{16pt}

\date{December 5, 2005}
\begin{titlepage}
\maketitle
\vspace{-36pt}

\begin{abstract}

\n Cosmic superstrings are introduced to non-experts.  First
D-branes and $(p,q)$ strings are discussed.   Then we explain how
tachyon condensation in the early universe may have produced F, D
and $(p,q)$ strings. Warped geometries which can render horizon
sized superstrings relatively light are discussed. Various warped
geometries including the deformed conifold in the Klebanov-Strassler
geometry are reviewed and their warp factors are calculated.  The
decay rates for strings in the KS geometry are calculated and
reasons for the necessity of orientifolds are reviewed. We then
outline calculations of the intercommuting probability of F, D and
$(p,q)$ strings and explain in detail why cosmic superstring
intercommuting probabilities can be small. We explore cosmic
superstring networks. Their scaling properties are examined using
the Velocity One Scale model and its extra dimensional extensions.
Two different approaches and two sets of simulations are reviewed.
Finally, we review in detail the gravitational wave amplitude
calculations for strings with intercommuting probability $P<1$.
\end{abstract}
\thispagestyle{empty} \setcounter{page}{0}
\end{titlepage}

\tableofcontents

\section{Introduction}

Cosmic strings and superstrings have been studied for more than 20
years. There has been some cross fertilization of ideas.  For
example, cosmic string theorists have constructed supersymmetric
cosmic superstrings which might model superstrings at low energy
\cite{susycs1, susycs2,susycs3,susycs4,susycs5}, and string
theorists often use the field theory langauge of effective theories
and topological defects to describe superstrings, D-strings and
D-branes \cite{stringsoliton1, ortin}. However in general,
interactions between the cosmic string and superstring communities
have been infrequent.

The reasons for this may be as follows.  (1) Superstrings have
Planckian tensions and observational data precludes such incredibly
heavy strings. (2) Twenty years ago Witten showed that long
fundamental {\em BPS} strings in the most phenomenologically
acceptable version of string theory at the time (the heterotic
theory) are unstable and hence would never be seen in the sky.
Four-dimensional BPS strings are axionic and assuming an axion
potential is generated (string theory abhors global continuous
symmetries) they bound domain walls which collapse very long string
loops.  This killed off interest in astrophysical superstrings
particularly because it was widely believed that non-BPS strings are
also unstable and could not grow to cosmic sizes. (3) Despite being
speculative, cosmic string theory is constrained by the latest
observational data. Since the observed world is four dimensional and
for example, the extra moduli fields originating from say stringy
compactifications of the extra dimensions are not seen, the cosmic
string community has understandably avoided superstrings. (4) Until
recently, string theorists have by and large avoided cosmological
issues - the favorite subject of many cosmic string theorists.

The climate has now changed.  Cosmic string theorists are now more
open to extra dimensions and the extra machinery of string theory
\cite{Davisreview,Kibblereview}. Also, string theorists are much
more interested in cosmology and possible string theory imprints in
the sky \cite{reviewpol}.

In fact the picture that is now emerging is that long superstrings
may be stable and may appear at the same energy as GUT scale cosmic
strings. These strings are similar to cosmic strings in that they
radiate, generate networks, lense distant objects, etc. From the
point of view of cosmic string theorists, this is interesting, since
much of the machinery and work from 15 years ago carries over to
these new stringy objects, albeit with some crucial differences.
From the string theory point of view, this is very exciting because
by positing stable cosmic superstrings which radiate in an
experimentally accessible band, they have stumbled upon a possible
string theory object which may be detected in our lifetime.   The
most general cosmic superstrings are $(p,q)$ strings which package
fundamental and solitonic strings into a single object
\cite{pqstring1,pqstring2,pqstring3}.

The technical developments which have led to this emerging picture
are the following. (1) The {\sc AdS-CFT} correspondence has taught
theorists that gauge strings and superstrings are two faces of the
same object \cite{adscft1,adscft2,adscft3}.  Thus the strings that
cosmic string theorists and superstring theorists have been studying
may be the same objects. (2) The discovery of D-branes as anchors
for open string endpoints and as possible hyperplanes where we may
live has made Type II and Type I string theories phenomenologically
much more attractive and has opened up many new avenues for string
model building, moduli fixing and string cosmology
\cite{firstdbranes1,firstdbranes2,firstdbranes3,firstdbranes4,firstdbranes5,fluxcompact}.
In these theories it is possible to construct long macroscopic
strings which are not BPS but are nonetheless stable  and
potentially observable \cite{copeland}. (3) The study of more
general extra dimensional compactifications has led to the
investigation of warped compactifications in which superstring
tensions can be reduced by an enormous factor of $\sim 10^{-8}$
\cite{fluxcompact}. Superstrings in such warped geometries will not
overclose the universe and are potentially cosmologically viable.
Furthermore, the warping can turn previously unstable non-BPS
strings into ``stable" non-BPS strings. (4) The realization that
gravitational waves from strings with cusps is very non-Gaussian
means that gravitational waves from superstrings in a warped
geometry may be observable by gravitational wave experiments like
LIGO and LISA \cite{damourvil1,damourvil2,damourvil3}.

Let us trace the recent history of cosmic superstrings to understand
which of their properties are model dependent and which are generic.

Type II or Type I cosmic superstrings recently appeared in brane
inflation models
\cite{costring1,costring2,costring3,costring4,costring5}. These
models, tried to use the flexibility of objects (branes) moving in
extra dimensions to produce inflation. Inflation in these scenarios
ended in a phase transition mediated by so-called open string
tachyons \cite{braneinf}. This violent phase transition left in its
wake daughter objects, as many field theoretic phase transitions do,
which are solitonic ``D-strings" and long fundamental  ``F-strings"
\cite{costring1,costring2}. This led to excitement that
string-theoretic brane inflation produces cosmic strings. However,
there were significant problems with these brane inflation models.
The tensions of the strings they produced were not necessarily small
and had to be finetuned \cite{costring4,costring5}. Eventually the
brane inflation picture was refined by models in which the extra
dimensions of the models are ``dynamically" fixed by using a more
general {\em warped} metric which depends on extra dimensional
coordinates \cite{kklmmt1,kklmmt2,kklmmt3,krause}. The warping
naturally leads to low tension  non-BPS cosmic strings which are
stable. The same trick of warping the large four dimensions was used
by Randall and Sundrum and these flux compactification models are
string theoretic realizations of the Randall Sundrum model
\cite{rs1,rs2}.

It might thus seem that cosmic superstrings are relevant only if
brane inflation occurred and  if our world is warped by extra
dimensions. However, we will take a more general point of view. If a
tachyonic (non-supersymmetric) phase transition ever happened in an
expanding universe (after inflation) and if our 4D metric contains
traces of the extra dimensions via some sort of warping- then cosmic
superstrings will inevitably appear. And while brane inflation,
though interesting may be farfetched,  in the author's point of view
it would not be surprising if some non-supersymmetric phase
transition like tachyon condensation on a space filling object
occurred and if our 4D metric contains traces of the extra
dimensions. Given those two ingredients, cosmic superstrings are
reasonably plausible and the brane inflation picture is not crucial
for their relevance.

In fact the problem might not be how string theory can be coaxed to
produce cosmic strings, but that it produces too many and too many
kinds of cosmic superstrings.   A cosmic superstring may be a $Dp$
brane wrapped on a $(p-1)$ dimensional compact cycle. Common
(Calabi-Yau)  compactifications often have thousands of 3D $S^3$ and
2D $S^2$ cycles and a string obtained by wrapping on one $S^3$ is
different from a string  obtained by wrapped on a different $S^3$.
Hence, there are thousands of kinds of cosmic strings in string
theory. Also, dimensional reduction of ten dimensional string fields
to four dimensions gives something like 70 scalar fields. In 4D, a
string can magnetically couple to a scalar and hence 70 types of
string can appear just from the existence of the extra dimensions
\cite{hanany}.

This review is a writeup from various lectures delivered at various
places from 2004-2005, for example at the Cosmology in Laboratory
Conference in Ambleside, UK in 2004, Imperial College, Dhaka
University and the University of Leiden.  Many of the most
contemporary remarks in the review originate from talks, discussions
and debates at a Cosmic Superstrings Workshop at the Institut Henri
Poincare in Paris from September 22-27, 2005.

The review is aimed at students. Hence, it is detailed and works out
some of the more important calculations in the subject. Also because
cosmic superstrings involve a wide array of tools, from boundary
conformal field theory to calculations of the dependence of a
gravitational wave's amplitude on the burst frequency -- it is felt
that others may also benefit from the detail. In particular, the two
specialist topics of string scattering and astrophysical traces of
cosmic strings are reviewed in detail.

The review is structured as follows.  First we try to introduce
D-branes to the novice and explain why the appearance of extended
objects like D-branes may be natural in higher dimensional theories.
Then we show how D-strings are topological defects of an effective
supergravity theory and discuss how string duality leads to the more
general $(p,q)$ strings.  We examine some properties of  $(p,q)$
strings like the junction conditions. We then explain one way to
think about  galaxy sized superstrings and discuss their links to
the more familiar cosmic strings. In the next section we discuss how
tachyon condensation in the early universe can produce objects like
$D3$ braneworlds and D-strings via the Kibble mechanism.  We try to
explain why tachyon condensation may be natural in the early
universe and how it can be thought of as another symmetry restoring
transition at high temperature. We then explain how long fundamental
strings are produced as remnants of tachyon condensation.  The
standard boundary conformal field theory calculation of the number
density of produced strings is briefly outlined. We then ask how can
one make such strings reasonable -- how can one suppress their
tension? Various warped geometries which lower the tension are
discussed and the popular Klebanov-Strassler deformed conifold is
reviewed and its warp factor is calculated. An element of such
models  confusing to cosmologists is orientifolding. Reasons why
orientifolds must appear are discussed. A by-product of the
orientifolding is that F and D strings are not BPS in these models
and hence are not axionic. We calculate the annihilation rates and
show that they are exponentially small. We also show why Type II
fundamental strings are axionic and how membrane instantons lead to
an axion potential. Next, we investigate, F, D and $(p,q)$ string
scattering and calculate their scattering amplitudes and the
probability that reconnection will occur.  We review D-D scattering
in some detail and discuss the effect of compactification if the
strings are free to move in the extra dimensions or are confined by
some potential to particular points. We explain how quantum
fluctuations  blur the positions of strings classically fixed in
space by a potential.

In the last third of the review we discuss observational issues.
Scaling for cosmic string networks is reviewed. We review a
simulation of a (3+1) dimensional $(p,q)$ string network and ask
what happens when strings can move in the extra dimensions. This
motivates our review of the generalized extra dimensional velocity
one scale model and its insights on the effects of extra dimensions
and an intercommuting probability $P<1$. In the final section, we
investigate gravitational wave signatures from cosmic superstrings.
First basic properties of cusps on cosmic strings are reviewed and
then the gravitational burst amplitude $h$ is calculated and its
dependence on the intercommuting probability $P$.

{\sc A note on the literature:} Other reviews of cosmic superstrings
are \cite{Davisreview,Kibblereview,reviewpol}. The central papers on
which this review is based are
\cite{copeland,fluxcompact,damourvil1,damourvil2,damourvil3,costring2,Lambert03,jonespolchinski,Tye05,
Avgoustidis04}.

\section{What are D-branes? What is their relation to cosmic strings?}

\subsection{Extra dimensions naturally give extra extended objects}

The variety of antisymmetric fields a theory can possess increases
with the spacetime dimension. Because gauge field strengths are
antisymmetric, the number of possible field strengths also increases
with dimension. Since field strengths give rise to gauge fields
which couple to objects carrying some sort of charge, as the variety
of field strengths increases with dimension so does the variety of
charge carrying objects.  Such objects are known as branes.  In
general a theory living in $d$ spacetime dimensions can have field
strengths with at most $d/2$ indices and gauge fields with at most
$d/2-1$ indices \cite{polchinskibook}.\footnote{If $S \sim \int d^d
x |F_{p+2}|^2$ then the e.o.m and Bianchi identity are $d F_{p+2} =
0$ and $d *F_{p+2} =0$. This hints of a symmetry between $F_{p+2}$
and $*F_{p+2}$ allowing us to replace $F_{p+2}$ by
$\tilde{F}_{d-(p+2)} \equiv *F_{p+2}$. The corresponding gauge field
$A_{p+1}$ then gets replaced by a new gauge field
$\tilde{A}_{d-p-3}$. } The upper bound appears because field
strengths with more than $d/2$ indices can be related to new field
strengths with less than $d/2$ indices by epsilon contraction.  For
example, in (3+1)D a 3-form field strength $F_{\mu \nu \lambda}$ can
be related to a 1-form field $\tilde{F}_{\rho}$ by contraction with
the (3+1)D epsilon tensor $\tilde{F}_{\rho} = {\epsilon^{\mu \nu
\lambda}}_{\rho} F_{\mu \nu \lambda}$ -- this is called Hodge
duality. The gauge field, $A_{\mu \nu}$  corresponding to $F_{\mu
\nu \lambda}$ is then mapped to a scalar gauge field $A$ such that
$\partial_{\rho} A = F_{\rho}$.

String theory which lives in ten dimensions by the same reasoning
can have field strengths with up to $\frac{10}{2} =5$ indices. Field
strengths with say 6 indices, such as $F_{\mu_1\cdots \mu_6}$ are
related to field strengths with four indices $\tilde{F}_{\mu_7
\cdots \mu_{10}}$ by contraction with a 10D epsilon tensor
$\epsilon_{\mu_1 \cdots \mu_{10}}$.  What is the interpretation of
such higher index fields? The natural thing to do to an
antisymmetric $p+1$-index field, in particular to the $p+1$ index
{\em gauge field} of a $p+2$ index field strength is to integrate
it. For example for $p=0$,

\beq \int A_{\mu} dx^{\mu} = \int A_{\mu} \left (\frac{dx^{\mu}}{d
\tau} \right )d\tau. \eeq

The integral of $A_{\mu}$ thus translates to the integral of
$A_{\mu}$ contracted with the tangent vector of some curve
$x^{\mu}(\tau)$, which we interpret as the worldline of a particle.
The worldline is parameterized by $\tau$. Thus given a vector
$A_{\mu} $ we get a particle.  More generally, integrating a $p+1$
index gauge field $A_{\mu_1 \cdots \mu_{p+1}}$ we get

\beq \int  A_{\mu_1 \cdots \mu_{p+1}} dx^{\mu_1} \wedge \cdots
\wedge dx^{\mu_{p+1}} = \int  A_{\mu_1 \cdots \mu_{p+1}} \left (
\frac{dx^{\mu_1}}{d\sigma^1} \wedge \cdots \wedge
\frac{dx^{\mu_{p+1}}}{d\sigma^{p+1}} \right ) d\sigma^1 \cdots
d\sigma^{p+1}. \label{current} \eeq

Thus the integral of  $A_{\mu_1 \cdots \mu_{p+1}}$ likewise
translates to the integral of $A_{\mu_1 \cdots \mu_{p+1}}$
contracted with some $p+1$ dimensional surface with tangent vectors
$dx^{\mu_i}/d\sigma^i$. The  integration is then over the
coordinates $\sigma^i$ of the surface.  Thus given an $A_{\mu_1
\cdots \mu_{p+1}}$ we find a $p+1$-dimensional surface, which we
interpret as the {\em worldvolume} of a $p+1$ dimensional surface.

So what field strengths/gauge fields does string theory possess?  To
answer this we must construct part of the superstring spectrum.

Since superstring theory is a supersymmetric theory we should not be
surprised that it possesses a spinor groundstate $|{\bf s}
\rangle_L$ for the left moving modes and a spinor groundstate
$|\tilde{{\bf s}} \rangle_R$ for the right moving modes. Thus the
total groundstate is $| {\bf s } \rangle_L \otimes |\tilde{{\bf s}}
\rangle_R$. A Dirac spinor in $d$ spacetime dimensions is $2^{d/2}$
dimensional, which for $d=10$ is a {\bf 32} dimensional spinor.
However, this {\bf 32} is reducible into two Weyl spinors {\bf 16}
and ${\bf 16'}$ which have opposite chiralities.  Thus, we can write
the ${\bf 32} \otimes {\bf 32}$ groundstate in terms of sixteen
dimensional spinors. A crucial ingredient in string theory is the
physical state condition which ensures that unphysical states
decouple.  This condition projects ${\bf 16} \rightarrow {\bf 8}$
and ${\bf 16'} \rightarrow {\bf 8'}$. Thus our groundstate can be
written as a representation of a product of two eight dimensional
spinors. To produce a chiral theory like the Type IIB theory, we
take the eight dimensional spinors to have the same chirality as in
${\bf 8} \otimes {\bf 8}$. To produce a non-chiral theory like the
Type IIA theory we form a product of eight dimensional spinors with
opposite chirality as in ${\bf 8} \otimes {\bf 8'}$.

Now, a tensor product of a spinor with another spinor will produce
states with integer spins.  Hence, we can decompose the spinor
product into a sum of tensor representations $[n]$, where $[n]$ is
an antisymmetric tensor with $n$ indices. We now state without proof
that

\beq {\bf 8} \otimes {\bf 8}  =  [0]\oplus [2] \oplus [4]_+ \ \ \ \
\ \ \ \ \  {\bf 8} \otimes {\bf 8'}  =  [1]\oplus [3]
\label{spectrum} \eeq

These antisymmetric tensors are the so-called Ramond-Ramond gauge
fields. Thus, the Type IIB theory possesses a scalar $A$, a two form
$A_{\mu \nu}$ and a four  form potential\footnote{The + in $[4]_+$
indicates that the corresponding 5-form field strength is actually
self/anti-self dual} $A_{\mu_1 \cdots \mu_4}$ and by epsilon
contraction: 6 form, 8 form and 10 form gauge fields. From our
previous discussion, we know that higher form gauge fields give rise
to/charge extended objects. Thus, these Ramond-Ramond gauge fields
charge 9,7,5,3, and 1 dimensional extended objects. (The scalar $A$
charges an instantonic object known as a $D(-1)$ brane.) These
9,7,5,3,1 dimensional objects are known as the $D9, D7,D5,D3$ and
$D1$ branes of Type IIB string theory. Because (\ref{spectrum})
states that no $[1]$ and $[3]$ gauge fields exist in the Type IIB
theory, the Type IIB theory possesses no {\em stable} $D0$ or $D2$
branes. These objects do exist but because there is no gauge field
to charge them, they are uncharged and hence non-BPS. Using
(\ref{spectrum}) we deduce in a similar way that the Type IIA theory
possesses BPS $D0,D2,D4,D6$ and $D8$ branes which are charged by  1
form, 3 form, and by epsilon contraction: 5 form, 7 form and 9 form
gauge fields.

\subsection{D-branes as solitons of the low energy
theory}

We can formally construct D-branes as extrema of a tree-level
supergravity effective action with $F_{p+2} = dA_{p+1}$ field
strengths. We can guess the effective action up to numerical values
of various coefficients (the coefficients are determined by
supersymmetry). For example, suppose we include a three form
Ramond-Ramond field strength, $F_3$. It will then appear in the
action as $|F_3|^2$ times a suitable coefficient. Additionally, we
have the usual Einstein-Hilbert term $\sqrt{|g|}R$. Tree-level
quantities in closed string theory are weighted by $g_s^{-2}$.
However, in string theory there are no tunable parameters like a
coupling constant. Instead the coupling constant $g_s$ is the field
$e^{\phi}$ where  $\phi$ is called the dilaton. Thus the effective
action also possesses a kinetic term for the dilaton $(\partial
\phi)^2$.    The effective action is thus

\beq S = \frac{1}{2\kappa^2_{10}} \int d^{10}x \sqrt{|g|} e^{-2\phi}
\left ( R + 4 \partial^{\mu} \phi \partial_{\mu} \phi - \frac{1}{12}
|F_3|^2 + \cdots \right  ) \label{effaction} \eeq

\n Note, because of the $e^{-\phi}$ weighting, the dilaton kinetic
term appears with wrong sign which can be reversed with a field
redefinition.

A $D1$-brane is a solitonic solution of this action which is
Poincare invariant in $1+1$ dimensions and isotropic in the
transverse $8$ directions. A suitable ansatz is

\beq ds^2_{D1} = \frac{-(dt)^2 + (dx^1)^2}{\sqrt{h(r)}} +
\sqrt{h(r)}(dr^2 + r^2d\Omega_{8-p}^2) \eeq

\n for some function $h(r)$. We then use the equation of motions
from (\ref{effaction}) and the requirement that the D-brane be
supersymmetric.   Since it is bosonic, all fermions should vanish,
and the susy variations of the dilatino $\delta \lambda$ and
gravitino $\delta \psi$ should also vanish. These give us the
conditions to solve for $H,  \phi$ and $A_2$:

\begin{eqnarray}
 h(r) = 1 + \left (\frac{r_p}{r} \right )^6 & e^{2\phi} = g_s^2 h(r) & A_{\mu \nu}  =
 -\frac{1 - h(r)}{g_s h(r)} \epsilon_{\mu \nu}
 \label{D1sugrasolution}
 \end{eqnarray}

\n Here $g_s$ is the string coupling $e^{\phi(\infty)}$ at spatial
infinity.

We identify these solutions as $D1$ branes because they possess unit
RR charge: $ Q_{D1} = \int_{S^7} * F_3   = 1$. Since they are
supersymmetric (BPS) solitons, their tension equals their charge
implying: $\mu_{D1} = Q_{D1} \sim \frac{1}{g_s}$ because $A_2 \sim
\frac{1}{g_s}$ from (\ref{D1sugrasolution}).

The relation $\mu_{D1} \sim \frac{1}{g_s}$ implies that they are
associated with ``open strings."  These solitons are unusual because
at small $g_s$, they do not gravitationally backreact. This is
because in string theory Newton's constant varies as $G_d \sim
g_s^2$. Thus the gravitational potential  vanishes: $\frac{G_d
\mu_{D1}}{r^6} \sim g_s^2 \cdot \frac{1}{g_s} \rightarrow 0$ for
$g_s \rightarrow 0$.  Hence, heavy D-branes decouple from bulk
gravity.

Another interesting property of the D-brane tension is that at $g_s
\rightarrow \infty$ the tension vanishes and D-branes become
``massless." Hence, at large $g_s$, D-branes may become the
fundamental excitations instead of fundamental strings.  This is
indicative of a much deeper SL($2,\IZ$) symmetry that interchanges
fundamental strings with D-branes when $g_s \rightarrow
\frac{1}{g_s}$ \cite{johnson,schwarzM}.

\subsection{SL(2,$\IZ$) symmetry and $(p,q)$
strings}\label{sectionpq}

 A general SL(2, $\IZ$) transform is of the form

\beq
\Lambda =
\left(
\begin{array}{cc}
a & b \\
c & d \\
\end{array}
\right); \ \ \ \  \ ab-cd =1 \ \ \ \ \ \forall \ a,b,c,d \in
\mathbb{Z} \eeq

In the string theory context an SL($2,\IZ$) transformation
interchanges the potentials which charge the D1 string and
fundamental F-string. The F-string is charged by an antisymmetric
2-index field, $B_{\mu \nu}$.  We have seen that D-strings are
charged by the 2-index gauge field $A_{\mu \nu}$. An
SL($2,\IZ$)transformation with $a=d=0, b = -c = 1$ acts on a doublet
of these two fields as

\beq
\left(
  \begin{array}{c}
    B_{\mu \nu}  \\
    A_{\mu \nu} \\
  \end{array}
\right) \longrightarrow
\left(
  \begin{array}{cc}
    0 & 1 \\
    -1 & 0 \\
  \end{array}
\right)
\left(
  \begin{array}{c}
    B_{\mu \nu} \\
    A_{\mu \nu} \\
  \end{array}
\right)
 \ \ = \ \
  \left(
    \begin{array}{c}
      -A_{\mu \nu} \\
       B_{\mu \nu}\\
    \end{array}
  \right).
  \label{sl2zonba}
\eeq

The same SL($2,\IZ$) transform acts on the scalar doublet $\tau$,
which is composed of the Ramond-Ramond scalar $A$ which charges a
$D(-1)$ brane (which is really just an instanton)  and inverse
dilaton $e^{-\phi}$. The doublet is defined as

\beq \tau = A + \frac{i}{g_s} \label{dilatonaxion} \eeq

\n The previous SL($2,\IZ$) transformation  in (\ref{sl2zonba}) acts
on $\tau$ as

\beq \tau \longrightarrow \frac{a \tau + b}{c \tau + d}
\longrightarrow -\frac{1}{\tau}  \label{sl2zdilaton} \eeq

If there are no $D(-1)$ instantons we can gauge $A$ to $A = 0$. Then
(\ref{sl2zdilaton}) implies that an SL($2,\IZ$) transformation takes
$g_s \rightarrow 1/g_s$ and from (\ref{sl2zonba}) that it
interchanges a fundamental string with a D string. Hence, the
fundamental string and D-string are the same under a nonperturbative
$g_s \rightarrow 1/g_s$ transformation. This is called {\em
S-duality}.

\begin{figure}
\begin{center}
\includegraphics[width=2.7in]{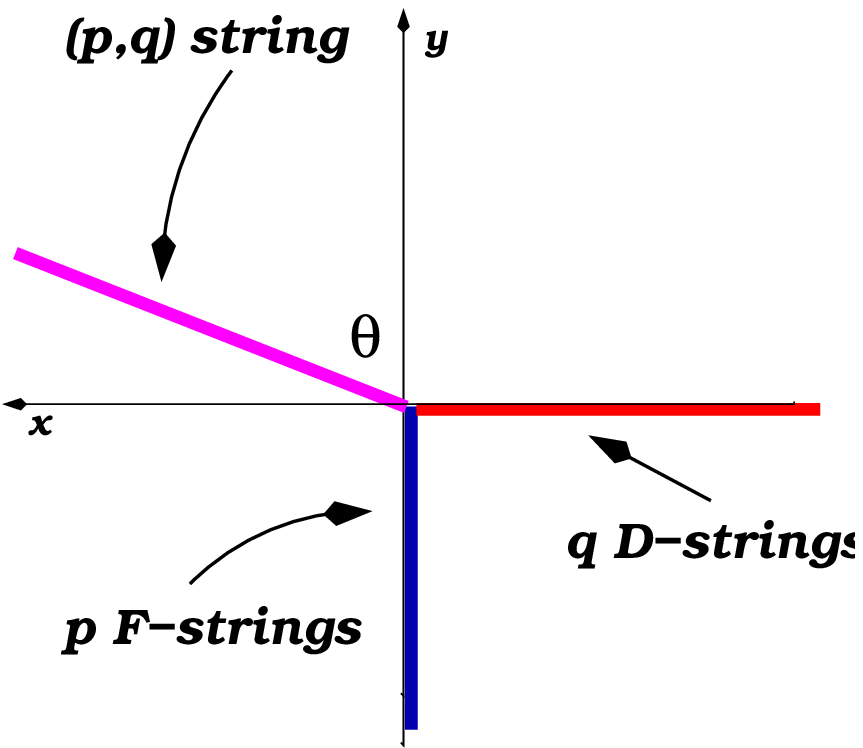}
\includegraphics[width=2.7in]{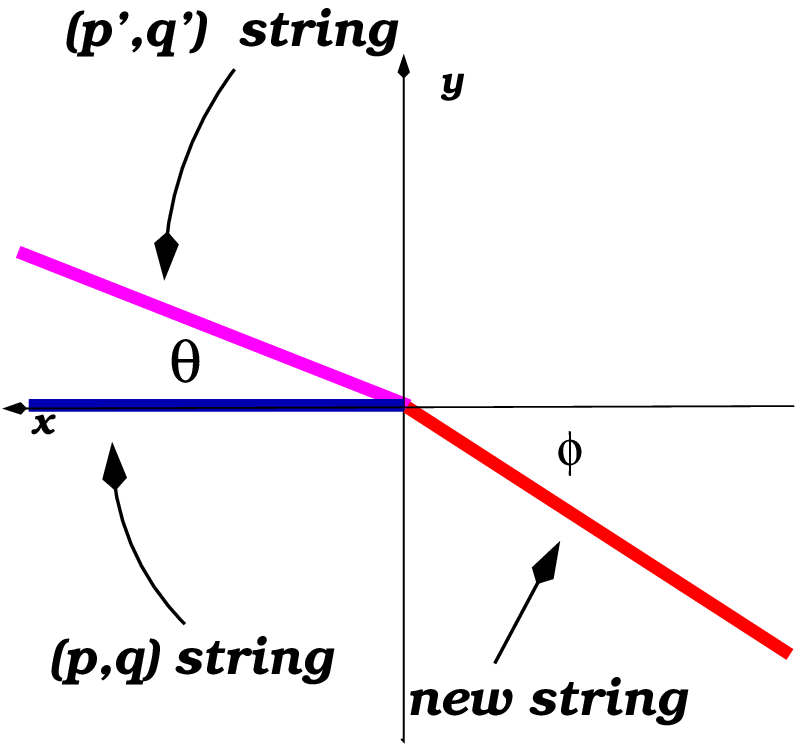}
\caption{(1) On the Zero force condition on $q$ D-strings $\perp$
$p$ F-strings gives a $(p,q)$ string. (2) On the right a $(p,q)$ and
$(p',q')$ string meet to form a new string -- a $(p \pm p',q \pm
q')$ string.} \label{pqstringfig}
\end{center}
\end{figure}

A striking reflection of this duality is that a D-string and a
fundamental string can form a bound-state.  We can guess the tension
of a bound state of a F and D string using a simple minded force
balance argument in the special case described in figure
\ref{pqstringfig}. In the figure $q$ D-strings and $p$ fundamental
strings meet at a right angle and produce a so-called ($p,q$) string
with tension $\mu_{(p,q)}$ at an angle $\theta$ in the $x,y$ plane.
The force balance equations are

\begin{eqnarray}
x:  & {} & \mu_{(p,q)} \sin \theta  = q \mu_{D1}  = \frac{q}{2 \pi \alpha' g_s} \nonumber\\
y:  & {} & \mu_{(p,q)} \cos \theta = p \tau_{F1}  = \frac{p}{2 \pi
\alpha' }
\end{eqnarray}

\n which yield

 \beq
 \mu_{(p,q)} = \sqrt{(p\mu_{(1,0)})^2 +
 (q\mu_{(0,1)})^2} = \frac{1}{2 \pi \alpha'} \sqrt{p^2 + q^2/g_s^2}
\label{tensionpq}
 \eeq

 Note, if the Ramond-Ramond scalar $A$ is nonzero
there is a more general formula: $\mu_{(p,q)} = \mu_{F} [(p-Aq)^2 +
q^2/g_s^2]^{1/2}$.  Note, for relatively prime $p$ and $q$ the
triangle inequality ($|x| + |y| \ge |x+y|$) states that a
$\mu_{(p,q)} < p \mu_{F1} + q \mu_{D1}$ as expected for a bound
state.

More generally the force balance condition  for a $(p_i,q_i)$ string
vertex is $\sum_i \mu_{(p_i,q_i)} \hat{n}_i = 0$, where $\hat{n}_i$
is the direction along which the $(p_i,q_i)$ string is aligned.  The
total charge entering and leaving a vertex must also be zero
implying $\sum_i p_i =0$ and $\sum_i q_i = 0$ at a vertex. This
implies that when a $(p,q)$ string and a $(p',q')$ string meet, that
either a $(p+p',q+q')$ string or a $(p-p',q-q')$ string will form.


Align the $(p,q)$ string along the $x$-axis and the $(p',q')$ string
at an angle $\theta$ as in the right side of figure
\ref{pqstringfig}. The forces on the string junction are then

\begin{eqnarray}
F_x(\theta, \phi) & = &  \ppm \cos \phi  - \mu_{(p,q)}  - \p' \cos \theta   \\
F_y(\theta,\phi)  & = &  \p' \sin \theta -\ppm \sin \phi
\end{eqnarray}

We now find the angle $\theta_{\pm}$ commensurate with $F_x = F_y =
0$. After squaring both sides, then adding them together, and using
$\ppm^2 = \pq^2 + {\p'}^2 \pm 2(pp' + g_s^{-2}qq')$ we find that the
critical angles $\theta_{\pm}$ are

\beq \cos \theta_{\pm} = \pm \frac{pp' + g_s^{-2} qq'}{\sqrt{p^2 +
g_s^{-2}q^2} \sqrt{p'^2 + g_s^{-2} q'^2}} \equiv \pm
\frac{{\mbox{\boldmath $\pq \cdot \p'$}}}{ {\mbox{\boldmath $| \pq |
| \p' |$ }}} \label{thetac} \eeq

\n where we have defined ${\mbox{\boldmath $\pq $}} = (p,g_s^{-1}q)$
and the inner product to be  ${\mbox{\boldmath $\pq \cdot \p'$}} =
pp' + g_s^{-2}qq'$ which  implies ${\mbox{\boldmath $|\pq| $}} =
\sqrt{p^2 + g_s^{-2} q^2}$.   Note, a configuration with both
strings pointing towards the vertex and meeting at an angle
$\theta_+$ is equivalent to a configuration where one string points
in and the other points out of the vertex and where the intersection
angle is $\theta_- = \pi - \theta_+$.



What happens when two strings meet and the angle is not $\theta_+$
or $\theta_-$? The intersection is not BPS and the strings will try
to move to a BPS angle.   If the heavier $(p+p',q+q')$ string
instead of a $(p-p',q-q')$ is to form then the $(p,q)$ and$(p',q')$
strings should be closer together (in terms of the angle $\theta$)
to balance the tension of the heavier $(p+p',q+q')$ string. If
initially $\theta < \theta_+$, the vertex will move into the second
quadrant of figure \ref{pqstringfig} and the $(p,q), (p',q')$
strings will curve so that near the vertex the angle $\theta $ grows
to $\theta_+$ to make the vertex stable.  A lighter $(p-p',q-q')$
string can only form if the strings  are far enough apart to balance
the lighter $(p-p',q-q')$ string. If suppose two strings (one
pointing inwards and one pointing outwards) meet at an angle $\theta
> \theta_+$. Then the lighter string can form if the vertex moves
such that the two strings curve near the vertex to decrease $\theta$
to $\theta_+$.


\subsection{Wrapped branes as 4D baryons}\label{sectionbaryon}

$(p,q)$ strings can end on branes which may be partially or
completely wrapped on compact cycles.  A completely wrapped brane
looks like a heavy particle from a (3+1)D point of view and is
sometimes called a {\em baryon}. At first glance this is confusing
as the endpoints of strings are charged and charge conservation
requires that a string ending on a brane must transfer its charge to
the brane \cite{strominger1,townsend1,marolf1}. This in fact happens
because the gauge invariant 2 form on the brane is not the
electromagnetic field strength $\cf_2 = d{\cal{A}}_1$, but rather
the combination $\cf_2+B_2$.  Thus if $B_2$ disappears on a brane
(i.e an F-string ends on a brane) then a non-zero flux of $\cf_2$ on
the brane is produced. Also interestingly, if say $M$ units of $\int
F_3$ flux threads a compact cycle ${\cal{K}}_3$ and a $D3$ brane
wraps the cycle then charge conservation requires that $M$ F-strings
end on the baryon. This effect arises because of the Chern-Simons
term $\int_{D3} \cf_2 \wedge A_2$ on the brane which allows
$F_3=dA_2$ to source $\cf_2 +B_2$.

The action of a \textsf{brane + fundamental string} is ignoring
various constants including  F-string and brane tensions

\begin{eqnarray}
 S & = &  -\frac{1}{2}
\int_{D3}  |\cf_2+ B|^2  -\frac{1}{2}  \int_{\cm}  |H_3|^2  +
\int_{F1} B_2 + \int_{D3} \cf_2 \wedge A_2 \label{d3action}
\end{eqnarray}

The  first term  comes from expanding the Born-Infeld action $\int
\sqrt{\eta_{\mu \nu} + \cf_{\mu \nu} + B_{\mu \nu}}$ of a $D3$
brane. In the second term $H_3$ is the field strength of the gauge
field $B_2$ and the integral is over spacetime $\cm$. The third term
$\int_{F1} B_2$ is the analog of (\ref{current}) for a string with a
2-form potential $B_2$ and can be converted to a 10D integral
$\int_{\cm} B \wedge * j$ by introducing a 2-form current $j$. In
our notation * is the 10D Hodge duality operator, and $\star$ is the
4D Hodge duality operator.  Thus $\int_{D3} |\cf_2 + B_2|^2 \equiv
\int (\cf_2 + B_2) \wedge \star (\cf_2 +B_2)$ and $\int_{{\cal{M}}}
|H_3|^2 \equiv \int H_3 \wedge *H_3$.  The last term is a {\em
Chern-Simons} term which reduces to an analog of (\ref{current}) for
the $A_2$ gauge field if $\int \cf_2 \in \IZ$. Then $\int \cf \wedge
A_2 \sim \int A_2$. Essentially, an integral $\int \cf_2$ induces
D-string charge.

The variations of (\ref{d3action}) with respect to $\delta B_2$ and
$\delta \ca_1$ where  $\ca_1$ is the gauge field of $\cf_2 = d
\ca_1$ are

\begin{eqnarray}
\delta S_{\delta B} & = &  - \int_{D3}  \delta B_2 \wedge \star
\,(\cf_2 + B_2) + \int_{\cm}  \delta B_2 \wedge d *H_3  + \int_{\cm}
\delta B_2
\wedge *j_{F1} \\
 \delta S_{\delta \ca} & = & -\int_{D3} \delta \ca_1 \wedge d\star(\cf_2
+ B_2) + \int_{D3} \delta \ca_1 \wedge F_3
\end{eqnarray}

\n which give rise to the field  equations
\begin{eqnarray}
d*H_3 & =  & -*j_{F1}  +\star \cf_2 \, \delta^6(x)   + \star B_2 \, \delta^6(x) \ \label{eomH} \\
d \star (\cf_2 +B_2)  & = &  F_3 \label{eomF}
\end{eqnarray}

\n We next integrate the L.H.S of (\ref{eomH})  over an  $S^8$ which
intersects the F-string at only a point.  Then $\int_{S^8} d * H =
0$ is the integral of a total derivative over a compact surface and
vanishes.  The $S^8$ intersects the $D3$ brane in a $S^2$.  Suppose
that $B_2= 0$ on the brane.  Then

\begin{eqnarray}
 Q_{F1} & = &  \int_{S^8} *j_{F1} = \sum_i \int_{S_i^2}  \star \cf_2
 \label{qf1}
\end{eqnarray}

\n where $S_i^2$ is the $S^2$ surrounding the endpoint of the $i$th
string ending on the $D3$ brane.  We sum over all string endpoints
$i$ in (\ref{qf1}).

Suppose the $D3$ brane wraps the compact 3-manifold ${\cal{K}}_3$.
For mathematical convenience we introduce punctures
 at the places where the strings end
on the ${\cal{K}}_3$. Then integrating (\ref{eomF}) over the
${\cal{K}}_3$ and setting $B_2=0$ we find


\beq \int_{{\cal{K}}_3} d \star \cf_2 = \sum_i \int_{S^2_i} \star
\cf_2 = \int_{{\cal{K}}_3} F_3 \label{intK3} \eeq

\n Thus combining (\ref{qf1}) and (\ref{intK3})

\beq Q_{F1} = \int_{{\cal{K}}_3} F_3 \eeq.

Thus if $\int_{{\cal{K}}_3} F_3 =M$, then $M$ F-strings must come
out of the wrapped $D3$ brane baryon.  One can analogously show that
if $\int_{{\cal{K}}_3} H_3 = K$ then $K$ D-strings must emerge from
the baryon.  For both $M$ and $K$ nonzero, a $(M,K)$ string or $M$
D-strings
 and $K$ F-strings must end on the baryon.

This also means that a $(p,q)$ string may break on a baryon. A
$(p,q)$ string will enter the baryon and a $(p-M,q-K)$ string will
exit the baryon if the $(p-M,q-K)$ string can suck enough energy out
of the $(p,q)$ string for baryon pair production to occur. Suppose
$K =0$. We can roughly argue that this will occur if the difference
in energies of the $(p,q)$ and $(p-M,q)$ strings is positive:
$\Delta E = E_{(p,q)} - E_{(p-M,q)} \sim \sqrt{p^2 + q^2/g_s^2} -
\sqrt{(p-M)^2 + q^2/g_s^2} \propto 2p -M >0$.  Here we have assumed
the strings have roughly the same length so that the energy
difference is measured by the difference in tensions. Thus, baryon
production can only occur if

\beq |p| \ge \frac{M}{2} \ \ \ \ \ \ \ \  |q| \ge\frac{M}{2}
\label{baryonppcond} \eeq

\n where we have also generalized to the $K\neq 0$ case.  See
(\ref{baryonrate}) for a more detailed explanation.

\subsection{Making sense of long superstrings}

Superstring theory possesses a single length scale $\ell_s =
\sqrt{\alpha'}$ which is related to the Planck scale by $\ell_P =
g_s \ell_s$.  Thus strings are typically of Planckian size. In order
to grow a fundamental string one can wrap it on a very large compact
cycle.  The energy of the string is  the tension, times the size of
the cycle, times the number of times it winds the cycle $w$. For
example if wrapped on a circle,  $E = 2 \pi R w\mu_F$.  Such a
string wrapped on a circle/torus is unexcited and its size solely
comes from the winding energy.  Another way to grow a string is to
heavily excite it. Quantization of the string gives $m \sim
\sqrt{N}$ where $N$ is the total excitation number of the string.
Since $m = \mu_F \ell$ where $\ell$ is the length of the string, a
highly excited string will have $\ell \sim \sqrt{N}$.

If the string is macroscopic then it can be described by a Brownian
random walk, as each bit of a long string seems to act independently
of other bits of the string.  One can calculate the mean squared
separation of two points on an open string as ${\cal{D}}^2 = d \int
\frac{d\sigma}{\pi}[ X^{\mu}(0,\sigma + \Delta \sigma) -
X^{\mu}(0,\sigma)]^2$. Here we have picked out some dimension $\mu$
among the $d$ spacetime dimensions and averaged over $\sigma$. Using
the mode expansion for $X^{\mu}$ we can show that ${\cal{D}}^2
\propto \Delta \sigma \propto \ell$
\cite{mitchellturok1,mitchellturok2,turok3}. A Brownian random walk
is characterized by the mean squared displacement between two points
on the random walk curve satisfying $|\bx_1 -\bx_0|^2 \propto
t_1-t_0$ where $(t_1-t_0)$ is essentially the number of random walks
steps taken from point $\bx_0$ to get to point $\bx_1$ if the time
step is $\Delta t =1$. Hence, $(t_1 -t_0)$ is also the arclength $L$
of the random walk curve between $\bx_0$ and $\bx_1$ and therefore
the random walk is characterized by $|\bx_1 -\bx_0|^2 \propto L$.
Thus a long superstring with ${\cal{D}}^2 \propto \ell$ can also be
thought of as a random walk (with step size $\ell_s$ and number of
steps $\sqrt{N}$).

Now cosmic strings can also be thought of as Brownian random walks
\cite{turok3,frieman,shellard}, and thus it seems reasonable to
identify  fundamental cosmic superstrings as highly excited
superstrings. Such strings can be thought of as classical
correspondence limits (with mode number $N\rightarrow \infty$) of
quantum strings. However, there are a number of issues associated
with such highly excited strings. First, they naturally appear only
at around the Hagedorn temperature which is not well understood. At
such high densities long open strings soak up all the extra energy
and tend to grow longer at the expense of smaller superstrings
\cite{mitchellturok1,mitchellturok2,turok3}. Second, although
interactions are generally ignored, in the presence of gravity
Hagedorn strings can encounter a Jeans instability
\cite{atickwitten}.

In fact, one might worry that the self-gravity of such massive
strings will cause them to collapse to black holes. A string can
smoothly turn into a black hole only when its entropy $S_{string}$
matches the black hole entropy $S_{bh}$. However,  the entropies
match only at a critical string mass $m_c \sim m_s/g_s^2$.  Above
$m_c$ the string entropy $S_{string} \sim m$ is less than the black
hole entropy $S_{bh} \sim m^{(D-1)/(D-2)}$ in $D$ spatial
dimensions. Only, when $m$ decreases to $m_c$ will string
self-gravity collapse a random walking string to a size equal to its
Schwarzschild radius, which in this case is $\ell_s$
\cite{bhcorr1,bhcorr2,damourveneziano}.  But, since $m \gg m_c$ for
horizon-sized cosmic strings with $g_s \sim {\cal{O}}(10^{-1})$ we
will not worry about this. A black hole can also form if a random
walking cosmic string with say $G\mu_F \sim 10^{-7}$ random walks
itself into a region less than its Schwarzschild radius $\sim
\ell_s$. However, it would then have to double back on itself about
$10^7$ times since the random walk step size is $\ell_s$ for a
superstring. Furthermore, if the string is stretched by the cosmic
expansion, then the step size will be stretched to the size of the
horizon making black hole formation even more improbable
\cite{polchinskiemail}.

The picture that was  previously put forward for fundamental cosmic
superstrings is that they form at very high densities at around the
Hagedorn temperature as perhaps in tachyon condensation. Then as the
universe expands and the temperature redshifts the energy density
drops and instead of long strings being entropically favored,
smaller loops become entropically favored. Hence, the long strings
tend to discharge part of their lengths in loops.  This allows the
strings to scale and eventually around $\sim \! \! 20\%$ of the
string density ends up in loops as opposed to initially being $\sim
\! \! 100\%$ in long strings \cite{turok3, shellard}.

The correspondence between a cosmic string and a D-string is more
straight-forward.  Since D-strings are topological defects they are
naturally very long (infinite in flat space) and there is no problem
in thinking about astrophysically large D-strings. In some sense,
with respect to closed string interactions, one can even think of
them as (non-normalizable) coherent states of the closed string
raising operators $\alpha_{-n}^{\mu}$. For example, for the bosonic
string by operating on the vacuum state $| 0 \rangle$ with an
exponential of operators $\alpha_{-n}$, which excites  mode $n$ of
the string, we get a D-string.

\beq | D1 \rangle  \sim  \delta(x^{m}) \left [ \prod_{n=1}^{\infty}
\exp \left ({-\frac{1}{n} \alpha_{-n}^{\mu}  S_{\mu \nu}
\tilde{\alpha}_{-n}^{\nu}}  \right ) \right ] | 0 \rangle
\label{boundarystate} \eeq

\n Where $S$ is a $D \times D$ matrix \cite{divecchia1,divecchia2,
polchinskibook,johnson}.  As described in the next section they are
dynamically formed in a relatively well understood tachyon
condensation process.

Another way to think about the cosmic string and superstring
correspondence is via the {\sc AdS-CFT} correspondence.  Cosmic
strings are gauge theory solitons and stringy objects have a gauge
theory description via the {\sc Ads-CFT} correspondence
\cite{adscft1, adscft2,marija}. For example the dynamics of a
D-string are governed by a $U(1)$ gauge theory with  8 scalars.  And
in the Klebanov-Strassler geometry (to be discussed later),
F-strings are described by confining flux tubes. D-strings appear as
Abrikosov-Nielsen-Oleson vortices in  an Abelian-Higgs system
\cite{Gubser04}.

A different method of identifying gauge theory strings with
D-strings has been the construction of Abrikosov-Nielson-Oleson
vortices in $N=1,\, d=4$ super-Yang-Mills + gravity.  In
\cite{fayetilio1,fayetilio2}, the authors tried to describe tachyon
condensation using supergravity and the ensuing D-string creation
using a $D$-term potential with a Fayet-Iliopoulos term. They
identified the cosmic strings of the theory as D-strings in the low
energy limit. If their identification is correct then it would
provide a non-stringy (field theoretic) description of stringy
objects and should be useful in examining the properties of stringy
cosmic strings such as their stability, etc..

\subsection{Cosmic string velocity}

Finally, because we will continually invoke this result we prove as
in \cite{Avgoustidis04,shellard} that the mean square average
velocity of a closed cosmic string in flat space is 1/2 and in
expanding spacetime is $\le 1/2$. Here we ignore the extra
dimensions and write $X^{\mu} = (X^0, \bx)$ where $\bx$ is the 3D
string position vector.

We define $\langle v^2 \rangle$ as

\beq \langle v^2 \rangle  = \int_0^T \frac{dt}{T} \int_0^L
\frac{d\sigma}{L} \bxd^2 \label{avevel} \eeq

If we use the gauge conditions described in (\ref{gaugeconstraints})
and the equation of motion $\ddot{\bx} = \bx''$ and integrate by
parts then we can write

\beq \int d^2 \sigma \bxd^2 = - \! \int d^2 \sigma \bx \! \cdot \!
\ddot{\bx} = - \!\int d^2 \sigma \bx \! \cdot \! \bx'' = \int d^2
\sigma (\bx')^2 = \int d^2\sigma (1- \bxd^2) \label{veleqn} \eeq

\n where the integral is over one string oscillation period and over
the length of the string.  Combining (\ref{avevel}) and
(\ref{veleqn}) we deduce that $\langle v^2 \rangle = 1/2$.

Now consider an expanding spacetime with scale factor $a$. If we
differentiate the generalized version of (\ref{avevel}) using the
averaging in (\ref{averaging}) and using the equation of motion of
$\bx$ we can find an evolution equation for $v$:

\beq \dot{v} = (1-v^2) \left (\frac{k}{R} - 2\frac{\dot{a}}{a} v
\right ) \label{eveq}\eeq

\n where $k(v)$ is given by the phenomenological formula

\beq k(v) = \frac{2\sqrt{2}}{\pi} \frac{1-8v^6}{1+8v^6} \eeq.

Thus for $v = 1/\sqrt{2} \Rightarrow k =0$ and from (\ref{eveq}),
$\dot{v} <0$.  Thus in an expanding spacetime $v^2 \le 1/2$.  This
result also applies to $3+n$ spatial dimensions where the $n$
directions are fixed.

\section{Dynamical production of D-branes}

%
%

We have seen that D-branes are nonperturbative (in the sense that
they cannot be obtained from the linearized field equations)
solutions of the gravity + supersymmetry equations of motion.
However, how does one dynamically produce a brane? Phase transitions
are well known mechanisms for producing solitons as topological
defects and are signalled by the presence of an unstable mode -- a
negative mass excitation, a tachyon.  We thus look for controllable
tachyonic string backgrounds.

\subsection{Tachyon condensation at zero and finite temperature}
\label{sectiontachyoncond}

Certain {\it open string} backgrounds are well known to be tachyonic
because they are not supersymmetric. (They are often represented by
coincident brane-antibrane pairs.)  On these backgrounds, the
tachyon field, $t$, possesses a potential $V(t)$ which depends on
only $|t|$ and has a double well shape.  Initially, the tachyon
starts at the top of the potential $t=0$, and then rolls down to the
bottom $t_0$. If no topological obstructions to tachyon condensation
exist, then at the bottom of the well $t= t_0$, the negative tachyon
energy cancels the energy of the open string background (the
tensions of the brane and antibrane) leaving a {\it closed string}
background  populated with heavy closed fundamental strings as decay
products.  However, if topological obstructions exist then solitonic
defects may also be produced.

In particular, if tachyon condensation in IIB string theory happens
in $D_t$ spatial dimension, then  branes with $D_t-2, D_t-4,
D_t-6,...$ spatial dimensions may be produced as topological
defects. A vortex with $p=D_t-2$ is the highest dimension D-brane
that can be produced.  It corresponds to $\pi_1({\cal{M}}) \neq 0$
where ${\cal{M}}$ is the vacuum manifold of the tachyon potential.
Hence if tachyon condensation occurs  in only (3+1) dimensions, only
D-strings and $D(-1)$ instantons may be produced.

How do we fit this into a cosmological scenario?  We give two
methods.

(1) If a brane collides with a parallel antibrane and forms a bound
state, i.e. doesn't simply pass through the antibrane -- then a
tachyonic background will form on the volume jointly occupied by the
brane and antibrane.  Most brane inflation mechanisms employ this
scenario to create tachyons to create topological defects. A variant
of this scheme is when two branes intersect each other in a
``non-BPS" way. Tachyons appear at the intersection.

\begin{figure}
\begin{center}
  \includegraphics[width=8cm]{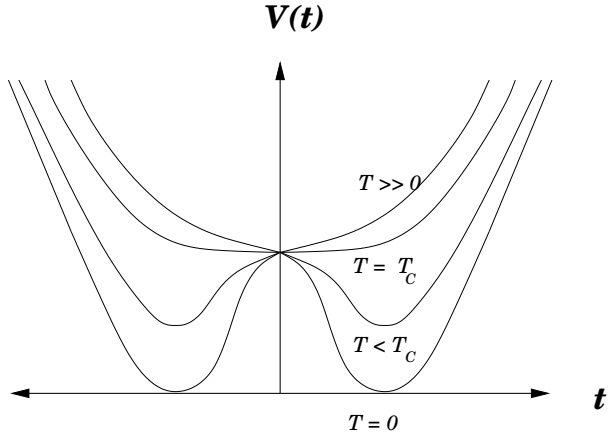}
\end{center}
  \caption{The tachyon potential
$V(t)$ above and below the critical temperature $T_c$ $\sim
T_H/\sqrt{g_s N}$\label{thermalpotential}}
\end{figure}

(2) Suppose the universe starts off in some non-supersymmetric open
string state like a state with $N$ spacefilling brane-antibrane
pairs. It will generally then be tachyonic at zero temperature. If
the temperature is sufficiently high, finite temperature corrections
may turn the negatively curved part of the tachyon potential into a
positively humped part, thus removing the tachyon
\cite{danielsson01}. See figure \ref{thermalpotential}. This will
occur at temperatures above $T>T_c=T_H/\sqrt{g_s N}$, where $T_H$ is
the Hagedorn temperature. Finite temperature effects shift the
equilibrium value of the tachyon field upward -- towards the
symmetric point. This decreases the tachyon mass. Although, it costs
energy to do this, a gas of smaller mass tachyons has a larger
entropy.   This phenomenon is familiar in finite temperature field
theory where symmetry restoration results from finite temperature
loop corrections \cite{kapusta}.

However, if the universe is adiabatically expanding, for example
expanding due to the positive vacuum energy of the $t=0$ vacuum
state, the temperature will redshift and drop.  Once it falls below
$T_c$, a tachyon will develop dynamically and destabilize the space
on which the tachyon field has support.  The tachyon will then
condense by rolling to the minimum of the potential which is
characterized by the vacuum manifold ${\cal{M}}$ of the tachyon
potential. On ${\cal{M}}$, $t$  will be characterized by a set of
phase angles, $\{ \theta_i \}$, and a modulus $|t_0|$. For example,
if $\cm = S^1$, the tachyon will receive the expectation value
$\langle t \rangle =|t_0|e^{i\theta}$. However, near the temperature
$T_c$, the tachyon field will randomly fluctuate, rolling down the
potential and then rolling up via thermal fluctuations. Hence $t$
will not take on any definite values and the phase angles will
fluctuate. Once the temperature falls below $T_c$ and reaches the
Ginzburg temperature $T_G$, thermal fluctuations will no longer be
able to push the tachyon up the potential again.  At this point,
once the tachyon rolls to the bottom of $V(t)$, its phase will be
frozen in. Note, the Ginzburg temperature is close to $T_c$ and can
be written as $h(g_s) T_c$, where $h(g_s)$ depends only on the
string coupling and is typically close to unity~\cite{shellard}.

\subsection{D-string creation by the Kibble mechanism}

In a second order phase transition such as the tachyonic transition,
the correlation length of the field is very large.  However,
expanding universes have causal horizons which bound the distance
over which causal processes can occur.  In a universe with a Hubble
parameter $H\sim 1/t$, causal processes can occur only within a
sphere of diameter $H^{-1}$.  Thus an expanding universe will have
regions which are causally disconnected from each other.

Suppose that no topological obstructions to tachyon condensation
exist. Then the tachyon field will have a magnitude $|t_0|$
everywhere. However, because Hubble volumes are causally
disconnected and since the tachyon's phase on $\cm$ is randomly
determined, the tachyon's phase will generally be different in
different Hubble volumes. Spacetime will thus possess a domain type
structure, with the expectation value $\langle t \rangle$ varying
from Hubble region to region in a relatively random way. The
question answered by Kibble about cosmological phase transitions
(like our tachyonic transition) was whether any residue of false
vacuum remains anywhere.  In particular can false vacuum be trapped
at the intersection points of Hubble regions like flux tubes are
trapped in a superconductor~\cite{Kibble76}? The answer is yes, but
depends on the shapes of the Hubble regions, how they intersect and
the number of expanding spatial dimensions $D_e$. In general since
one intersection can roughly be associated to each Hubble volume, a
{\em lower bound} on the number of branes formed by tachyon
condensation in an expanding universe is: one brane per Hubble
volume.

\begin{figure}
\begin{center}
\includegraphics[width= 2.7in]{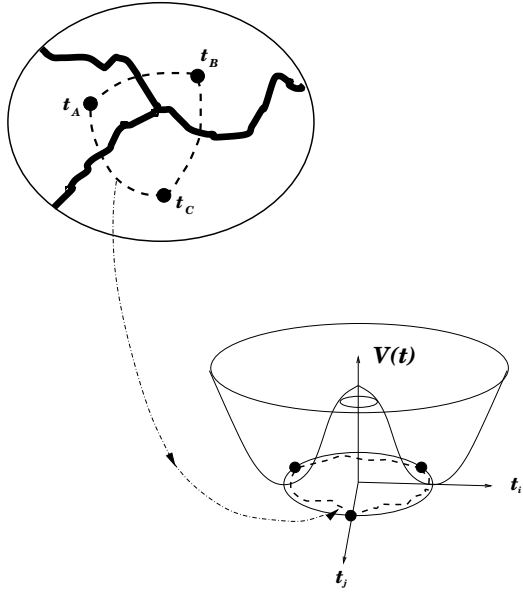}
\end{center}
  \caption{Three Hubble volume sized
regions $A,B,C$, intersect at a point $P$, which is actually a 7D
edge.  The tachyon field takes the values $t_A, t_B,t_C$ on $A,B$
and $C$ respectively.  The tachyon field maps the upper loop in
spacetime, $D$, which encloses $P$ to the lower loop on the vacuum
manifold, $E$, and is an element of $\pi_{1}(\cm)$.}\label{homotopy}
\end{figure}

Suppose that all the directions that the tachyon has support on are
expanding such that if $D_e$ is the number of expanding directions:
$D_t = D_e$. Also, suppose that three cells meet along an edge as in
figure \ref{homotopy}. The edge is $D_e-2$ dimensional and coming
out of the paper.  The phase change around the closed curve $\gamma$
enclosing the edge at $P$ will be $2\pi$. The tachyon field, $t(\x)$
maps $\gamma$ to the locus $\Gamma$ which winds $\cm$ and is an
element of $\pi_1(\cm)=\IZ$. Conversely, non-triviality of
$\pi_1(\cm)$ implies that there exists a configuration, notably the
3 intersecting cells, for which a $S^1$ in spacetime can be mapped
to a locus winding the vacuum manifold.

Attempts to shrink the curve in spacetime will cause the path
$\Gamma$ to move off of $\cm$ and upwards to the false vacuum $t=0$.
Thus along the edge, which is the intersection of the cells, a line
defect of false vacuum will be trapped.  For a tachyon solely
residing in the expanding $D_e$ dimensions this corresponds to a
$(D_e -2)$-brane. For example, if the tachyon has support on 5
spatial dimensions the trapped false vacuum will correspond to a
$D3$ brane -- a braneworld.  If $D_e \ge 5$ then $D3$ branes must
form at intersections of more than three Hubble regions. For
example, if nine dimensions expand then $D3$ branes will  form at
the intersections of 7 different Hubble regions.   This is because
the charge of a $D3$ brane is an integral of the five form field
strength $F_5$ over an $S^5$: $Q_{D3} = \int_{S^5} *F_5$.  An $S^n$
sphere is determined by $n+2$ points, and an $S^5$ is determined by
seven points belonging to different Hubble volumes\footnote{Use
induction and the fact that a $S^1$ is determined by an inscribed
triangle (three points). To add an extra dimension to a polyhedron
add a point in the extra dimension. Alternatively, $S^n$ is the
locus $\sum_1^{n+1} (x_i - \bar{x}_i)^2 = R^2$.  All the parameters
$(\bar{x}_1,\ldots,\bar{x}_{n+1},R)$ can be determined by $n+2$
equations, i.e. $n+2$ points.}. Alternatively, the vacuum manifold
in this case is $\cm = U(2^2) \sim S^5$.  Hence a topological defect
will exist if $\pi_5(\cm) \neq 0$, which requires a mapping of an
$S^5$ in spacetime to an $S^5$ on $\cm$.

Suppose now that $D_e =3$ and that the number of spatial dimensions
in which tachyon condensation occurs is $D_t = 3+ n$ and  the $n$
dimensions are compact. Then only 3D regions lying along the
expanding directions are causally separated. This trivial fact means
that in 3+1 spacetime dimensions, the Kibble mechanism cannot
populate the extra $n$ dimensions with topological defects. The
Kibble mechanism will operate only in the expanding directions. I.e.
the Kibble mechanism will produce only D-strings in large numbers.
Monopole-like $D0$ branes or higher dimensional branes will not be
Kibble produced.

\begin{figure}
\begin{center}
\includegraphics[width=2.7in]{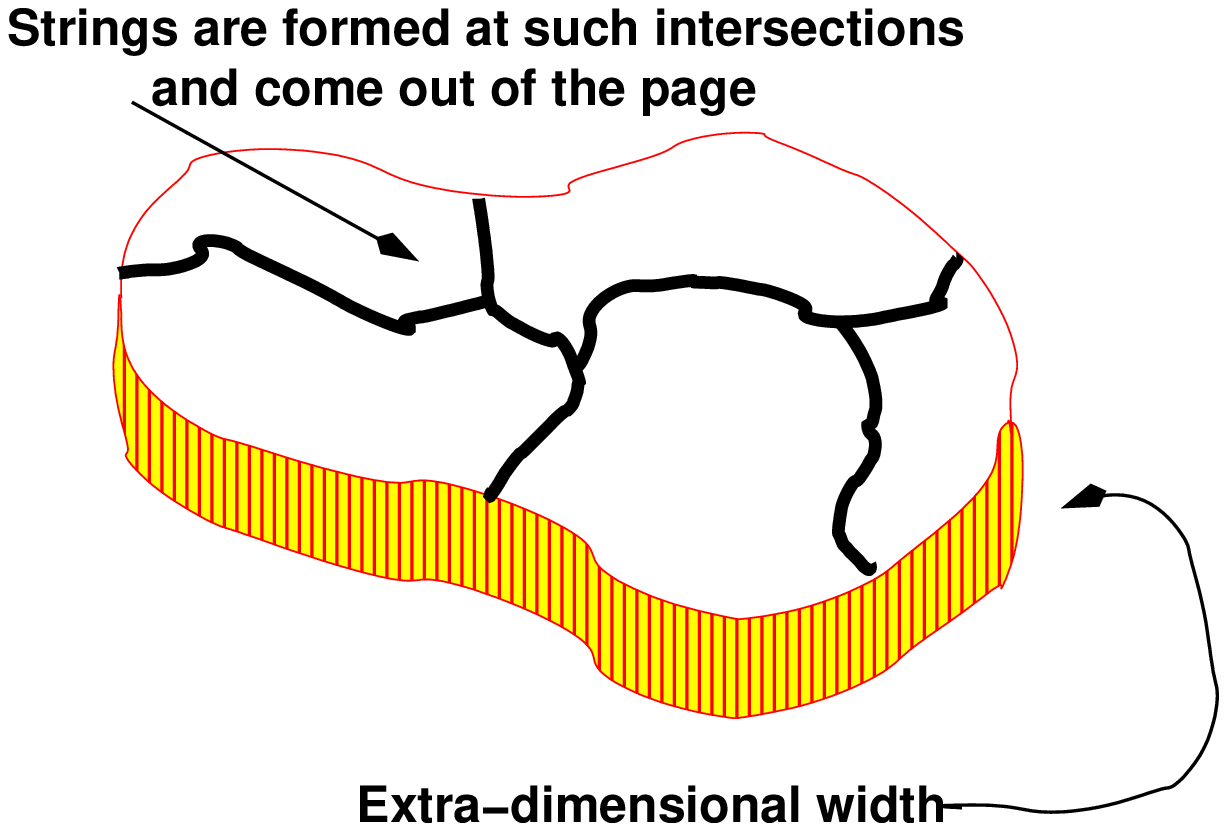}
\end{center}
  \caption{4D strings formed at the intersections of different Hubble volumes
  may actually be higher dimensional branes wrapped on cycles in the extra dimensions.
  Their extra-dimensional part can be thought of as their 4D ``thickness.''}\label{thickness}
\end{figure}

For example, suppose tachyon condensation occurs in 7 spatial
dimensions ($n=4$). Then $D5$ branes, $D3$ branes, D-strings and
D-instantons may be produced. However, only D-strings, or objects
which look like strings to a 4D observer will generically be mass
produced by the Kibble mechanism\footnote{There are however, some
interesting questions regarding D-instantons}. To produce a $D5$
brane two spatial dimensions of the 7D space must be cut away. If
these two directions are in the expanding directions then one of the
$D5$'s dimensions will be in the expanding directions and the other
four directions of the $D5$ will be wrapped on a 4D compact cycle
${\cal{K}}_4$ and the Kibble mechanism can mass produce them.
However, from a 4D point of view these $D5$ branes look like strings
with a small thickness -- which is their spread in the $n$
nonexpanding directions. See figure \ref{thickness}. $D3$ branes
cannot be produced in large numbers because four dimensions of the
7D space must be cut away. Because there are only three expanding
directions, one of those dimensions must be a nonexpanding
direction. The Kibble mechanism cannot operate in that dimension,
hence the Kibble mechanism will not produce 1 $D3$ brane/Hubble
volume if $D_e \le 3$. For $D3$ production $D_e$ must be at least 4,
corresponding to a 4+1 dimensional expanding spacetime.

\section{Production of F and $(p,q)$ strings}

We learned that open string tachyon condensation can lead to
D-string production.   We now show how the phase transition produces
F-strings.  Once F-strings and D-strings are produced collisions  of
$p$ F-strings and $q$ D-strings can  produce  $(p,q)$ string bound
states as described in section $\S$\ref{sectionpq}

\subsection{F-strings as confining flux tubes}

One argument put forward for the appearance of F-strings is the
following.  Suppose a $Dp-\bar{D}p$ brane anti-brane pair annihilate
to form a lower dimensional $D(p-2)$ brane.  A brane  has a $U(1)$
gauge symmetry.  Thus the gauge group of the brane-antibrane system
consists of the two $U(1)$'s on each brane and is $U(1) \times
U(1)$. The daughter brane possesses a $U(1)$ group which is
identified as the linear combination $U(1)_-$ of the original two
$U(1)$s.  The other linear combination $U(1)_+$ must disappear
because only one brane remains. The $U(1)_+$ is thought to disappear
by having its fluxes confined by confining strings which are thought
to be F-strings.

\subsection{Closed string production using boundary CFT}

A more technical argument which produces {\em closed} strings and
should carry over to open string production goes as follows.
Tachyons cause the production of D-strings and F-strings. Thus we
should understand how perturbations due to tachyons change the
string worldsheet action. Perturbations due to tachyons $t(X)$ and
open string fields like  $A_{\mu}$ can be calculated by adding to
the action a deformation $\delta S(t,A_{\mu})$.  The deformation
must preserve conformal invariance. Deformations preserving scale
invariance are by definition called {\em marginal}. A suitable
marginal deformation is

\eqn{marginaldef}{ \delta S = \int_{\partial D^2} ds \biggr
[t(X(\sigma)) +
\partial X^{\mu}(\sigma)A_{\mu}(X(\sigma)) + \cdots \biggr ]}

\n where the integration is over the boundary of the disk $\partial
D^2$ which is parameterized by $s$.

We are only interested in the effects of the tachyon $t(X)$ and thus
will set $A_{\mu} =0$.  The linearized string field theory equation
of motion for the tachyon  is $(\partial^2 -m^2) t(X) = 0$,
\cite{Sen02,Sen04}. For spatially homogeneous tachyons it is
satisfied by

\eqn{tachyonsol}{ t(X) = A e^{-X^0} + B e^{X^0} }

\n where we have set $m^2 = -1$.  For the boundary conditions
$t(X^0=0) = \lambda$ and $\dot{t}(X^0=0) =  0 $ (\ref{tachyonsol})
gives

\eqn{tachdef}{\delta S = \frac{\lambda}{2}\int_{\partial D^2} ds
e^{X^0}}

\n $\lambda/2$ measures the strength of the perturbation; it need
not be small.

Particle production in field theory can be described by an
interaction  $\int \! J(x) \phi(x)$ where the source  $J(x)$
satisfies $(\partial^2 -m^2)\phi(x) = J(x)$ .  For example, the
source  $J= g \chi \chi^*$ corresponds to an interaction like $g
\chi \chi^* \phi$.  In analogy, brane annihilation will produce
closed strings because D-branes source closed strings.   The source
$J_s$ which couples to a closed string state $|\phi_s \rangle $ is
the overlap $J_s = \langle \phi_s |J \rangle$ (up to multiplication
by ghosts). Up to multiplication by ghosts, $|J \rangle$ is the
boundary state $|J \rangle \sim | Dp \rangle$ representing the
$Dp$-brane, which is the $Dp$ version of (\ref{boundarystate}). The
analogous Klein-Gordon equation is then

\eqn{eqforsource}{ (\partial^2 - m^2_s ) \phi_s  = J_s}

\n The task thus boils down to calculating the source $J_s = \langle
\phi_s | Dp \rangle$.  The state  $\langle \phi_s | $ can be created
from the vacuum by the (vertex) operator $V_s$ as $\langle \phi_s |=
\langle 0 |V_s $.  Thus the item to calculate becomes

\beq J_s = \langle 0| V_s |Dp \rangle_{S + \delta S} \label{j_s}
\eeq

\n where we have made explicit that the overlap is to be calculated
with the modified action $S+ \delta S$.

The number density of emitted closed strings and their average
energy can be calculated as

\eqn{dense}{\frac{N}{V} = \int \frac{dE}{2E} \rho(E) |J_s|^2;
\;\;\;\;\; \;\frac{E}{V} = \int \frac{dE}{2E} \cdot E \cdot
\rho(E)|J_s|^2}

\n where $\rho(E)$ is the density of string states.   $N/V$ and
$E/V$ diverge unless gravitational backreaction of daughter strings
is taken into account \cite{Lambert03,Naqvi02,Gomis05}.
Unfortunately taking backreation into account has been possible only
in 2D string theory where gravity is much more benign
\cite{Wadia03,Klebanov03,Ambjorn04}. Therefore we do not present the
explicit calculation of (\ref{j_s}). Nevertheless, as $N/V$ and
$E/V$ are non-zero, we expect that fundamental strings are produced
by tachyon condensation/brane-anti-brane annihilation.

\subsection{How long are the F-strings?}

The fundamental strings which are produced are long heavy strings
much like field theory cosmic strings. The density of states
$\rho(E)$ of a string grows exponentially with the mass as $\rho(m)
\sim m^{a} e^{\alpha' m}$ \cite{GSWvolI}. Thus  the number of
available massive states far outnumbers the number of low mass
states. Note that an $|out \rangle $ state with a $\int J \phi$
interaction term is

\eqn{out}{|{\rm out} \rangle = T e^{-i \int H_I(t) }| {\rm in}
\rangle = T e^{-i \sum_s \int \rho(s) J_s \phi_s } |{\rm in} \rangle
\sim \sum_{n,m} (\rho J_s)^n (a^{\dagger}_s)^n | {\rm in} \rangle +
\cdots }

\n On the R.H.S.  we  used the quantization $\phi \sim \sum_m f_s
a_s + f^*_s a^{\dagger}_s$. For point particle theories the
$J^n(a^{\dagger}_s)^n |{\rm in} \rangle$  components of the final
state correspond to a final state with $n$ particles. However, in
string theory $a^{\dagger}_s$ corresponds to the mode operator
$\alpha_{(s<0)}^{\mu}$ which instead of creating strings, excites
massive modes.  Now $J_s$ is expected to fall exponentially with
energy while $\rho$ rises exponentially with energy
\cite{Lambert03}. If $\rho$ wins or if $\rho J_s$ is not steeply
suppressed tachyon condensation will lead to very massive strings.


The liberated energy density from $Dp$ brane annihilation is $\Delta
\rho \sim (2 \pi)^{-p} g_s^{-1}$ in string units.  Now, the Hagedorn
energy density $\rho_H$ is approximately $\sim 1$ in string units
\cite{Bowick92, Zwiebach}. Thus for $g_s <1$, $\Delta \rho \sim
\rho_H$. In the Hagedorn regime long strings are predominant. This
provides another hint that long strings are produced.

We can estimate the length of the F-strings as follows. A $Dp$-brane
is expected to decay inhomogeneously to $D0$ branes and then to
closed strings \cite{Lambert03}. A $D0$-brane has an energy $E_{D0}
= m_s/g_s$. If each $D0$ decays to a single string, the length of
the string would be $\ell = E_{D0}/\mu_F$.  Using $g_s = 0.1 $, we
find
$\ell = 60 \ell_s$.  If however  clusters of say $10^8$ D-particles
condense to form one F-string then instead $\ell \approx 10^{10}
\ell_s$.  This would compare quite favorably with the horizon size
$\ell_H$ at the brane inflation energy scale $m_{inf}$ since $\ell_H
\sim (\frac{m_P}{m_{inf}})^2 \ell_P$ \cite{Kolb}. (The Friedmann
equation gives $t \sim m_P/T^2$ where the temperature scales as $T
\sim m_{inf}$.)  If $m_{inf} \approx 10^{-4} m_P$ as in the KKLMMT
model and $\ell_p \sim \ell_s$ then the length of such strings is
around the size of the horizon, $\ell \sim \ell_H$ and they are
effectively infinite.


\section{How to make cosmologically viable cosmic superstrings}\label{suppression}

If the string scale mass $m_s \equiv 1/ \sqrt{\alpha'}$ is of order
the Planck mass $m_P$ and $g_s$ is not very small (say $g_s \sim
1/4$), then the four dimensional F and D string self gravity is $G_4
\mu_{F} \sim {\cal{O}}(g_s^2)$ and $G_4 \mu_{D1} \sim {\cal{O}}(g_s)
$.  However, current cosmic microwave background measurements have
placed an upper bound on the self gravity of line-like defects of
$G\mu < 10^{-6} $. Thus, without new ideas, networks of cosmic sized
F or D strings are thought to be unrealistic.

\subsection{Warped geometries suppress the string tension}

As is now familiar from the Randall Sundrum story one way to make
superstrings much lighter is to make the 4D constants dependent on
the extra dimensions \cite{rs1,rs2}. An innocuous way to do this is
by warping the 4D metric as follows

\beq ds^2 = G_{MN} dx^M dx^N = e^{2A(y)} g_{\mu \nu}(x) dx^{\mu}
dx^{\nu} + e^{-2A(y)}\tilde{g}_{mn}(y) dy^m dy^n
\label{warpedansatz} \eeq

\n where $g_{\mu \nu}(x), \tilde{g}_{mn}(y)$ are the metrics of the
4 large dimensions $x^{\mu}$, and the 6 extra dimensions $x^m$
respectively. The four dimensional warp factor is $e^{2A(y)}$. The
measure factor $\sqrt{|G_{MN}|}$ in the action then breaks up as
$\sqrt{|G_{MN}|} = e^{-2A(y)} \sqrt{|g(x)|} \cdot
\sqrt{|\tilde{g}(y)|}$. We then integrate over $y$. Then $m_P^2$ in
$S = m_P^2\int \sqrt{|G|}d^{10}x(R + \cdots)$ becomes $ m_P^2 \int
d^6y e^{-2A(y)} \sqrt{|\tilde{g}(y)|} \approx m_P^2$ if $A(y)$ is
linear in $y$ and $\sqrt{|\tilde{g}(y)|} \sim {\cal{O}}(1)$. Thus
warping doesn't appreciably change the four dimensional metric
determinant. However, warping changes the 4D metric from $g_{\mu
\nu}$ to $e^{2A(y)} g_{\mu \nu}$. Thus quantities depending on the
metric like the energy momentum tensor $T_{\mu \nu}$ for a domain
wall/brane/D-string change as \cite{fluxcompact}

\begin{eqnarray}  T_{\mu \nu}  = -\mu_{D1} g_{\mu \nu} \delta^8(x,y)
 \quad \Longrightarrow  \quad -\mu_{D1} e^{2A(y)}g_{\mu \nu} \delta^8(x,y)
\label{EMtensor}
\end{eqnarray}

 We
see that the tension is redshifted by the factor $e^{2A(y)}$.  At
certain points $\{y^m\}$ on certain compact manifolds we can
engineer  $A \sim -9$. At such points the tension of a F or D string
is reduced to

\begin{eqnarray} \mu_{(p,q)} \simeq  10^{-8} m_p^2 \ \ \ \ & \Longrightarrow &  \ \
\ G_4 \mu_{(p,q)} \sim 10^{-8} \end{eqnarray}

\begin{figure}
\begin{center}
\includegraphics[width= 2.7in]{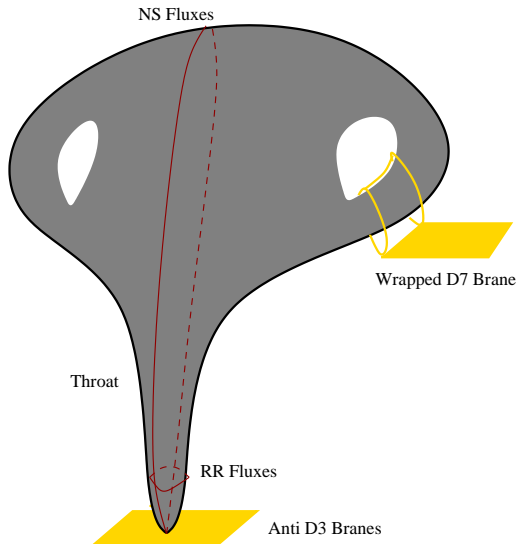}
\end{center}
  \caption{A finite throat is ``attached'' to a 6D Calabi-Yau.  Branes or antibranes can
  sit at the bottom of the throat, and other branes like $D7$'s can wrap other cycles of the CY.
  ({\em From} \cite{quevedo}.)}\label{throat}
\end{figure}

Such large redshifts can occur at the bottom of a gravitational
potential like a throat in a compact manifold, see figure
\ref{throat}.

Manifolds with throats like these can be engineered in a variety of
ways.  We now give several examples.

\subsection{Warped example 1: Anti-de Sitter Space}

Anti de Sitter space has large warping near $r \sim 0 $
\cite{adscft1,johnson}

\beq ds^2_{AdS} = \frac {r^2}{R^2} (-dt^2 + d \vec{x}^2) +
\frac{R^2}{r^2} dr^2 \eeq

\n where $R$ is the curvature radius of AdS.  The standard Randall
Sundrum approach uses this space and cuts off AdS at some $r_0$ and
$r_{max}$, such that  $r_0 <r < r_{max}$.  The cutoff is performed
by placing a brane at $r_0$ -- the "standard model" brane, and by
placing a brane at $r_{max}$ -- the "Planck brane." Mass scales on
the standard model brane are severely redshifted by
$(\frac{r_0}{R})^2  \ll 1$.

\subsection{Warped example 2: near horizon limit of coincident branes}

Earlier we wrote down the metric of a $D1$ brane in
(\ref{D1sugrasolution}).  $N$ coincident extremal (i.e.
supersymmetric) D3-branes have a similar looking metric
\cite{johnson,ortin}

\begin{eqnarray} ds^2 & = & h(r)^{-1/2} (-dt^2 + dx_i dx^i) + h(r)^{1/2} (dr^2 +r^2
d\Omega_{8-p})\nonumber\\
h(r) & = & 1 + \frac{4 \pi N g_s \alpha'^2}{r^4} \label{d3brane}
\end{eqnarray}

We can study the small $r$ region of (\ref{d3brane}) by taking the
``near horizon limit." This is obtained by taking the $\alpha'
\rightarrow 0$ limit (which decouples the brane from the bulk) while
scaling $r$ at the same time such that the variable $u \equiv
r/\alpha'$ is meaningful. If we regard $r$ as the distance between
parallel branes the mass of the strings connecting the branes is $u$
which also controls the values of gauge theory quantities on the
brane like the expectation value of the Higgs. Thus the latter
condition keeps the mass of these strings and Higgs vevs finite as
the branes become coincident at $r=0$.

In this limit as $r \rightarrow 0$ we can drop the "1+" in the
harmonic function $h(r)$. The $N$ D3 branes' metric then becomes
warped  AdS times a sphere $S^5$.  In the new coordinate $u$ the $N$
coincident $D3$s' metric is

\beq \frac{ds^2}{\alpha'} =  \frac{u^2}{L^2} (-dt^2 + dx_i dx^i) +
\frac{L^2}{u^2}( du^2 + u^2 d\Omega_5^2). \label{nearhorizon} \eeq

\n The radius of the sphere in string units is $ L = (4\pi N
g_s)^{1/4}$.

\begin{figure}
\begin{center}
\includegraphics[width= 2.7in]{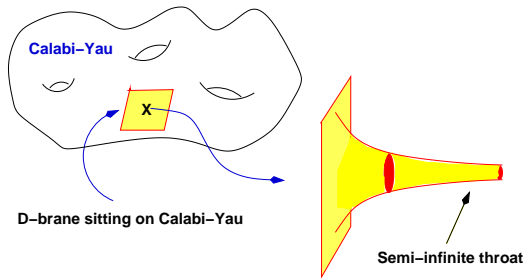}
\end{center}
  \caption{A brane sits on the Calabi-Yau at X. A semi-infinite throat branches off
  at X.}\label{verlindefig}
\end{figure}

The massive warping as  $u \rightarrow 0 $ means the brane sits at
the bottom of an infinite throat and all energies are infinitely
redshifted there, see figure \ref{verlindefig}.  Chan, Paul and
Verlinde (CPV) and others implemented this model in a real
compactification \cite{verlindemodel}. CPV placed a brane  at a
point on a compact Calabi-Yau. From the point of view of the
transverse directions, near the brane a semi-infinite throat appears
off the Calabi-Yau, see figure \ref{throat}.

However, this model is unsatisfactory because one would like a
finite throat producing severe but finite warping.  Also, in the CPV
model, the D3 brane could be anywhere. No potential fixing its
position on the Calabi-Yau appears.  This adds arbitrariness to the
model.

\subsection{Warped example 3: flux compactifications}

 Giddings, Kachru and
Polchinski (GKP) attempted to fix the problems of the Verlinde model
by transplanting a result of Klebanov and Strassler (KS) into the
Verlinde model. Klebanov and Strassler showed how to produce heavily
warped finite throats. Specifically, they showed that if certain
compact cycles, notably $S^3$'s and $S^2$'s, degenerated at certain
points, some of those cycles could be blown back up by threading
flux through the cycles. The geometry near the resolved singular
points is warped and the throat resulting from the warping is
finite. The fluxes act as a positive pressure expanding the
degenerating cycles and reduce the amount of supersymmetry. They
also generate a potential for various scalar fields which
parameterize the size and shape of the compact manifold.

\subsubsection{The conifold and resolution of its conic singularity}

\begin{figure}
\begin{center}
\includegraphics[height=2.7in]{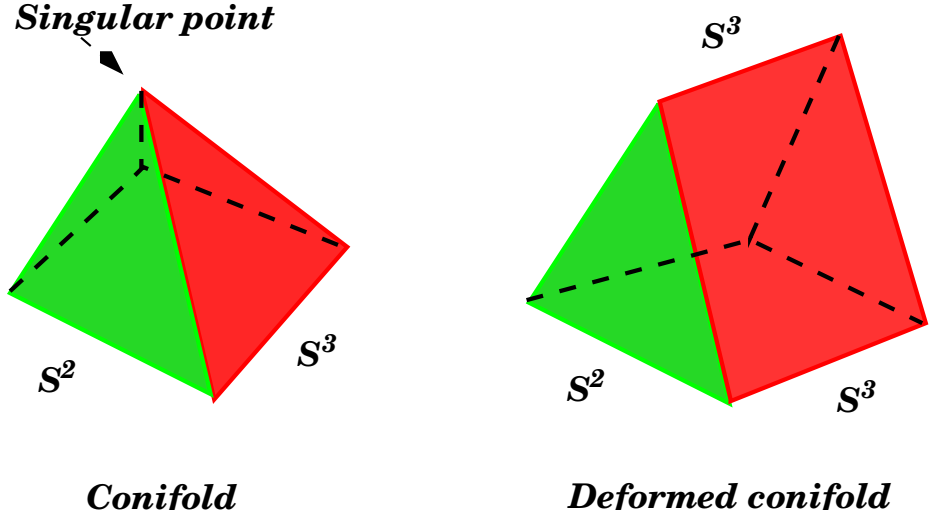}
\caption{A {\em conifold} can be de-singularized by blowing up the
$S^3$. The new space is the {\em deformed conifold}.} \label{coni}
\end{center}
\end{figure}

 A {\em conifold} is a  singular manifold which is topologically a cone
 and is an
  $S^3\times S^2$.  It becomes singular because  the $S^3$ and
$S^2$ shrink to zero size \cite{candelas1,candelas2, polchinskibook,
randjbar}. See figure \ref{coni}. If the singularity is
desingularized by blowing up the $S^3$ it becomes a {\em deformed
conifold}.

Conifold singularities are the most generic singularities in Calabi
Yau compactifications.  Calabi-Yau manifolds are often defined by
complex algebraic curves like $f(w_1,...,w_4) = 0$. The conifold is
singular because  $\partial_{w_i} f(w_1,...,w_4) = 0$ for all $i$.
However, it is not  ``that singular" because the matrix of second
derivatives is nonzero, $\partial_{w_i} \partial_{w_j}
f(w_1,...,w_4) \neq 0$. Near the singularity  $f(w_1,...,w_4)$ can
be written as

\beq f(w_1,w_2,w_3,w_4) =  w_1^2 + w_2^2 +w_3^2 + w_4^2 = 0
\label{conifold} \eeq.

\n The origin (0,0,0,0) is singular. Note (\ref{conifold}) defines a
6D manifold and is a cone because if {\mbox{\boldmath $w$}} is on
the conifold then so is $ \lambda {\mbox{\boldmath $w$}}$ for
$\lambda \in \IC$. To understand what the base of the cone looks
like, we first write the $w_i$ in terms of real coordinates $w_i =
x_i + i y_i$.  Equation (\ref{conifold}) becomes
${\mbox{\boldmath$x$}}^2 - {\mbox{\boldmath $y$}}^2 = 0 $ and $
{\mbox{\boldmath$x$}} \cdot {\mbox{\boldmath$y$}} = 0 $.  We then
intersect (\ref{conifold}) with a 7-sphere centered at the apex of
the cone: $|w_1|^2 + \cdots + |w_4|^2 = r^2$ which is equivalent to
${\mbox{\boldmath$x$}}^2 + {\mbox{\boldmath $y$}}^2 = r^2 $.  At the
intersection of $f(w_1,w_2,w_3,w_4)$ and the $S^7$ we find

\begin{eqnarray}
x_1^2 + x_2^2 + x_3^2 + x_4^2 =\frac{r^2}{2} \;\;& x \cdot y = 0 &
\;\;y_1^2 + y_2^2 + y_3^2 + y_4^2 = \frac{r^2}{2}
\end{eqnarray}

The first of these defines an $S^3$.  The second defines a surface
through $(x,y)$ space, implying $y$ is orthogonal to $x$, or
equivalently that $\{y\}$ is the coordinate of a fiber over the base
spanned by $\{x\}$.  The last equation  defines an $S^3$.  However
the plane $ x\cdot y$ slices the $S^3$, picking out an $S^2$.  Thus
(\ref{conifold}) describes an $S^2$ fiber over an $S^3$.  However, a
useful fact is that all such bundles over $S^3$ are trivial
\cite{candelas1}.  Thus the bundle is globally a product and the
conifold is simply an $S^3 \times S^2$, as shown in figure
\ref{coni}.

Suppose we now deform the conifold by a real parameter $z$.

\beq f(w_1,w_2,w_3,w_4) = z . \eeq

\n Then $z$ controls the size of the $S^3$ cycle which we denote by
$A$. $z$ can be defined by the integral of an analytic function
$\Omega$ with 3 indices (better known as the holomorphic 3-form)
over the $A$ cycle

\beq z  =  \int_{A} \Omega \eeq

\n Thus $z$ vanishes when the $S^3$ cycle $A$ vanishes.  Turning on
a finite $z$ turns the singular conifold into the less singular
deformed conifold \cite{polchinskibook,greene}.

When defining a basis of 3-cycles, we can restrict ourselves to
non-intersecting cycles $A_I$ and cycles called $B_I$ which
intersect $A_I$ once such that $A_I \cap B_J = \delta_{IJ}$. Note
the $B_J$ are nonintersecting, $B_J \cap B_K = 0$.  In this case
with a single $A$ cycle, a basis of 3-cycles spanning the 6D
extra-dimensional space ${\cal{M}}_6$ is formed by an $A$ cycle and
a $B$ cycle which intersects $A$ once. The $B$ cycle defines a
coordinate $G$ conjugate to $z$

\beq G = \int_{B} \Omega \eeq

If we go around the singularity $z = 0$ in a circle there is no
reason to expect that $G$ should return to itself, $G\rightarrow G$.
In fact if we go around the singularity along the curve $B$ we
actually end up traversing the curve  $B + A$. Hence,

\beq G \rightarrow  G + z \eeq

\n Such behavior can be mimicked by

\beq G = \int_B \Omega = \frac{z}{2 \pi i} \ln z + \cdots {\rm
nonsingular\, terms} \eeq

\n The branch cut of the log means

\beq G  \rightarrow \frac{z}{2 \pi i} \ln z  + \frac{z}{2 \pi i}
 \cdot 2 \pi i = G + z
\eeq

 \n  as desired if $z= |z| \exp{(i \phi)} \rightarrow |z| \exp{i(\phi + 2\pi)}$.

\subsubsection{Calculation of the warp factor of the deformed
conifold} \label{sectionfluxcompact}

The superpotential, from which the scalar potential emerges, is the
integral of the three form $G_3 = F_3 - \tau H_3$ over ${\cal{M}}_6$
\cite{fluxcompact}.  As usual, $\tau$ is the dilaton-axion  in
(\ref{dilatonaxion}), $F_3$ is a 3-form field strength for D-strings
and $H_3$ is the 3-form field strength for fundamental strings.

\begin{eqnarray}
W & = & \int_{{\cal{M}}_6} (F_3 - \tau H_3) \wedge \Omega \nonumber\\
& = & \left ( \int_A F_3 \right ) \wedge \left (\int_B \Omega \right
)  + \left ( \int_B \tau H_3 \right ) \wedge \left (\int_A\Omega
\right ) \label{superpot}
\end{eqnarray}

Note the sign change $\int_A \wedge \int_B = -\int_B \wedge \int_A$.
Suppose we place $M$ units of $F_{3}$ flux on the $A$ cycle and $-K$
units of $H_{3}$ flux on the $B$ cycle such that

\begin{eqnarray}
\frac{1}{2 \pi \alpha'}\int_A F_{3} = 2 \pi M  & {}& \frac{1}{2 \pi
\alpha'}\int_B H_{3} = -2 \pi K \label{flux}
\end{eqnarray}


\n Plugging (\ref{flux}) and the definitions of $z$ and $G$ into
(\ref{superpot}), the superpotential becomes

\beq W = (2 \pi)^2 \alpha' ( M G - K \tau z) \eeq

Thus a nonzero superpotential and nonzero flux quanta $M,K$ can
deform the conifold by turning on a finite deformation parameter
$z$.

Supersymmetry requires that the ground state energy be zero. The
scalar potential, $V$ is proportional to $(D_z W)^2$, where  $D_z$
is the covariant derivative, $D_z =
\partial_z +
\partial_z {\cal {K}}$.\footnote{The potential is actually
$V = \kappa^2_4 e^{{\cal{K}}}( {\cal{K}}^{i \bar{j}} D_i W
\bar{D}_{\bar{j}} \bar{W} - 3 |W|^2)$ where ${\cal{K}}_{i \bar{j}} =
\partial_i \partial_{\bar{j}} {\cal{K}}$ and the $(i, {\bar{j}})$ sum
is over all moduli fields. In this model there is only one volume
modulus  ({\em Kahler modulus}) $\rho$ whose Kahler potential is
${\cal{K}}_{\rho} = -3 \ln (i(\bar{\rho} -\rho))$.  Then
${\cal{K}}^{\rho \bar{\rho}} D_{\rho} W \bar{D}_{\bar{\rho}} \bar{W}
$ cancels the $-3|W|^2$. Thus $V = \kappa^2_4 e^{{\cal{K}}}
{\cal{K}}^{i \bar{j}} D_i W \bar{D}_{\bar{j}} \bar{W}$ where in the
$(i, {\bar{j}})$ sum, $(i,{\bar{j}}) \neq (\rho, \bar{\rho}) $.}
  Here ${\cal{K}}$ is the so-called Kahler
potential and will be irrelevant here because $\partial_z {\cal{K}}$
is much smaller than the other terms near the conifold point $z \sim
0$ \cite{freythesis}.  Thus using  $\tau = i/g_s + \cdots$

\beq 0 =\sqrt{V} \sim  D_z W \sim \frac{M}{2 \pi i } \ln z - i
\frac{K}{g_s} + {\cal{O}}(1) \eeq

\n which yields

\beq z \sim e^{-\frac{2\pi K}{M g_s}} \eeq

We can now estimate the warp factor $e^{2A}$.  The $H_3$ and $F_3$
fluxes can be related to $D3$ brane charge by $Q_{D3} \sim  - \int
H_3 \wedge F_3$.\footnote{The generalized field strength
$\tilde{F}_5$ which appears in the IIB supergravity action
(\ref{iibaction}) has a Bianchi identity $d\tilde{F} = H_3 \wedge
F_3 + 2g_s \kappa_{10}^2 \mu_{D3} \rho_3$ where $\rho_3$ is the $D3$
brane charge density due to $D3$ branes, O3 planes (which possess
negative $D3$ charge) and induced $D3$ charge via wrapped $D7$
branes, etc.. The integrated Bianchi density is then
$\int_{{\cal{M}}_6} H_3\wedge F_3 + 2 g_s \kappa^2_{10} \mu_{D3} Q_3
= 0$.}  Thus the threading of flux through the $A$ and $B$ cycles
can equivalently be thought of as the positioning of $N = MK$ $D3$
branes at the bottom of the conifold throat.   The warping can then
by thought of as due to $N$ $D3$ branes just as in
(\ref{nearhorizon}) but with distance to the branes $r$ cut off at
some minimum (because of the resolution of the conifold
singularity). The warping due to the branes $\sim 1/\sqrt{h(r)} \sim
r^2$, is then equivalent  to some $e^{2A}$ warp factor $\grave{a}$
$la$ the flat space version of (\ref{warpedansatz}).  It turns out
that $r$ is related to the conifold coordinates
$|{\mbox{\boldmath$w$}}|$ as $r \sim |{\mbox{\boldmath$w$}}|^{2/3}$.
Our resolution of the conifold singularity cuts the conifold off at
$w_i^2 \sim z$. Thus the minimum value of the warp factor is

\beq e^{2A}  |_{conifold \,\, tip}  \sim r^2 \sim
|{\mbox{\boldmath$w$}}|^{4/3} \sim z^{2/3} \sim e^{-\frac{4 \pi
K}{3M g_s}} \label{warpattip} \eeq

For reasonable values of $K$ and $M$ and $g_s$,  a suppression of
the string tension by at least $10^{-8}$ is possible.

One important point to note is that there is a minimum value of $M$
if brane-antibrane annihilation at the bottom of the conifold throat
is to occur and produce $(p,q)$ strings. It was shown in
\cite{kachruinstanton} that unless

\beq M \gsim 12 \label{minM} \eeq

\n  that a antibrane part of the brane-antibrane pair at the end of
the throat would be unstable and dissolve into flux.  Thus if $M <
12$ there would be no brane-antibrane pair whose annihilation would
produce $(p,q)$ strings.

\subsection{Why do warped models require orientifolds?}

Orientifolds are unfamiliar objects to cosmologists.  They are fixed
planes of negative tension and hence are admittedly, bizarre. Below,
without explaining why they are reasonable objects to work with, we
review GKP's argument of why they are generically required by warped
compactifications.

The relevant part of the Type IIB supergravity action is

\beq S_{IIB} = \frac{1}{2\kappa_{10}^2} \int d^{10} x \sqrt{|g|}
\left ( R - \frac{\partial_M \tau \partial^M \bar{\tau} }{2 ({\rm
Im} \; \tau )^2} - \frac{|G_{3}|^2}{12 \; {\rm Im} \;  \tau} -
\frac{|\tilde{F}_5|^2}{4 \cdot 5!} \right ) + S_{CS} + S_{loc}
\label{iibaction}
\eeq

\n Conceptually the above action is very similar to
(\ref{effaction}). However we have written it in the Einstein frame.
We have added the F-string field strength $H_3$, 5-form field
strength $F_5$ of $D3$ branes, and the D-instanton field strength
$\partial A$ and combined all of the fields in a very elegant
SL($2,\IZ$) way \cite{polchinskibook}.  A price of this elegance is
that Im$\, \tau$ appears in the denominator of 2 terms. Here
$S_{CS}$ is the Chern-Simons part of the action and $S_{loc}$ is the
local part due to the presence of matter sources like branes,
orientifold planes, etc.

In (\ref{iibaction}), $|G_3|^2$ means $G_3 \wedge * G_3$.
 Also in (\ref{iibaction}) $\tilde{F}_5$ is self-dual, $\tilde{F}_5 =
* \tilde{F}_5$, and is defined by  $\tilde{F}_5 = F_5 -\frac{1}{2}
A_2 \wedge H_3 + \frac{1}{2} B_2 \wedge F_3$.  An ansatz for
$\tilde{F}_5$ satisfying the Bianchi identity and self-duality is
$\tilde{F}_5 = (1 + *) d\alpha \wedge dx^0\wedge dx^1\wedge dx^2
\wedge dx^3 $.  Here $\alpha$ is a scalar function on the compact
space.

The trace-reversed Einstein equation is

\beq R_{M N} = \kappa_{10}^2 \left ( T_{MN} - \frac{1}{8} g_{MN} T
\right ) \label{tracerev} \eeq

\n Recall that $M,N$ are the 10D indices, $\mu,\nu$ are the
noncompact indices and $m,n$ are the compact directions' indices.
The noncompact part of the R.H.S. of (\ref{tracerev}) is

\beq T_{\mu\nu } - \frac{1}{8} g_{\mu \nu} T = - g_{\mu \nu} \left (
\frac{G_{mnp} \bar{G}^{mnp}}{48 {\rm Im} \tau} - \frac{e^{-8A}}{4}
\partial_m \alpha \partial^m \alpha \right ) + \kappa_{10}^2 \left (
T_{\mu \nu}^{loc} - \frac{1}{8} g_{\mu \nu} T^{loc} \right ) \eeq

\n  The Ricci scalar, $R_{\mu \nu}$ for the warped compactification
(\ref{warpedansatz}) is

\beq R_{\mu \nu} = -\eta_{\mu \nu} e^{4A} \tilde{\nabla}^2 A .
\eeq

\n Thus,

\beq \tilde{\nabla}^2 A = e^{-2A} \frac{G_{mnp} \bar{G}^{mnp}}{48
{\rm Im} \tau} + \frac{e^{-6A}}{4}
\partial_m \alpha \partial^m \alpha + \frac{\kappa_{10}^2}{8} e^{-2A} (T_m^m - T_{\mu}^{\mu})
\label{match} \eeq

If we integrate the above equation over the compact manifold
${\cal{M}}_6$, then the left hand vanishes because it is a total
derivative.  The right side apart from the last term is positive
semi-definite.  Thus for a nonvanishing warp factor $A$, the only
way of matching both sides of (\ref{match}) is if $T_m^m -
T_{\mu}^{\mu} < 0$. Now using (\ref{EMtensor}), if $T^{loc}$
originates from $Dp$ branes which fill the four non-compact
directions and wrap a cycle $\Sigma$ composed of $p-3$ compact
directions, then

\begin{eqnarray}
 T^{loc}_{\mu \nu} & = & -e^{2A} T_p g_{\mu \nu} \delta(\Sigma) \; \;
\Rightarrow  \; \;T_{\mu}^{\mu} = -4 T_p \delta(\Sigma)
\nonumber\\
 T^{loc}_{m n} & = &- \Pi_{mn}^{\Sigma}  T_p  \delta(\Sigma) \; \;
\ \ \; \Rightarrow \; \; T_{m}^{m} = -(p-3) T_p \delta(\Sigma)
\end{eqnarray}

Here $\Pi_{mn}^{\Sigma}$ is the projection of the compact
directions' metric $e^{-2A} \tilde{g}_{mn}$ onto the directions
spanned by $\Sigma$.  Thus,

\beq  0 = \int \tilde{\nabla}^2 A = {\rm positive\; terms} + (7-p)
T_p \label{intwarp} \eeq

If $p<7$ then the only way to satisfy (\ref{intwarp}) is if $T_p
<0$.  Thus negative tension objects like O3 planes are generically
required by warped compactifications.\footnote{If higher derivative
terms are ignored.}

\subsection{Consequences of orientifolds: unstable F and D strings}

The closed string field $X^{\mu}$ can be decomposed into a
right-moving part $X_+$ with worldsheet coordinate $\sigma_+$ and a
left-moving part $X_-$ with worldsheet coordinate $\sigma_-$.  Thus
$ X^{\mu}(\sigma_+,\sigma_- ) = X_+(\sigma_+) + X_-(\sigma_-)$. A
worldsheet parity operation $\sigma_+ \leftrightarrow \sigma_-$,
which is mediated by the operator $\Omega$ exchanges the right and
the left-moving parts and is a symmetry of bosonic and Type IIB
string theory because both are right-left symmetric. When this
symmetry is combined with a spacetime symmetry like a spacetime
reflection $R$ taking $X^{\mu} \rightarrow -X^{\mu}$, then at the
fixed points of the combined \textsf{world sheet parity $\times$
spacetime symmetry}, we find solitonic kink-like objects.  A
scattering calculation shows that they have negative tension and are
not dynamical.  These are known as  {\em orientifold fixed planes}.
For example, in 10D string theory, if $X^9$ is noncompact and we
identify  by the $\IZ_2$ symmetry, $X^9 \cong -X^9$, then a 8D
domain wall orientifold will be located at $X^9 =0$.  A string at
positive $X^9$ will be identified with an orientation-reversed
string at negative $X^9$. This general reflection property of
orientifolded string theories implies that there is a mirrored
reflection of every string or brane. In also means for example, that
if a geometry has a throat like the conifold warped throat, that
there exists a mirror throat.  Thus if there exist strings and
D-strings sitting at the bottom of a conifold throat, then there
exist mirror states -- strings with opposite orientation and
anti-D-strings sitting at the bottom of a mirror conifold throat.

In supersymmetric theories the orientifold projection $R \Omega$ is
modified by fermions to $R \Omega (-1)^{F_L}$. $F_L$ is the fermion
number of left-handed fermions. Thus $R \Omega (-1)^{F_L}$ projects
onto states with an even number of left-handed fermions.

$\Omega$ will project out various massless string modes. Left-handed
level-1 creation operators are denoted by $\alpha^{\mu}_{-1}$ and
right-handed level-1 creation operators by
$\tilde{\alpha}^{\nu}_{-1}$.   Then we can form the string state
$B_{\mu \nu}\alpha^{\mu}_{-1} \tilde{\alpha}^{\nu}_{-1} | 0 \rangle
$ where  $B_{\mu \nu}$ is the F-string antisymmetric gauge field.
Then under worldsheet parity, ${\sf right} \leftrightarrow {\sf
left}$ exchanging $\tilde{\alpha}^{\nu}_{-1} \leftrightarrow
\alpha^{\nu}_{-1}$.  But $B_{\mu \nu}$ is antisymmetric under $\mu
\leftrightarrow \nu$ and thus the state's $\Omega$ eigenvalue is -1:

\beq \Omega: B_{\mu \nu}\alpha^{\mu}_{-1} \tilde{\alpha}^{\nu}_{-1}
| 0 \rangle  \rightarrow - B_{\mu \nu}\alpha^{\mu}_{-1}
\tilde{\alpha}^{\nu}_{-1} | 0 \rangle  \eeq

The reflection eigenvalue of $B_{\mu \nu}$ is $(-1)^2 = +1$ since
$B_{\mu \nu}$ has two spacetime indices.  The $(-1)^{F_L}$
eigenvalue is +1 since the state $B_{\mu \nu} \alpha^{\mu}_{-1}
\tilde{\alpha}_{-1}^{\nu} | 0 \rangle$ involves no fermionic
operators.  The orientifold projection $R \Omega (-1)^{F_L}$ keeps
states with +1 eigenvalue.  Thus $B_{\mu \nu}$ is projected out.
However, its field strength $H_{\lambda \mu \nu}
 = \partial_{[\lambda} B_{\mu ] \nu}$ has reflection eigenvalue $(-1)^3$ and
thus is retained.


With no gauge field $B_{\mu \nu}$ to charge the F-strings it would
seem that F-strings are not allowed.  This is not correct. All it
means is that the net F-string charge vanishes.  For every
positively oriented string there must be a negatively oriented
string.  Thus the net string orientation vanishes. This is expected
since orientifolded theories are unoriented.  String ``orientation"
can disappear when positive and negative oriented strings combine.


The D-string gauge field $A_{\mu \nu}$ has $\Omega$ eigenvalue +1
because the $A_{\mu \nu}\psi_0^{\mu} \tilde{\psi}_{0}^{\nu} | 0
\rangle $ state is constructed with right and left-handed  fermionic
zero-mode operators $\tilde{\psi}_0^{\nu}$ and $\psi_0^{\mu}$ which
anticommute.  The $R$ eigenvalue is +1 because $A_{\mu \nu}$ has two
indices.  But the $(-1)^{F_L}$ eigenvalue is -1 since the state has
one left-handed fermion.  Thus $A_{\mu \nu}$ is projected out but
its field strength $F_{3} = dA_{2}$ is retained.  Again this only
means that the net D-string charge is zero and that  $F_{3}$ fluxes
can for example thread suitable 3-cycles of the compact space.


Thus in flux compactifications, F-strings and D-strings are
unstable.  However, to annihilate they must combine with their
mirror images (anti-D-strings and oppositely oriented F-strings) in
a mirror throat.  One might expect the whole system to be unstable
and that fundamental strings and branes in one throat will attract
and annihilate anti-fundamental strings and anti-branes in the
mirror throat -- leading to no cosmic strings. However, the heavy
warping in the throats severely hampers annihilation.

\subsection{Why stable long strings must be non-BPS}

Suppose now that $A_{\mu \nu}$ and $B_{\mu \nu}$ were not projected
out.  Then any strings coupling to these gauge fields will be axion
strings as axion strings couple to form fields.  Axion strings are
known to bound domain walls and once loops of axion string become
sufficiently large the domain wall energy dominates the string
energy and it is energetically favorable for the strings to stop
growing. If the domain wall tension is $\sigma$, the energy of a
$(p,q)$ string which bounds a domain wall plus the energy of the
domain wall is $\sim 2  \pi R \mu_{(p,q)} + \pi R^2 \sigma$. The
domain wall term dominates when $R \gsim \mu_{(p,q)}/\sigma$.  This
is the maximum size a string can grow to and is tiny compared to
astrophysical sizes unless the domain wall tension is
extraordinarily small. Hence the presence of gauge fields charging
the strings precludes long strings. Below we briefly show how
axionic domain walls arise in type II string theories and why Type
II strings are axionic.

A Type II string is charged by the 2-form potential $B_2$. In 4D
this is dual to a 4D axion $\phi$ such that $H_3 = dB_2 = \star d
\phi + \cdots$. The line integral of $d\phi$ around the contour
$\gamma$ which circles around a string and bounds a 2D surface $S$
is then

\beq \int_{\gamma} d\phi  = \int_{\gamma}  \star H_{3}   = \int_S d
\star H_{3} = 2 \pi \int_S  \delta^2(x_{\perp}) = 2 \pi
\label{axion} \eeq

\n where we used Gauss' law and the fact that the string is a source
for $H_{3}$ and thus $d\star H_{3} = 2 \pi \delta^2(x_{\perp})$.
Here $x_{\perp}$ are the coordinates perpendicular to the string.
Thus (\ref{axion}) implies that the axion $\phi$ is multivalued.
Because $\phi$ changes by $2 \pi$  around a string the curve
$\gamma$ must pierce a domain wall if the an axion potential is
generated by instantons/susy breaking. We can understand this as
follows.

Recall that for a Yang Mills instanton:  $\frac{1}{4} \int \tr(F
\wedge \star F) = 8 \pi^2n$ and $\frac{1}{4} \int \tr(F \wedge F) =
64 \pi^2n$ and the partition function is  $Z = \int [dA d\psi]
e^{-i(\frac{1}{4g^2} \int \tr(F\wedge *F) + \frac{\theta}{64 \pi^2}
\int \tr(F \wedge F) + \cdots )}$ where $S_{\psi}$ is the action of
the $\psi$ fields. When Euclideanized: $t \rightarrow -it$ and
$F_{0i} \rightarrow i F_{0i}$ and the partition function becomes $Z
= \int [dAd\psi] e^{-(\frac{1}{4g^2} \int \tr(F\wedge *F) +
i\frac{\theta}{64 \pi^2} \int \tr(F \wedge F) + S_{\psi} )} = \sum_n
e^{-8\pi^2 n/g^2-in\theta} \int [d\psi]e^{-S_{\psi}}$.  The least
action $\theta$ configuration is when $\theta=0$.  The instanton
generates a potential for the axionic $\theta$ angle which therefore
has a minimum at $\theta =0$. But because the partition function is
periodic, $\theta \cong \theta + 2 \pi$, the potential is periodic
with minima at $2n\pi$.

In the Type II fundamental string case, $B_2$ couples magnetically
to a five brane known as the NS5 brane which is charged by the gauge
field $\tilde{B}_6$ which has a field strength $\tilde{H}_7 = *H_3 =
*dB_2$. If a {\em Euclidean} NS5 brane wraps the six compact
directions then it will act like an instanton and produce a periodic
potential for the 4D axion $\phi$ in (\ref{axion}).  A Euclidean NS5
brane which wraps the 6D internal geometry will not modify the
internal geometry and can be characterized by the action
\cite{kiritsis}

\begin{eqnarray}
S  & = & \mu_{NS5} \int_{NS5}\! \! \! \sqrt{g_{\mu \nu} + \cdots}
+ \mu_{NS5} \int_{NS5} \tilde{B}_6  \nonumber \\
 & \rightarrow &  \mu_{NS5} \int_{NS5} \sqrt{g_{\mu \nu} + \cdots}
 + i \mu_{NS5} \int_{NS5} \tilde{B}_6  \nonumber \\
 &   = & \frac{2 \pi |m|}{g_s^2}\textsf{vol}_{NS5} + 2 \pi i m \phi.
 \label{NS5instanton}
 \end{eqnarray}

\n In the second line we Euclideanized ($\tilde{B}_{012345}
\rightarrow i \tilde{B}_{012345}$).  We used $\mu_{NS5} =(2\pi)^{-5}
\alpha'^{-3}g_s^2$ and  $\textsf{vol}_{NS5}$ is the 6D wrapped
volume in string units. Also $m$ is the number of times the NS5
wraps the six extra dimensions.  Since $d\tilde{B}_6  = *dB_2$ and
$d \phi = \star d B_2$, if $B_2$ has only 4D functional dependence
$\phi$ and $\tilde{B}_6$ are linearly related. In fact
$\epsilon^{\mu_1\cdots \mu_6} \tilde{B}_{\mu_1\cdots \mu_6}/6! =
\phi$ allowing us to write the last line of (\ref{NS5instanton}).

Therefore as in the Yang-Mills case, the wrapped Euclidean wrapped
NS5 brane will produce instanton corrections generating a periodic
potential as $\phi \cong \phi + 2 \pi$ with vacua at $\phi = 2n
\pi$.  Adjacent vacua separated by $\Delta \phi = 2 \pi$ in the
spacetime picture correspond to a domain wall.  Hence, since $\Delta
\phi = 2 \pi$ in a circuit around a Type II string a domain wall
appears.  More generally, if a $p+1$ dimensional brane wraps a
compact $p-1$ dimensional cycle ${\cal{K}}_{p-1}$ and looks like
string in 4D, closed loops of the string will bound a domain wall.
In this case the form charging the 4D string is $A_{[2]} =
\int_{{\cal{K}}_{p-1}} A_{p+1}$, and in 4D this will be dual to some
axion $\varphi$.  A $(6-p)+1$ dimensional Euclidean brane
magnetically charged by $A_{p+1}$ if wrapped around a $(7-p)$ cycle
which intersects ${\cal{K}}_{p-1}$ will then in the same way produce
instanton corrections and a periodic potential for $\varphi$. Hence
it will produce a domain wall.

\subsection{Annihilation probability of F \& D strings in orientifolded theories}

We now show how warping in orientifolded theories can make strings
stable even though $A_{\mu \nu}$ and $B_{\mu \nu}$ are projected
out.

 The semiclassical amplitude
for string annihilation is given by a Euclidean {\it worldsheet
instanton}. As we discussed in the previous section, Euclidean
instantons are the leading term of the Euclidean partition function
when the action is expanded around a local minimum. The action is
first Wick rotated to ensure that the path integral converges and is
then expanded about a local minimum ($\delta S/\delta \phi =0$ -- a
classical solution) so that

\beq S[\phi] = S[\phi_0] + \frac{\delta S}{\delta \phi} \delta \phi
+ \frac{\delta^2 S}{\delta \phi^2} (\delta \phi)^2 + \cdots \eeq

\n The partition function is then

\beq Z = \int D \phi e^{-S} = e^{-S[\phi_0]} \int D\phi
e^{-S''[\phi_0](\delta \phi)^2}(1 + \delta \phi \cdots )  =
e^{-S[\phi_0]} (\det S''[\phi_0])^{-1/2} + \cdots \eeq

We calculate the D-string anti-D-string annihilation probability.
The calculation of fundamental and $(p,q)$ annihilation amplitudes
are very similar.

\begin{figure}
\begin{center}
\includegraphics[width=3in]{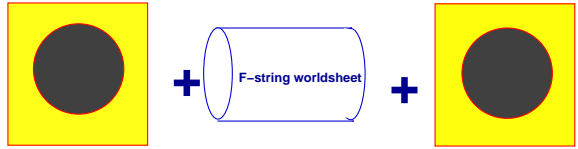}
\end{center}
\caption{For D-strings to annihilate,
  F-strings must connect the D and anti-D strings.}\label{wsinstanton}
\end{figure}

 A D-string can annihilate an anti-D-string if the
worldsheets of the two merge. This can happen if open strings appear
connecting the two branes and  "pull" the two together.  This will
happen if a hole on the D-string worldsheet, a hole on the
anti-D-string worldsheet, and tube of open string world sheet are
created, see figure \ref{wsinstanton}.

The probability to go from a \textsf{D-string worldsheet +
anti-D-string worldsheet} to \textsf{2 punctured worldsheets + a
tube of fundamental string worldsheet} can be thought of as the
conditional probability $P({\rm annihilation}|D1, \bar{D}1)$. Using
the definition of a conditional probability and probability in
quantum theory

\beq P({\rm annihilation}|D1, \bar{D}1) = \frac{P({\rm
annihilation})}{P(D1,\bar{D}1)}  \approx \frac{|e^{-S_2} \sqrt{\det
{S_2^{ \prime \prime} }}^{-1}|^2}{|e^{-S_1} \sqrt{\det {S_1^{ \prime
\prime} }}^{-1}|^2} = A e^{-2B}\eeq

\n where $B = S_2-S_1$ and $A$ is the determinant prefactor
providing the first quantum corrections.  Then difference in action
$B$, of the two configurations is the loss due to the 2 holes of the
D1,${\rm \bar{D}}$1 worldsheets which are cut out and the addition
of a tube of fundamental string worldsheet

\beq  B = \mu_F \cdot L \cdot 2\pi R - 2 \mu_{D1} \cdot \pi R^2 \eeq

\n where $R$ is the radius of the hole and $L$ is the length of the
tube. $R$ is chosen to minimize the tunneling action $B$.

\beq \frac{\partial B}{\partial R} = 0 \;\;\;  \Rightarrow \;\;\;
R_{min} = \frac{L}{2} \frac{\mu_F}{\mu_{D1}} \;\;\; \Rightarrow
B(R_{min}) = \frac{\pi L^2 \mu_F}{2} \left ( \frac{\mu_F}{\mu_{D1}}
\right ) \eeq

Because the D-strings sit at the severely redshifted bottoms of
resolved conifold throats, $\mu_{D1} = \frac{1}{2\pi \alpha'g_s}
e^{2A}$. However, because the F-strings connecting the $D1$s and
$\bar{D}1$s pass through the bulk and much of their length is in the
bulk their tensions are not redshifted. Hence $\mu_F = \frac{1}{2
\pi \alpha'}$ and $\mu_{D1} = \mu_Fg_s^{-1} e^{2A} \sim
\mu_Fg_s^{-1} \cdot 10^{-8}$. If $L \sim 1$ in string units then
\cite{copeland}

\beq P({\rm annihilation}|D1, \bar{D}1) = A e^{-g_s\cdot 10^{8}} \ll
1 \eeq

Hence, even if the $D1$ and its mirror ${ \bar{D}}1$ are separated
by only a few string lengths but are at the bottoms of mirror
throats, the probability of annihilation is extraordinarily small.
Likewise, it is very improbable for fundamental strings to emerge
from one throat and interact with fundamental strings in the mirror
throat. Thus strings and/or D-strings at the bottom of a heavily
warped throat are largely decoupled from what happens in another
throat or the unwarped part of the compactification manifold.

\begin{figure}
\begin{center}
\includegraphics[width=3in]{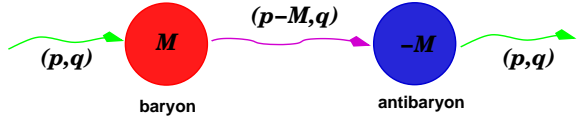}
\end{center}
 \caption{Baryon-antibaryon nucleation. If a $(p,q)$ string ends on $M$, then a $(p-M,q)$
  string must connect the baryon and antibaryon.}
  \label{baryonfig}
\end{figure}

Long strings can also disappear by the pair production of the 4D
baryons introduced in $\S \ref{sectionbaryon}$. This is analogous to
how cosmic strings can break apart by pair production of monopoles
and anti-monopoles. If a 3-cycle $A$ has $M$ units of flux through
it as in (\ref{flux}) and is wrapped by $D3$ branes, then a $(p,q)$
string can break on $A$ into a $(p,q)$ string ``entering" the
baryon, and a $(p-M,q)$ string leaving the baryon. See figure
\ref{baryonfig}.

A baryon and an antibaryon will be connected by a $(p-M,q)$ open
string as shown in figure \ref{baryonfig}.  The endpoint of a string
which ends on the baryon will trace out the boundary of a string's
2D worldsheet.  Because the string worldsheet is invariant under
Lorentz transformations along its surface, the string endpoint will
trace out a curve  (i.e. for a point particle theory this would be
the hyperboloid $x^2 + y^2 + z^2 - t^2 = R^2$).

\beq \sigma^2 - \tau^2 = R^2\eeq

\n where $R$ is some constant.  Wick rotation, $\tau \rightarrow
-i\tau$ turns the non-compact string worldsheet into a circle, since

\beq  \sigma^2 + \tau^2 = R^2 \label{traj} \eeq

 Thus the baryon and antibaryon
will move along a circle just as in the well known Schwinger pair
production process.  The force on a baryon being tugged by two
strings is $F = \mu_{(p,q)} - \mu_{(p-M,q)}$. The baryon action is
then

\begin{equation}
S_B  =  -m\int ds - \int V ds
\end{equation}

\n where $V$ is the potential energy.  After a Wick rotation, $S_B$
picks up a minus sign,

\begin{eqnarray}
S_B^E& = & m \int ds + \int V ds \nonumber \\
& = & m \int ds + \int \partial V \cdot dA \nonumber \\
& = & m \int ds  - (\mu_{(p,q)} - \mu_{(p-M,q)})  \int dA \nonumber \\
& = & m \cdot 2 \pi R - (\mu_{(p,q)} - \mu_{(p-M,q)}) \cdot \pi R^2
\label{baction}
\end{eqnarray}

\n  In the last line we have used the Euclidean baryon trajectory
(\ref{traj}). $m$ is the baryon mass which is the mass of the $D3$
brane wrapped on the $S^3$ \cite{klebanovreview}

\beq m = \mu_{D3} \textsf{Vol}(S^3)  =  \beta \frac{M^{3/2}}{2 \pi
\alpha'^2 g_s} \eeq

\n where \textsf{Vol}$(S^3)$ is a constant $\beta$ times $M^{3/2}$.
Minimizing (\ref{baction}) with respect to $R$ gives,

\beq R = \frac{ m }{(\mu_{(p,q)} - \mu_{(p-M,q)})} \eeq

Thus \cite{copeland}

\begin{eqnarray}
S_B^E & = &  \frac{\pi m^2}{\mu_{(p,q)} - \mu_{(p-M,q)}} \nonumber \\
& = & \frac{\pi [\mu_{D3} \cdot \textsf{Vol}(S^3)]^2}{\mu_F
(\sqrt{p^2 +
q^2/g_s^2} - \sqrt{(p-M)^2 + q^2/g_s^2})} \nonumber \\
& \sim &   \frac{qM^2}{2p-M} \label{baryonrate}
\end{eqnarray}

\n where in (\ref{baryonrate}) we obtained the last line by taking
the $g_s \ll 1$ limit.  As $S_B^E$ can be very small for large $p$,
$(p,q)$ strings can break apart on baryons. However, if $|p| < M/2$
then $S_B^E$ becomes negative. Since $S_B^E = S_2-S_1$ which is the
action of the final state  with a $(p-M,q)$ string and baryon pair
production minus the action of just a $(p,q)$ string, if $|p| <
M/2$, the initial state has larger action than the final state.
Hence baryon pair production does not occur for small $p$ as argued
in (\ref{baryonppcond}).

Now $M$ is constrained by (\ref{minM}) so that $M \gsim 12$.  Thus
low $p$ and $q$ strings will be stable. $\S$\ref{sectionpqnetwork}
discusses how a $(p,q)$ string network ends up being composed of the
lightest strings which are $(\pm 1, \pm 1)$ and $(\pm 1,0)$ and
$(0,\pm 1)$ strings. Breakage on baryons  aids this migration
towards lowest mass strings.

\section{String reconnection probability} \label{reconnection}

The probability that two colliding field theory strings reconnect is
essentially one \cite{shellard} (although see \cite{tong}).
However, the same probability for superstrings is suppressed by a
factor of $g_s^2$ and can thus be much less than one. Thus a
reconnection probability $P < 1$ is one of the distinguishing
features of superstrings. Below we give a simplified account of the
reconnection probabilities calculated in \cite{jonespolchinski}.

The following is a rather interesting application of tree-level
string scattering.  The most important string theory  factors which
show up in $P$ are  $g_s^2$ and a velocity function. The $g_s^2$
factor is the reason why superstrings intercommute much more
infrequently than field theory strings.  The velocity function for
the F-string goes to infinity for small and large $v$ hinting that
$F$ strings tend to intercommute at extreme $v$.  For the D-strings,
the velocity function blows up at high $v$ and reconnection is also
very probable for $ v \sim 0$. Another very important factor in $P$
is a geometrical factor measuring how often ``strings miss each
other'' when moving in the extra dimensions.

\subsection{Reconnection of F-strings}\label{Freconnection}

Two strings which collide will exchange gravitons and massless
particles.  In the $t$ channel and forward scattering small $t$
limit we expect the tree level scattering amplitude to have a pole
at $t=0$. Because we are calculating graviton exchange, the
amplitude should be weighted by Newton's constant $G_d \sim \kappa^2
\sim g_s^2$.
We should also have a factor of $V$ in the denominator if we box
normalize the string wavefunctions. If for convenience we put the
strings on a large 2-torus and a 6D small compact manifold with
volume $V_{\perp}$, then $V = V_{\perp} V_{T^2}$.  Now in the large
energy $s \gg 0$ (and fixed $t$) limit, string amplitudes display
Regge behavior -- i.e. ${\cal{A}} \sim s^{\alpha_0(t)}$ where
$\alpha_0(t)$ is a linear function of $t$.  In this case
$\alpha_0(t) = 1 + \frac{\alpha' t}{4}$. When we analytically
continue from Euclidean to Lorentzian momentum the $s^{\alpha' t/4}$
generates a multiplicative factor of $e^{-i \pi \alpha' t/4}$. Thus
in the small $t$ and large $s$ limit

\beq {\cal{A}} \sim -\frac{g_s^2}{V} \frac{s^{1+
\frac{\alpha't}{4}}}{t} e^{-i \pi \alpha't/4} \label{amplitude} \eeq

We can calculate the total probability of intercommutation using the
optical theorem

\begin{eqnarray}
 P & = & \frac{1}{4 E_1 E_2 v} 2 \;{\rm Im} \; {\cal{A}}|_{t
\rightarrow 0} \label{opticalthm} \end{eqnarray}


The imaginary part of ${\cal{A}}$ comes from $\textsf{Im}(e^{-i \pi
\alpha' t/4}) =  -\sin \pi \alpha't/4 $, which combined with the
$1/t$ pole factor gives the finite result $-\pi \alpha'/4$ as
$t\rightarrow 0$. Now because of the $V_{\perp}$ in $V$ in
(\ref{amplitude}) we expect $P \sim 1/V_{\perp}$.  We expect the
volume of the torus $V_{T^2}$ to disappear. Otherwise our results
will depend on the size of the large wrapped dimensions. In the
$t\rightarrow 0$ limit, $s^{1+ \frac{\alpha' t}{4}} \rightarrow s$.
The $E_1E_2$ in ({\ref{opticalthm}) will then cancel out the energy
dependence coming from the $s$ in ${\cal{A}}|_{t\rightarrow 0}$.
Hence, only geometrical factors depending on the angle $\theta$ and
velocity $v$ will be left over. Thus we expect

\beq  P = g_s^2 \frac{V_{min}}{V_{\perp}} f(\theta, v)
\label{probability} \eeq

\n $V_{min} = (2 \pi \sqrt{\alpha'})^6$ is a constant factor which
appears to make $P$ dimensionless, and represents the volume of a
T-dual 6D torus -- the minimum volume torus in string theory.
$f(\theta, v)$ is an ${\cal{O}}(1)$ function of $\theta$ and $v$ as
long as $v$ is not too relativistic or not too small. We would
expect that static F-strings intersecting in a non-BPS way will
reconnect and hence it is not surprising that  blows up as $f \sim
1/v$ as small $v$.  At high velocity $f$ also diverges.  This is
coming from the distinctive Regge effect of strings.  Strings tend
to interact more strongly at higher energies thus $P$ grows with
increasing $v$.

The probability that a moving F-string breaks and connects with a
$(p,q)$ string  can be estimated in a similar way. The result is
identical to (\ref{probability}) with $f(\theta, v)$ replaced by a
different ${\cal{O}}(1)$ $p,q$ dependent function $h_{p,q}(\theta,
v)$ which reduces to $f(\theta, v)$ for $p=1,q=0$ ($h_{1,0} = f)$.

\beq h_{p,q} = \frac{1}{g_s^2 v} \frac{q^2 v^2+ g_s( p -\cos \theta
\sqrt{(1-v^2)(p^2 + q^2/g_s^2)})}{8 \sin \theta \sqrt{(1-v^2)(p^2 +
q^2/g_s^2)}} \eeq

\subsection{Reconnection for strings with D-string charge}
\label{Dreconnection}

The intercommutation of D-strings is more complex.
Reconnection/intercommutation is a tree-level process for F-F  and
$F-(p,q)$ interactions.  The end-state of the tree-level
supergravity solution of colliding F-strings, or F-strings colliding
with a $(p,q)$ string is a reconnected string.  However, D-D and
$(p,q)-(p',q')$ string intercommutation is a 1-loop open string
quantum process because open string (tachyons) must be nucleated to
glue the D-strings together. The classical solution corresponds to
D-strings passing through each other \cite{khashimoto05}.

Because D-strings are composite objects in the sense that they may
be surrounded by a halo of open strings, D-D and $(p,q)-(p',q')$
reconnection is also qualitatively different from F-F reconnection.
For example, D-strings at high energy are surrounded by a halo of
energetic open strings.  When a D-string passes another D-string,
strings in the halo of one D-string may connect with strings in the
halo of the onrushing D-string, or equivalently open strings may be
nucleated between the two branes.  These open strings can then pull
the D-strings toward one another and eventually cause reconnection.
The energy to pair produce the open strings is interestingly
recaptured from the work done to stretch the nucleated strings by
the moving D-strings. Thus the faster the strings move past each
other the heavier the nucleated strings may be and the longer they
may be \cite{bachas, arfaei}.  This unusual feature simply reflects
the Regge nature of string scattering -- that like fundamental
strings, their scattering cross-section grows with velocity. In 6
compact directions, for $v \lsim 1$, the D-strings act as 6D black
disks of area $\sigma \sim \alpha'^3 (-\ln(1-v^2))^3$
\cite{jonespolchinski,bachas}. However, as the average $v^2$ for a
cosmic string is $< \frac{1}{2}$, we will concentrate on much
smaller velocities.   Note however, that as two branes approach each
other light closed strings are radiated by the branes and this gives
rise to an inter-brane potential of $V \sim -v^4$ \cite{lifschytz,
macallister}.  However, open strings are needed to glue the two
branes together for reconnection.
\begin{figure}
\begin{center}
\includegraphics[height=2.7in]{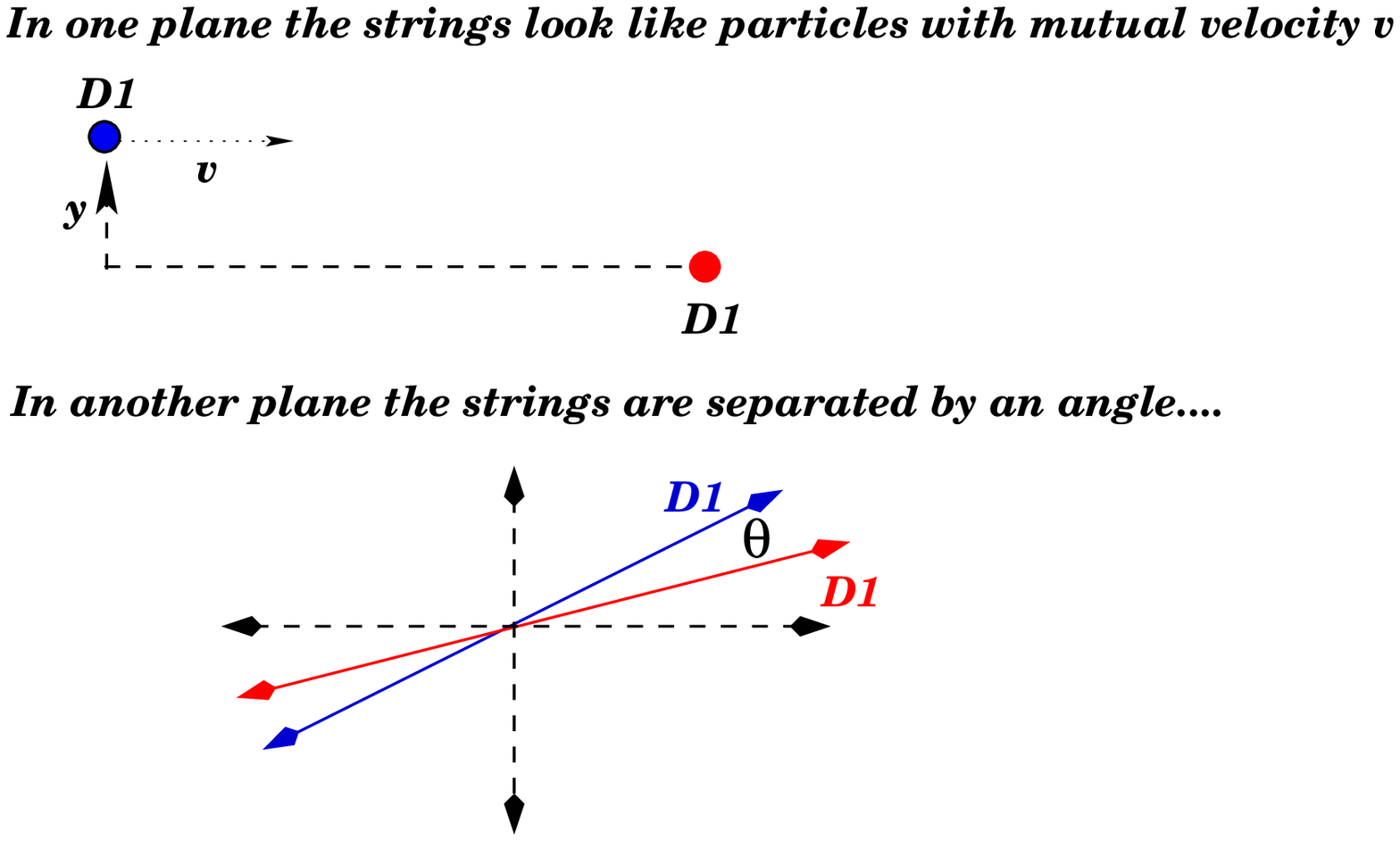}
\leavevmode
\end{center}
  \caption{$y$ is the impact parameter and $\theta$ the angle between 2
D-strings
  in a particular 2D plane. }
 \label{tilt}
\end{figure}

For $v\ll 1$ only the lowest mass string state will contribute to
the string scattering/reconnection amplitude.  If the branes are
tilted by an angle $\theta$ in the transverse directions, then for
generic values of $\theta$ the lowest mass state is a tachyon with
mass $m^2 = \frac{y^2}{(2\pi \alpha')^2} -\frac{\theta}{2 \pi
\alpha'}$. Here, $y$ is the impact parameter. See figure \ref{tilt}.
A tachyon will appear once $y < \sqrt{ 2 \pi \alpha' \theta}$.  Once
the tachyon appears the branes are almost sure to reconnect for
small $v$. (Note, there is an interplay between $\theta$ and $y$. As
$\theta$ decreases, it becomes harder to excite tachyons if $y\neq 0
$ because the branes become more parallel and parallel branes are
not tachyonic.)  For $y \le \sqrt{ 2 \pi \alpha' \theta}$,
reconnection occurs and the low $v$ cross-section for tachyon pair
production/reconnection  is the black sphere cross section, $\sigma
= \textsf{Ball}_6 = \frac{\pi^3}{6} y^6  = \frac{1}{6}(2 \pi^2
\alpha' \theta)^3$. Thus, the probability of tachyon pair production
$P_{pp}$ which in this case is the same as the reconnection
probability $P$ for $v\rightarrow 0$ is (if we ignore the factor of
$\frac{1}{6}$ as JJP do)


\beq P_{pp} = P(v\rightarrow 0) = \frac{(2 \pi^2 \alpha'
\theta)^3}{V_{\perp}} = \frac{V_{min}}{V_{\perp}} \left (
\frac{\theta}{2} \right )^3 \label{nonrelDD} \eeq

This is very similar to (\ref{probability}).  In the small $\theta$
and small $ v$ limit, $f(\theta,v) = h_{1,0}(\theta, v) =
\frac{\theta^3}{8v}$. Thus apart from a factor of $\frac{1}{v}$, the
F-F reconnection probability is the same as the tachyon pair
production probability for D-D reconnection.


For $v^2 \sim \frac{1}{2}$, higher mass states, in particular states
whose mass vanishes with vanishing $\theta$ will contribute to the
open string pair production probability $P_{pp}$. Using the standard
formalism of Schwinger pair production in \cite{schwinger51},
$P_{pp}$ can be related to the scattering amplitude ${\cal{A}}(y)$
as

\begin{eqnarray}
 1-P_{pp}(y)  & = & |\; e^{i{\cal{A}}(y)} |^2 =  \; e^{-2 \,{\rm
Im}\, {\cal{A}}(y)}
 = {\prod_{i, {\rm fermions}} (1 - x_i)}{{\prod_{j, {\rm
bosons}} \frac{1}{(1 + x_j)}}} \label{pprate}
\end{eqnarray}

\n where $x_k = e^{-2 \pi \alpha' m^2_k/\epsilon}$. Here $m_k$ is
the mass of the string state  $k$ and the velocity is $v = \tanh \pi
\epsilon$ where $\pi \epsilon$ is the rapidity.

We can understand (\ref{pprate}) as follows. The first two equations
on the L.H.S of (\ref{pprate}) follow from Schwinger's prescription.
The R.H.S results from evaluating ${\cal{A}}$  which is essentially
the 1-loop vacuum energy of two branes with relative velocity $v$.
The 1-loop vacuum amplitude can be calculated using the
Coleman-Weinberg formula: ${\cal{A}} = E_{vac} \sim \int
\frac{dt}{t} Z(t)  \sim \int \frac{dt}{t} \sum_i e^{-t m_i^2/2}$
\cite{johnson,colemanweinberg}. \footnote{The Coleman-Weinberg
formula gives the 1-loop vacuum amplitude ${\cal{A}} = \ln Z_{vac}$.
Using $\ln (k^2 +m^2) = \int_0^{\infty} \frac{dt}{t}e^{-t(m^2
+m^2)/2}$ we can write \cite{johnson} \beq {\cal{A}} = \ln Z_{vac}
=-\frac{1}{2} \tr \ln(\Box^2 + m^2) = -\frac{V_D}{2} \int \frac{d^D
k}{(2 \pi)^D} \ln (k^2 + m^2) = V_D \int \frac{d^D k}{(2 \pi)^D}
\int_0^{\infty}\frac{dt}{2t} e^{-t(k^2+m^2)/2} \eeq} Now the sum
over $i$ is over all bosonic and fermionic open string states
connecting the two branes. The sum can be written in terms of Jacobi
$\vartheta_a$ functions; in particular $Z(t) \propto
1/\vartheta_1(\epsilon t | i t)$. This has poles at the zeros of  $
\vartheta_1(\epsilon t | i t)$ which are at $\epsilon t = 1 ,
2,3,...$.  Because of these poles, ${\cal{A}}$ has an imaginary part
which is the sum of the residues of the poles of the partition
function: $\sum_{n=1}^{\infty} \textsf{
Res}[Z(\frac{n}{\epsilon})]$.

We now define the fermionic partition function $Z_f(t) = \sum_i
e^{-2 \pi \alpha' t m_i^2}$ which sums over all fermionic open
string states $i$ and define the bosonic partition function $Z_b(t)
= \sum_j e^{-2 \pi \alpha' t  m_j^2}$, which sums over all bosonic
open string states $j$.  Perhaps unsurprisingly the sum over
residues $\sum_{n=1}^{\infty} \textsf{Res}[Z(\frac{n}{\epsilon})]$
can be related to individual partition functions $Z_b$ and $Z_f$ as

\begin{eqnarray}
 -2 {\rm Im} {\cal{A}}(y) & = &  - \sum_{n=1}^{\infty} \left [\frac{(-)^{n+1}}{n}
Z_b \left (\frac{n}{\epsilon} \right ) + \frac{1}{n} Z_f \left
(\frac{n}{\epsilon} \right ) \right ]
\nonumber \\
& = & \sum_{j} \left ( \sum_{n=1}^{\infty} (-)^n \frac{x_j^n}{n}
\right )   - \sum_i \left (\sum_{n=1}^{\infty} (-)^n
\frac{(-x_i)^n}{n} \right ) \nonumber \\
& = & \ln \prod_i \frac{1}{1+x_i} +  \ln \prod_j (1-x_j)
\end{eqnarray}

\n finally yielding the R.H.S of (\ref{pprate}).

The bosonic states whose mass vanishes as $\theta, y \rightarrow 0$
have $m^2_j = \frac{y^2}{(2\pi \alpha')^2} - \frac{(3-2j) \theta}{2
\pi \alpha'}$ with multiplicities of state $\{j\} = \{1,2,3...\}$
being $\{1,7,8,8,8,8,...,8,...\}$.  The analogous fermionic states
$\{ i \} = \{1,2,3...\}$ have $m_i^2 = \frac{y^2}{(2\pi \alpha')^2}
- \frac{(2-2i) \theta}{2 \pi \alpha'}$ with multiplicities
$\{4,8,8,8,8,...,8,...\}$.  For, $v^2 \sim \frac{1}{2}$, the
rapidity parameter is $\epsilon \approx 0.3$. For an angle $\theta$
of order 1, we then find $e^{-\theta/\epsilon} \simeq 0.04$. Thus
ignoring the $y$ dependence we have $x_j = e^{-2\pi \alpha'
m^2_j/\epsilon} \propto e^{-2j\theta/\epsilon}$ and $x_i = e^{-2\pi
\alpha' m^2_i/\epsilon}\propto e^{-2i\theta/\epsilon}$ and $x_i$ and
$x_j$ decrease exponentially with increasing $i$ and $j$
respectively. (Note, here $i$ is not the imaginary unit
$\sqrt{-1}$.) Thus we can ignore states with high $i$ or $j$ in
(\ref{pprate}). If fact if we restrict to $j=1$ which corresponds to
the bosonic tachyon, and $i=1$ which corresponds to the lightest
fermion, then (\ref{pprate}) is

\beq 1 -P_{pp} \simeq \frac{(1-e^{-{y^2/ 2 \pi \alpha'
\epsilon}})^4}{1 + e^{-{y^2/ 2 \pi \alpha' \epsilon} + {\theta
/\epsilon}}} \label{DDprob}\eeq

The open string pair production probability $P_{pp}$ rapidly decays
to 0 as $y$ increases.  See figure \ref{probabilitiesfig}. However,
if the strings do collide then $y=0$ at the moment of collision and
$P_{pp} \simeq 1$. This implies that open string pairs are always
produced for $y=0$. These open string pairs will always lead to
reconnection unless $g_s \ll 1$. For $\theta$ small, more open
string states must be included in (\ref{DDprob}).  However, JJP
claim that $P_{pp} \simeq 1$ in most cases.

\begin{figure}
\begin{center}
\includegraphics[height=2.7in]{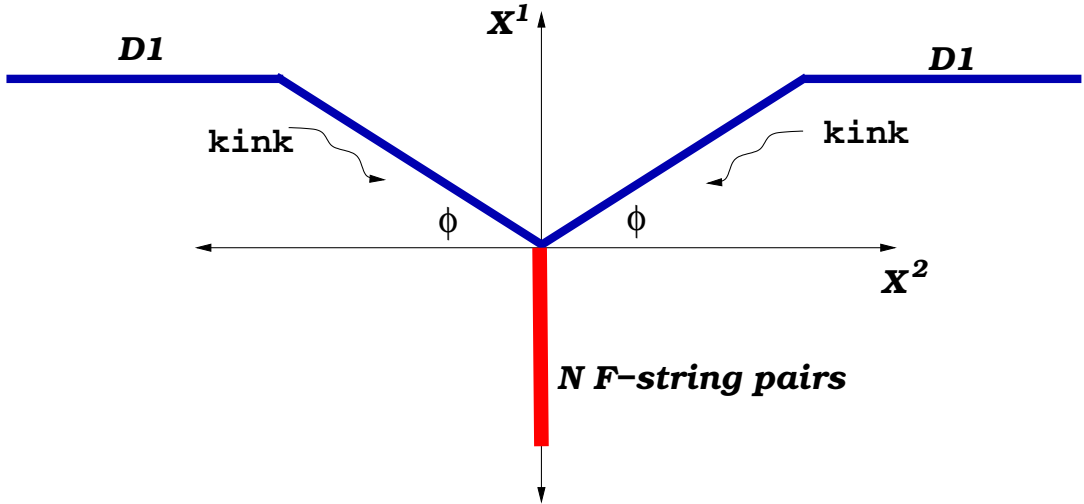}
\leavevmode
\end{center}
  \caption{A moving D-string nucleates
  $N$ F-string pairs. Two kinks are created which move with velocity 1 in the $X^2$ and
  $-X^2$ directions. The D-string moves upward with velocity $u$. The strings are shown
  in the rest frame and for a stable intersection the string intersection
  angle is $\phi$.}
 \label{ystringfig}
\end{figure}

For $g_s \ll 1$ the colliding strings are very massive since
$\mu_{(p,q)} \sim \mu_F\frac{1}{g_s}$.  Thus one might expect that
the colliding strings will pass through each other unless
${\cal{O}}(\frac{1}{g_s})$ open strings are nucleated gluing the
D-strings together. JJP argue that at least

\beq N > \frac{1}{g_s}\sinh \frac{\pi \epsilon}{2} \label{numpairs}
\eeq


\n open string {\em pairs} must be produced for intercommutation to
occur. This essentially comes from doing a force balance calculation
as in $\S\ref{sectionpq}$ at some time $t$ as in figure
\ref{ystringfig}. Figure \ref{ystringfig} shows  a string vertex at
the origin of the $(X^2,X^1)$ plane with $2N$ F-strings lying along
the negative $X^1$ axis attached to a  D-string which is shown in
blue in figure \ref{ystringfig}. The kinked parts of the D-string
make an angle $\phi$ with the $X^2$ axis. Force balance determines
$\phi$ to be $2\mu_{D1} \sin \phi = 2N \mu_F$. When the two
D-strings scatter two kinks are created on each D-string which move
in opposite directions along the strings at the speed of light. The
speed at which D-strings are moving past each other in the $X^1$
direction is the center of mass speed $u = \tanh \frac{\pi
\epsilon}{2}$. JJP argue that if $u
> \tanh \frac{\pi \epsilon}{2}$ reconnection will not occur because of the following.

Go to the rest frame to avoid worrying about the angle $\phi$ being
boosted.  Then each kink will be moving horizontally in the $+X^2$
or $-X^2$ direction with velocity 1 (since relativistic cosmic
string kinks move at the speed of light) and will be moving upwards
in the $X^1$ direction with velocity $u$ and thus will effectively
be moving at an angle $\arctan u$. The vertex will be stable if
$\arctan u = \phi$. However, if the D-strings are moving very fast
past each other then $\arctan u > \phi$ and the vertex will be
unstable and the strings will pass by/through each other. Thus the
nucleated $F$ strings will glue the moving D-strings together and
cause D-string reconnection only if $u < \tan \phi$. This condition
then gives (\ref{numpairs}).

 JJP argue that the open string
pairs are produced in a squeezed state and that these strings will
be the lightest possible strings -- tachyons
\cite{Epple03,Hashimoto03,Gomez02,Huang03}.  If $p$ is the
probability of pair producing a pair of strings in a squeezed state
of one oscillator, then $p^N$ is the probability of producing $N$
pairs. For a tachyon $p \simeq (1-e^{-\theta/\epsilon})$, thus $p^N
\simeq \exp(-N e^{-\theta/\epsilon})$.  Using the value of $N$ in
(\ref{numpairs}) the probability of D-D reconnection is

\beq P(y=0) \simeq \exp \left({-\frac{1}{g_s} \sinh \frac{\pi
\epsilon}{2} e^{-\theta/\epsilon}} \right ) \label{pdd}\eeq

\n which interestingly is a non-perturbative result because of the
$1/g_s$ and was qualitatively verified by \cite{khashimoto05} using
an effective theory approach. (Note, in addition to tachyons, 4
fermionic pairs are also produced with essentially unit probability.
JJP's version of (\ref{pdd}) is the probability of producing $N-4$
tachyons and 4 fermions.  We have focused only on the tachyons
because as soon as $y \neq 0$ fermion production is suppressed
relative to tachyon production.)

\begin{figure}
\includegraphics[width=2.7in]{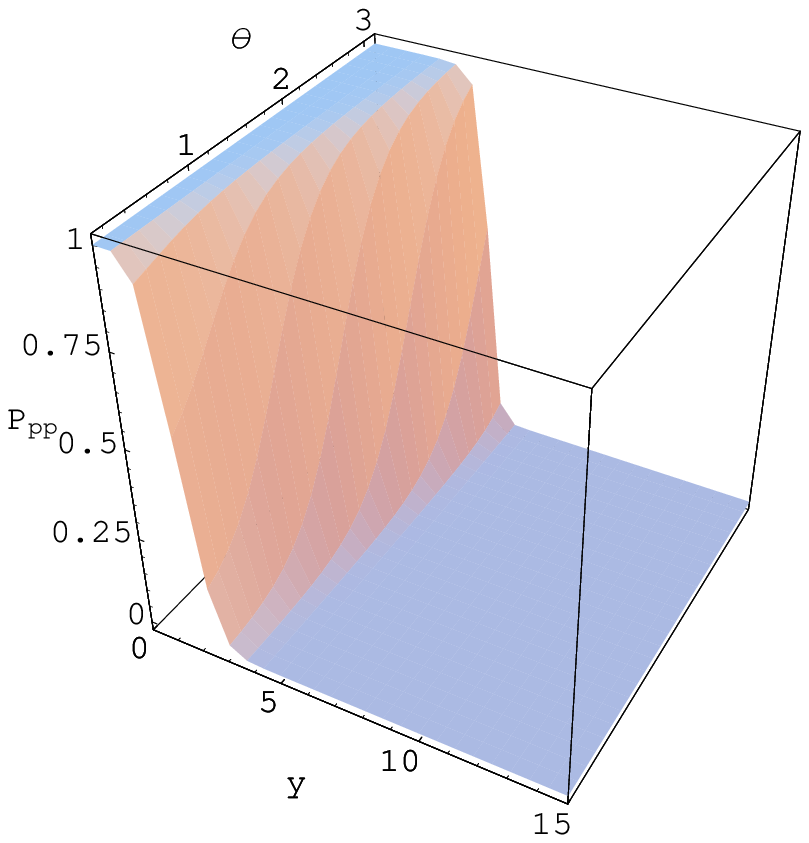}
\includegraphics[width=2.7in]{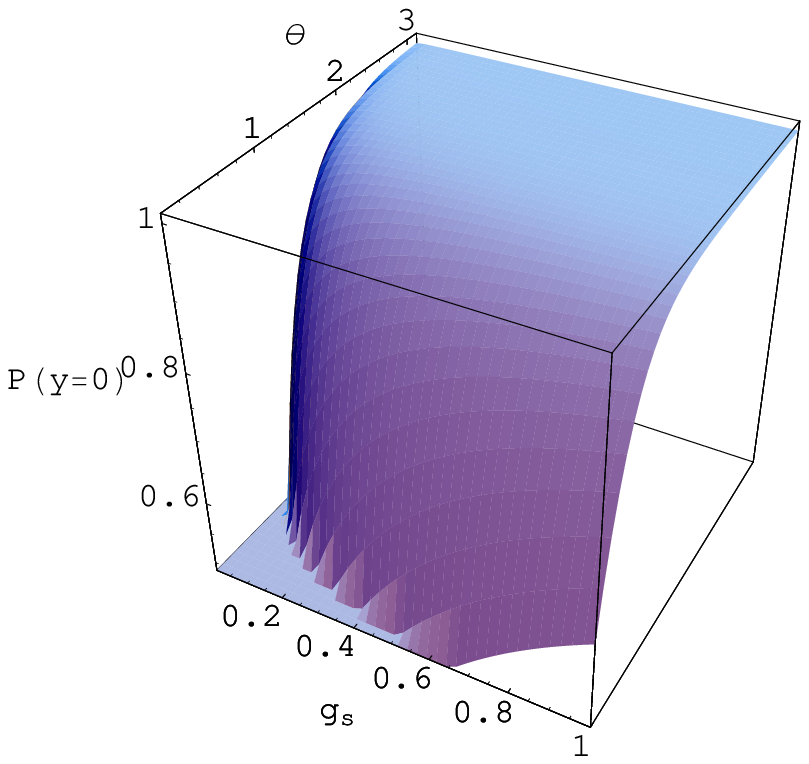}
\leavevmode
  \caption{(1) The left figure shows that the pair production
  probability $P_{pp}$, falls rapidly with $y$. (2) The
  figure on the right shows that the probability that colliding D-strings
   will reconnect $P(y=0)$ is
  $\simeq 1$ for $\theta \sim 1$ and $g_s \gsim 0.2$. For small values of $g_s$ and/or
   $\theta$, $P(y=0) \ll 1$. Both calculations only take the 2
   lightest string modes into account.}
 \label{probabilitiesfig}
\end{figure}

For $g_s \rightarrow 0$ but $v\neq 0$, the reconnection probability
vanishes, $P \rightarrow 0$. But it doesn't start to nosedive until
$g_s$ becomes small. For example, if $g_s \sim 0.1$ and  $\theta
\sim 1, \epsilon =0.3$ we find $P = 0.7$. For smaller angles $\theta
< 1$, higher mode states must be taken into account in (\ref{pdd})
and $P$ can decrease significantly, see figure
\ref{probabilitiesfig}. Thus, although $P \simeq 1$ generally, there
is a range of small angles where there is a significant probability
of D-strings passing through each other.

Equation (\ref{pdd}) gives the probability that colliding D-strings
will reconnect.  The probability that a random D-string will collide
with another random D-string and reconnect is:  the probability that
the two strings will run into each other: $\sigma(v)/V_{\perp}^{DD}$
where $\sigma(v)$ is the scattering cross-section,  times
(\ref{pdd}) extended to finite $y$. However, $P(y) \sim 0$ if $y$ is
much larger than a few string lengths. Also, for $v$ not extremely
relativistic, $\sigma(v)$ doesn't grow much larger than the black
sphere $\sigma(0)$ in (\ref{nonrelDD}) since $\sigma(v)$ grows
logarithmically with $v$ \cite{jonespolchinski,bachas}. Thus we can
approximate the reconnection probability for random D-strings  to be
the probability that two strings come within $\sqrt[6]{\sigma(0)}
\sim \sqrt{2
 \pi \alpha' \theta}$ of each other such that tachyons are excited,
  times the probability that enough open
strings are pair produced to induce reconnection.  Thus

\beq  P \sim \frac{V_{min}}{V_{\perp}} \left (\frac{\theta}{2}
\right )^3 \exp \left({-\frac{1}{g_s} \sinh \frac{\pi \epsilon}{2}
e^{\frac{y^2}{2 \pi \alpha' \epsilon} -\frac{\theta}{\epsilon}}  }
\right )  \simeq \frac{V_{min}}{V_{\perp}} \left (\frac{\theta}{2}
\right )^3 \Theta \biggl (\sqrt{2 \pi \theta \alpha'} - y \biggr )
\label{combinedDDprob} \eeq

The exponential factor in (\ref{combinedDDprob}) is $\simeq 1$ for
$y < \sqrt{2 \pi \theta}$ and vanishes for $y \gsim \sqrt{ 2 \pi
\theta \alpha'}$.  Thus we replaced the exponential factor by the
Heaviside function on the R.H.S. which is zero if $y$ is too large
to excite any tachyons.  Note the exponential inside the exponential
is simply  $\exp[-(\textrm{tachyon mass})^2/(2 \pi \alpha'
\epsilon)]$. The fermions don't contribute much to
(\ref{combinedDDprob}) because their pair production probability $p$
vanishes much faster than $p$ for tachyons once $y\gsim \sqrt{2 \pi
\theta \alpha'}$.

 For  $(p,q)$ and
$(p',q')$ strings, the nucleated open strings have $q$ ways of
connecting to the $(p,q)$ string and $q'$ ways of connecting to the
$(p',q')$ string.  Thus the multiplicities of the $i$ and $j$ states
will change by a multiplicative factor of $|qq'|$.  Also since
adding F-string flux to a string may be thought of as boosting the
string, the rapidity $\epsilon$ and angle $\theta$ will change to
some new $\tilde{\epsilon}$ and $\tilde{\theta}$.  The formulae are
given on page 22 of \cite{jonespolchinski}.  But after substituting
the new $\tilde{\epsilon}$  and $\tilde{\theta}$ and increasing the
degeneracy we find for just tachyon pair production at $y=0$

\beq (1-P_{pp})  = \frac{1}{(1 +
e^{\tilde{\theta}/\tilde{\epsilon}})^{|qq'|}} \simeq
e^{-|qq'|\tilde{\theta}/\tilde{\epsilon}} \eeq

\n and thus  $p$ in the paragraph above (\ref{pdd}) becomes $p
\simeq 1-e^{-|qq'|\tilde{\theta}/\tilde{\epsilon}}$, and
$\frac{1}{g_s} e^{-\tilde{\theta}/\tilde{\epsilon}}$ in (\ref{pdd})
will be replaced by $\frac{1}{g_s}
e^{-|qq'|\tilde{\theta}/\tilde{\epsilon}}$. Hence a small $g_s$ can
be compensated  by a large $q$ or $q'$, allowing $P \simeq 1$ for a
much larger range of $g_s$.  However, if $y \neq 0$, then  $
e^{\frac{y^2}{2 \pi \alpha' \epsilon} -\frac{\theta}{\epsilon}}$ in
(\ref{combinedDDprob}) is replaced by $ e^{|qq'|(\frac{y^2}{2 \pi
\alpha' \tilde{\epsilon}}
-\frac{\tilde{\theta}}{\tilde{\epsilon}})}$ allowing the middle
expression of (\ref{combinedDDprob}) to possibly decay to zero even
faster for suitable values of $\tilde{\epsilon}, \tilde{\theta}$
once $y \gsim \sqrt{2 \pi \tilde{\theta} \alpha'}$.

\subsection{What is $V_{\perp}$ if the strings' position moduli are
fixed?} \label{volume_perp}

Cosmologists tend to regard the spatial positions of objects like
cosmic strings, domain walls, and monopoles as free. They are free
to move, collide and do whatever external forces push them to do.
However, in string theory the positions of strings, branes, etc.,
are scalar fields.  They are unprotected by any symmetry. Hence,
there is nothing to {\em prevent} their positions from being fixed
and prevent the generation of a potential fixing their positions. In
fact, it is almost a mantra of faith among string theorists that
unprotected fields must surely become fixed; else string theory
would predict too many massless fields and violate the equivalence
principle, etc.

In this section we will assume that the positions of  cosmic strings
are somehow fixed.  If the positions are not fixed then $P$ can be
hugely suppressed if the radius of the extra dimensions is say 100
string lengths or more ($R \sim 100 \ell_s$), since then
$V_{min}/V_{\perp}\sim (\ell_s/R)^6 \sim 10^{-12}$.

Classically, there are several reasons to believe that strings will
be localized in the extra dimensions.  (1) Warped geometries
naturally generate a potential by localizing objects at the bottom
of a throat. I.e. it is very difficult to emerge from a heavily
warped throat because of extreme redshifting.  (2) Supersymmetry
breaking generically generates potentials. (3) Tachyon condensation
in the classical limit $g_s\rightarrow 0$ posits that all decay
products like F, D and $(p,q)$ strings lie along the plane of the
original hypersurface on which the condensation took place.  This
means such strings' extra dimensional positions are fixed to lie on
the hyperplane.

Note, we will assume the dilaton is constant throughout and hence
that $\mu_{(p,q)}^2(Y) = \frac{1}{2\pi \alpha'} (p^2 + q^2
e^{-2\phi(Y)}) $ is a constant. JJP write $\mu_{(p,q)}(Y)
=\frac{1}{2 \pi \alpha'} \nu(Y)$ to allow for a varying dilaton.
Thus, wherever one sees $\nu(0)$ one should replace it by

\beq \nu(0) \rightarrow 2 \pi \alpha' \mu_{(p,q)} \eeq

 Let $Y^i$ be the coordinates of a string in the extra dimensions.
We will assume that the potential $V(Y)$ fixing the strings'
positions comes from the warped metric. If the string is fixed at
$Y^i=0$ we would like to calculate the quantum spread in position
$\langle Y^i Y^i \rangle$. For a scalar $\phi$ in 2D with action $S
= -\frac{Z}{2}\int d^2 \sigma [(\partial_a \phi)^2 + m^2 \phi^2]$ as
JJP mention a Feynman diagram calculation for the spread $\langle
\phi^2 \rangle$ yields

\beq \langle \phi(0) \phi(0) \rangle = \frac{1}{Z} \int_0^{\Lambda}
\frac{d^2k}{(2 \pi)^2} \frac{1}{k^2+m^2}  = \frac{1}{4\pi Z} \ln
\frac{\Lambda^2 + m^2}{m^2} \label{qfluctuation} \eeq

\n The UV cutoff is the string scale, $\Lambda^2 \sim 1/\alpha'$.
But a 4D observer at the bottom of a warped throat will instead see
the warped cutoff $\Lambda^2 \sim m_s^2 e^{2A(0)} = 1/(\alpha'
e^{-2A(0)})$.

In flat space the worldsheet action for the position fields $Y^i$
contains a kinetic piece $\frac{\mu_{(p,q)}}{2} \int d^2 \sigma
(\partial_a Y^i)^2$ and a potential piece $\int d^2 \sigma V(Y)$
which we calculate later. Expanding the potential about the minimum
$Y^i =0$ and assuming $V(0) = 0$, we have $S \approx -
\mu_{(p,q)}\int d^2\sigma \frac{1}{2}(\partial_a Y^i)^2 - \int
d^2\sigma \frac{1}{2} V_{,ij}(0) Y^i Y^j$. Therefore, $k$ in
(\ref{qfluctuation}) maps to $Z= \mu_{(p,q)}$ and $m^2$ in
(\ref{qfluctuation}) maps to $m^2 = \frac{2 \pi \alpha'
V_{,ii}(0)}{\mu_{(p,q)}}$ where we have diagonalized the $V_{,ij}$
matrix by rotating the coordinates $Y^i$ among themselves. Then the
analog of (\ref{qfluctuation}) is

\beq \langle Y^i Y^i \rangle =  \frac{\omega_i}{4 \pi \mu_{(p,q)}};
\ \ \ \ \ \ \ \omega_i = \ln \left ( 1 + \frac{e^{2A(0)}
\mu_{(p,q)}}{ \alpha' V_{,ii}} \right ) \label{ysquared} \eeq

The localizing effects of warped geometries can be described by a
potential as follows. The Nambu-Goto action for a string in the
warped metric (\ref{warpedansatz}) is $-\mu_{(p,q)} \int d^2 \sigma
\sqrt{-h_{ab}}$ where $h_{ab} = e^{2A(y)} g_{\mu \nu} \partial_a
X^{\mu} \partial_b X^{\mu} + e^{-2A(y)} \tilde{g}_{mn} \partial_a
Y^i \partial_b Y^j$.  Choosing $g_{\mu \nu} = \eta_{\mu \nu}$ and
$X^0 = \sigma^0$ and $X^1=\sigma^1$ the Nambu-Goto action becomes
$\int d^2 \sigma V(Y)$ where $V(Y)$ is the potential

\beq V(Y) = -{\cal{L}}  = \mu_{(p,q)} e^{2A(Y)}  = \mu_{(p,q)}
e^{2A(0)} \left (1 - \frac{r^2}{R_3^2} \right )^{-1/2} \eeq

\n On the R.H.S. we specialized to the special case of flux
compactifications discussed in $\S \ref{sectionfluxcompact}$.  $R_3
\approx \sqrt{g_s M \alpha'} $ is the radius of the $S^3$ being
blown up and $r$ is the parameter parameterizing how `long' the
throat is.  The blown up $S^3$ is at $r=0$.  Recall that $e^{2A(0)}
\sim 10^{-8}$ from (\ref{warpattip}).  Note, $V''(0) = V(0)/R_3^2$.
This means that the spread of a string's position in the directions
along the throat and transverse to the $S^3$ will be $\langle Y^i
Y^i \rangle = \frac{1}{4 \pi \mu_{(p,q)}} \ln(1+ g_s M)$ and
$\omega_i = \ln(1 + g_s M)$

Also note that $V(0)$ is constant on the $S^3$ ($r=0$).  Hence $V$
doesn't localize the strings on the $S^3$ and the strings are free
to move all over the $S^3$. The occurs because of an $SU(2) \times
SU(2)$ symmetry of the deformed conifold geometry (i.e of the throat
geometry). However, once the throat is connected to a larger
Calabi-Yau, this symmetry is expected to be broken because the
overall CY lacks this $SU(2) \times SU(2)$ symmetry.  This breaking
presumably generates a potential {\em on the $S^3$}. The
dimensionless throat length is approximately $e^{-A(0)}$.  The
further away the CY is the smaller its effect on the $S^3$. Thus its
influence can be measured by the distance $e^{-A(0)}$ which
translates to a mass scale $\alpha' m^2 \sim e^{2A(0)}$.  Thus
instead of being able to roam around the entire $S^3$, a string will
we able to spread over a (distance)${}^2$ of $\langle Y^i Y^i
\rangle = \frac{1}{4 \pi \mu_{(p,q)}} \ln (1 +
\frac{\Lambda^2}{m^2}) = \frac{1}{4 \pi \mu_{(p,q)}} \ln (1 +
e^{-2A(0)})$ since the redshift factors $e^{2A(0)}$ of $\Lambda^2$
and $m^2$ cancel.
 Since $e^{2A(0)} \sim 10^{-8}$, $\langle Y^i Y^i \rangle \sim \frac{1}{4 \pi \mu_{(p,q)}}
\ln e^{-2A(0)}$ and $\omega_i \sim \ln e^{-2A(0)} = -2A(0)$.

Thus we can estimate $V_{\perp} \sim (\langle Y^iY^i\rangle)^3$.
This increases logarithmically with the mass of the $Y^i$ fields,
$V_{,ii}$.  This method gives the correct dependence on the various
parameters but may miss out some numerical factors.  A more clever
way to estimate the volume is to use a delta function $\delta^6(Y)$
as an inverse volume operator since $\delta(x) =\frac{1}{V}$  and
evaluate $\langle \delta^6(Y) \rangle$. Note the $V_{\perp}$
appearing in the reconnection probability will depend on the kinds
of string being scattered and we will label it accordingly.

 Suppose F and D strings are localized in the same area.
Fluctuations in F-string positions will be much larger than D-string
position fluctuations unless, $g_s \simeq 1$.  Thus we can
approximate the D-strings as fixed and find the F-string fluctuation
about a fixed center -- the D-strings' position, $Y^i =0$. This
means we take $\mu_{(p,q)}$ in (\ref{ysquared}) to be $\mu_F $.

\beq \frac{V_{min}}{V_{\perp}^{FD}} = V_{min} \langle \delta^6(Y)
\rangle = V_{min} \int \frac{d^6 k}{(2 \pi)^6} \langle e^{ik\cdot Y}
\rangle = \int \frac{d^6 k}{\alpha'^{-3}}  e^{-k_ik_j \langle
Y_i^2\rangle/2} = \prod_{i=1}^6 \sqrt{\frac{4 \pi}{ \omega_i}}
\label{FDfluct} \eeq

The expectation value $\langle e^{ik\cdot Y} \rangle $ was
calculated using a Gaussian wavefunction since to 2$nd$ order the
strings are confined by a harmonic oscillator potential, $V =
V_{,ij} Y^i Y^j$. Thus $\langle e^{ik\cdot Y } \rangle $ is just the
Fourier Transform of the normal distribution and
${\cal{F}}(\frac{e^{-x^2/2a^2}}{\sqrt{2 \pi a^2}}) = e^{-a^2
k^2/2}$.

If the $SU(2) \times SU(2)$ of the $S^3$ is unbroken such that
strings can freely wander all over the $S^3$ then (\ref{FDfluct})
breaks up into a part measuring the fluctuations over the $S^3$
$V_{min}(3D)/V_{\perp}(S^3)$ times a part giving the fluctuations
over the other 3 compact directions $\prod_1^3 \sqrt{4
\pi/\omega_i}$. Here, $V_{\perp}(S^3)$ is the volume of the $S^3$
and $V_{min}(3D) = (2 \pi\sqrt{\alpha'})^3$ is the minimal (T-dual)
3D volume in string theory.  The second factor arises because the
strings are confined by the warped factor potential in the 3 compact
directions transverse to the 3 directions of the $S^3$. Now
$V_{\perp}(S^3) = 2 \pi^2 R_3^3 =  2 \pi^2 \ell_s^3(g_s M)^{3/2}$
and $\omega_i = \ln(1 + g_s M)$ for $i$ in the non-$S^3$ compact
directions. This gives the first expression of the R.H.S. of
(\ref{KSforFD}).

If an $SU(2)\times SU(2) $ breaking potential is generated, then as
discussed above, $\omega_i$ in the $S^3$ directions is $\ln
e^{-2A}$, or equivalently $\ln H^{1/2}(0)$ in JJP's notation where
$H^{1/2}(Y) = e^{-2A(Y)}$. Then (\ref{FDfluct}) becomes the second
expression of (\ref{KSforFD}). For some choice of parameters as the
fluctuations spread out over the $S^3$, the second expression
(\ref{KSforFD}) becomes smaller than the first expression.  This is
due to limitations of the calculation. We then choose the maximum of
the two expressions.

 \beq \frac{V_{min}}{V_{\perp}^{FD}} =  \max \left
[\frac{4\pi}{(g_sM)^{3/2}} \frac{(4 \pi)^{3/2}}{\ln^{3/2} (1 + g_s
M)} \ , \ \frac{(4 \pi)^3}{\ln^{3/2} (e^{2A(0)}) \ln^{3/2} (1 + g_s
M)} \right ] \label{KSforFD}  \eeq

Suppose the F strings are localized at a distance $(Y_D^i
Y_D^i)^{1/2}$ away from the D-strings.  Then in for example the
unbroken $SU(2) \times SU(2)$ case, $1/V_{\perp}^{FD}$ is not
suppressed by $(Y_D^i Y_D^i)^{-3/2}$. Rather it is exponentially
suppressed as

\beq \frac{V_{min}}{V_{\perp}^{FD}}|_{Y_D^i} = V_{min} \langle
\delta^6(Y-Y_D) \rangle = \prod_{i=1}^6 \sqrt{\frac{4 \pi}{
\omega_i}}  e^{-\sum_1^6 \frac{Y_D^i Y_D^i}{\alpha' \omega_i}} =
\frac{V_{min}}{V_{\perp}^{FD}}   e^{-\sum_1^6 \frac{Y_D^i
Y_D^i}{\alpha' \omega_i}} \label{distancesuppression} \eeq

\n because the exponent $\langle Y^i Y^j \rangle$ in (\ref{FDfluct})
is replaced by $\langle (Y^i -Y^i_D)(Y^j-Y^j_D) \rangle = \langle
Y^i Y^j \rangle  + Y_D^iY_D^j$ for  fixed $Y^i_D$.

Suppose all the F-strings are localized in the same region. If $Y^i$
is the coordinate of one of the colliding F-strings and $Y^{\prime
i}$ is the coordinate of the other colliding F-string, then for F-F
collisions we have

\beq \frac{V_{min}}{V_{\perp}^{FF}} = V_{min} \langle \delta^6(Y-Y')
\rangle = \prod_{i=1}^6 \sqrt{\frac{2 \pi}{ \omega_i}}  = \lambda
\frac{V_{min}}{V_{\perp}^{FD}}\eeq

\n the factor $\lambda$ is $\prod_i^3 \frac{1}{\sqrt{2}} =
\frac{1}{2^{3/2}}$ for fluctuations covering the $S^3$ and
$\prod_1^6 \frac{1}{\sqrt{2}} = \frac{1}{2^3}$ for $SU(2) \times
SU(2)$ breaking. It arises because the exponent in (\ref{FDfluct})
$\langle (Y^i-Y^{\prime i})(Y^j -Y^{ \prime j})\rangle = \langle
Y^iY^j \rangle + \langle Y^{ \prime i} Y^{\prime j} \rangle = 2
\langle Y^i Y^j \rangle$ since $Y^i$ and $Y^{\prime i}$ are
independent variables.  The Gaussian integral then brings down a
factor of $\frac{1}{\sqrt{2}}$ for every $i$ in the $\prod_i$ in
(\ref{FDfluct}).

 For D-D string collisions we find

\beq \frac{V_{min}}{V_{\perp}^{DD}} = V_{min} \langle \delta^6(Y-Y')
\rangle = \frac{1}{g_s^3} \prod_{i=1}^6 \sqrt{\frac{2 \pi}{
\omega_i}}  = \lambda' \frac{V_{min}}{V_{\perp}^{FD}} \label{DDvol}
\eeq

\n Here $\lambda' = (2g_s)^{-3/2}$ if the fluctuations fill the
$S^3$.  $g^{-3/2}$ appears because D-string fluctuations are smaller
than F-strings fluctuations, i.e. $\langle Y^i Y^i \rangle =
\frac{\omega_i}{4\pi \mu_{(p,q)}}$ which is $\frac{g_s \omega_i}{4
\pi \mu_F}$ for a D-string.  For broken $SU(2) \times SU(2)$,
$\lambda' = (2g_s)^{-3}$ since the $i$ in $\prod_i\sqrt{2
\pi/\omega_i}$ of (\ref{DDvol}) ranges from $i=1$ to $i=6$.

If all the D-strings are localized around the same point, then
classically the impact parameter $y$ in (\ref{DDprob}) is zero. But
because of quantum fluctuations, $y^2$ will have a quantum spread
$\sum_i \langle (Y^i -Y^{\prime i})^2\rangle = 2 \sum_i  \langle Y^i
Y^i \rangle$ where $Y$ and $Y'$ are the positions of two different
colliding D-strings. Thus, even if $e^{-y^2/(2 \pi \alpha'
\epsilon)} =1$ classically, $P_{pp}$  will be quantum suppressed by

\beq e^{-y^2/(2\pi \alpha' \epsilon)} \sim \exp\left(-
\frac{g_s\sum_i \omega_i}{2 \pi  \epsilon}\right) \sim \exp \left
(-\inf \left [\frac{g_s M}{\pi \epsilon}, \frac{3g_s \ln
e^{-2A(0)}}{2\pi \epsilon} \right ] \right ) \label{ey2} \eeq

For unbroken $SU(2) \times SU(2)$, $y^2 \simeq R^2_3 = \alpha' g_s
M$.  There is an extra factor of 2 from the identical string effect
mentioned in the previous paragraph.  This gives the first
expression of the R.H.S of (\ref{ey2}). For the broken case
$\sum_1^6 \omega_i \sim \sum_1^3 \omega_i(S^3) \simeq 3\ln
e^{-2A(0)}$.  The usual maximum of (\ref{KSforFD}) converts to a
minimum because of the minus sign of (\ref{ey2}). Note the R.H.S. of
(\ref{ey2}) differs from the analogous equation (7.21) of JJP which
is $\exp(-\inf [\frac{g_s M}{2\pi \epsilon}, \frac{3g_s \ln
e^{-2A(0)}}{8 \pi \epsilon}])$.

The $\Theta(\sqrt{2 \pi \alpha'} \theta - y)$ in
(\ref{combinedDDprob}) can be written as $\Theta(\theta - y^2/(2
\pi\alpha'))$. Then using the expression for $y^2$ in (\ref{ey2}),
we find for the {\em no} $SU(2) \times SU(2)$ breaking case we have
$\Theta(\theta - g_s M/\pi)$ and for the $SU(2)\times SU(2)$
breaking case we have $\Theta(\theta - \frac{3 g_s}{2 \pi} \ln
e^{-2A(0)})$.

\subsection{Probabilities of reconnection}

We now combine the results of sections $\S\ref{Freconnection},
\S\ref{Dreconnection},\S\ref{volume_perp}$.  From the previous three
sections we learned the following lessons. \\

\n {\bf (A)} F-string reconnection is suppressed by $g_s^2$
essentially because there are 2 {\em in} states and 2 {\em out}
states. I.e., the emission amplitude of a closed string $\sim g_s$.
However, the normalization of the partition function $\sim 1/g_s^2$.
Thus the reconnection amplitude $\sim g^4_s \cdot 1/g_s^2\sim
g_s^2$. F-$(p,q)$ reconnection is suppressed by only $g_s$ because
open string amplitudes are weighted differently (open string
emission amplitude $\sim \sqrt{g_s}$ and normalization of the
partition function $\sim 1/g_s$). Because D-strings are
non-perturbative there is no simple perturbative explanation of the
$g_s$ dependence of $P$ as for F-F reconnection. However, because
the tensions of the branes $\sim 1/g_s$ and because one might expect
$P$ to be proportional to the product of the two D-string tensions,
the reconnection probability for D-strings is significantly greater
than for F and F-$(p,q)$ reconnection. In fact it is ${\cal{O}}(1)$
for a significant range of parameter values. (One might have
expected to be able to treat D-D reconnection by truncating to the
lowest open string states and then using perturbative theory.
However, as is a common theme in string theory, one must include the
infinite tower of open string states and thus things are more
involved.)

\n {\bf (B)} Reconnection of all strings is suppressed by
$1/V_{\perp}$.

\n {\bf (C)} If the strings are free to move around on the $S^3$,
then $1/V_{\perp} \sim [(g_s M) \ln (1 + g_s M)]^{-3/2} \sim (g_s
M)^{-3/2} = 1/R^3_3$. The $g_s$ dependence of $1/V_{\perp}$ comes
the $S^3$ radius: $R_3^2 = g_s M$.

\n {\bf (D)} If the strings are confined on the $S^3$ then,
$1/V_{\perp} \sim [|A(0)| \ln (1+g_sM)]^{-3/2}$, where from
(\ref{warpattip}), $|A(0)| \simeq 9$ for flux compactifications.

\n {\bf (E)} Quantum fluctuations suppress $P_{pp}$ by
$e^{-{\cal{O}}(g_s/\epsilon)}$ for D-D interactions -- see
(\ref{ey2}).

\n {\bf (F)} For very small $g_s$, the probability of D-D
reconnection is exponentially suppressed because of the need to pair
produce ${\cal{O}}(1/g_s)$ open string pairs to glue  the two
D-strings together.

We now assemble all the results together. The kinematic factor
$f(\theta, v)$ averaged over all velocities and angles is around
$0.5$. Thus

\begin{eqnarray}
\begin{tabular}{|c|c|c|}
  \hline
{}  &  \textsf{No ${S^3}^{\phantom{1}}$ Potential} &  $S^3$
\textsf{Potential}
\\ \hline
 FF & $  \! \! \! P  \sim 100 {\sqrt{g_s/M^3}^{}}^{\phantom{1}}\Gamma $ &
 $ \! \! \! P \sim 1.5 g_s^2  \Gamma$
\\
 FD  & $ \! \!  \! P \sim 280 {\sqrt{1/g_sM^3}}^{\phantom{1}} \Gamma  $  &
$ \! \! \! \! P \sim 12 g_s \Gamma $
\\
DD &   $  P  \sim 13 \Gamma {\left (\frac{\theta}{g_s
\sqrt{M}}\right)^{3}}^{\phantom{1}}_{\phantom{1}} \! \!
\Theta(\theta - 0.3 g_s M  ) $
 & $ \, \, \, P \sim 0.2
\Gamma \left (\frac{\theta}{g_s} \right )^{3} \Theta(\theta - 8.8
g_s )$
\\
  \hline
\end{tabular}
\label{allprobs}
 \end{eqnarray}

\n where  $\Gamma = \ln^{-3/2}(1+g_sM) $. The results for F-$(p,q)$
string are essentially the same as for F-D interactions in
(\ref{allprobs}). Also the results for $(p,q)-(p',q')$ interactions
are essentially the same as for D-D interactions, except the
D-string tension is replaced by the tension of the lighter tension
($p,q$) string, and is a factor of 2 greater because the $(p,q)$ and
$(p',q')$ strings are not identical. Note that if the different
strings are localized on different parts of the compact manifold,
then the probability $P$ for reconnection of different strings is
exponentially suppressed as in (\ref{distancesuppression}).

Because $\textrm{Vol}(S^3) \propto R^3 \propto (g_s M)^{3/2}$,  $P$
in column 1 of (\ref{allprobs})  doesn't vary as $g_s^2$. Also as
$M$ grows, $P$ falls, since threading flux through the $S^3$
increases its size.

Notice that $P$ for D-D interactions goes as $g_s^{-3}$ and thus
$P_{DD} \gg P_{FF}$ if  $g_s <1$ and $\theta$ is large enough.  This
is because D-string fluctuations are smaller. Hence,
$1/V^{DD}_{\perp}$ is much larger.  It is also because $P_{DD}$ is
not weighted by $g_s^2$.  From $\S \ref{Dreconnection}$ we learned
that D-D string collisions and $(p,q)$ and $(p',q')$ string
collisions generically lead to reconnection, $P \simeq 1$. Here, we
learn that because of their smaller fluctuations, even if they can
roam around their reconnection probability is much higher than that
of F-strings.

However collisions that would classically occur if strings are
localized at the same region such that $y=0$, don't necessarily
happen quantum mechanically. That is because D-strings' coordinates
acquire a spread of $\langle Y^2 \rangle$. Hence, the
$(\textrm{mass})^2$ of the lowest mass open string connecting the
scattered objects: $\langle Y^2 \rangle/(2\pi \alpha'^2)^2 -
\theta/(2 \pi \alpha')$, is tachyonic only if $0 \le  \theta \le
\pi$ is sufficiently large. In that case the exponential in
(\ref{combinedDDprob}) is $\simeq 1$ and the Heaviside functions of
(\ref{allprobs}) is one. For $(p,q) -(p',q')$ string collisions
where, $|(p,q)| < |(p',q')|$ because $\langle Y^2 \rangle$ is a
factor of two smaller than for the identical string case and because
$\langle Y^2 \rangle \propto g_s/\sqrt{g_s^2 p^2 + q^2}$, for
sufficiently large $p$ or $q$, the range of angles allowing
reconnection can be significantly larger than for DD-interactions.

Note, the previous calculations did not explicitly take the strings'
small scale structure into account.  Intercommutation probabilities
for wiggly strings is expected to be larger by a factor of 2-3 than
for smooth strings. This is because in a single string collision
several parts of a wiggly string may come into contact with several
parts of another wiggly string \cite{Avgoustidis04}.

Finally we comment on the meaning of the probabilities of
(\ref{allprobs}).  They are the probabilities that can be given to a
computer program which performs simulations in 4D. They encapsulate
the effects of the extra dimensions and stringy effects on string
collisions. (However, note that there is an additional
extra-dimensional velocity effect discussed in the next section.)

\section{Scaling of strings}

Having described how galaxy sized superstrings may form we must face
up to the cosmological consequences of long lived string remnants.
Remnant extended objects usually dominate the universe's energy
density and cause it to become overdense. Fortunately, this is not
necessarily so for strings.

In the absence of interactions the string energy density $\rho$,
will redshift as $1/a^2$. For a string network with a correlation
length $L$, there is about 1 string per volume $L^3$.  Thus the
energy density will be $\rho = \mu L /L^3 = \mu a L_0/a^3 L_0^3 \sim
\mu/a^2$. However, strings will collide and self-intersect.  This
will lead to loop formation. Loops redshift as $1/a^3$ because they
are smaller than the horizon and are not conformally stretched like
long strings whose energy grows as $\sim a$ (they don't feel the
expansion). Loops oscillate, lose energy by emitting gravitational
radiation and eventually collapse. Strings will remain a constant
fraction of the matter density of the universe if they discharge
enough of their length in loops such that the network's size remains
a constant fraction of the horizon -- i.e. the network is scale
invariant. This is called {\it scaling} as the correlation length
scales as $L= \gamma t$ and remains a fixed fraction of the horizon
size $\sim t$. For $L = \gamma t$, we find $\rho t^2 =
\textsf{constant}$. Using the Friedmann equation, the matter density
$\rho_m$,  grows as $\rho_m \sim H^2 \sim 1/t^2$. Hence $\rho_m/\rho
= \textsf{constant}$, and strings do not dominate the energy density
of the universe.

\subsection{The velocity one scale  model}

Because the energy density of long strings stretched by cosmic
expansion varies as $ \rho \sim \mu/L^2 \sim \mu /a^2$, we find
$\dot{\rho}|_{dilution} \sim - (2 \dot{a}/a) \rho$ from cosmic
dilution. In a time $\Delta t$ a string will travel $v \Delta t$ and
the number of string collisions will be $v \Delta t/L$. Let $P$ be
the probability that colliding non-wiggly strings will intercommute.
Then the probability that a string collision/self-intersection leads
to loop formation will be proportional to some function $f(P)$.  The
number of loops formed  can then be written as  $v \Delta t f(P)/L$.
This corresponds to a loss in long string energy of $\Delta E
\approx v \Delta t f(P)/L \cdot \mu L$ if each loop has
circumference $\sim L$. The rate of loss  is then
$\dot{\rho}|_{loop} \approx - v f(P)\rho/L$. If $\dot{\rho} =
\dot{\rho}|_{dilution} + \dot{\rho}|_{loop}$, then

\beq \dot{\rho} = -2\frac{\dot{a}}{a} \rho -  v f(P) \frac{\rho}{L}
\label{dotrho} \eeq

\n This is called the {\em one-scale model}
\cite{shellard,Kibble84}. To show that the energy density scales we
plug $L= \gamma(t) t $, where $\gamma(t)$ is time dependent, and
$\rho = \mu /L^2$ into (\ref{dotrho}).  We find for the case $a \sim
t^{\beta} $,

\beq \frac{\dot{\gamma}}{\gamma} = - \frac{1}{2t} \left (2(1-\beta)
- \frac{vf(P)}{\gamma} \right ) \label{gammaeq} \eeq

For a radiation dominated universe ($\beta = 1/2$) equation
(\ref{gammaeq}) has the attractor solution $\gamma = vf(P)$. Thus
$\gamma(t)$ eventually converges to the fixed point $\gamma(t) = v
f(P)$ unless $f(P)$ or $v$ is extremely small.  If $f(P) \sim P$
then the cosmic string fraction of the energy density of the
universe varies as $\Omega_{cs} \sim \rho \sim L^{-2} \sim
\gamma^{-2} \sim P^{-2}$. Thus while a very small $P$ may allow a
scaling solution, string domination may still occur and cause the
universe to close ($\Omega_{cs} > 1$).

However, this is only part of the story and leads us to the {\em
velocity one scale} (VOS) model
\cite{Martins95,Martins96,Martins00}. The velocity of a moving
object redshifts as $\dot{v} = - (\dot{a}/a) v$ leading to $v(t)
\sim 1/\sqrt{t}$ in the radiation dominated era. As the velocities
of strings decrease they will collide less frequently and loop
production will cease. We can crudely estimate the velocity effect
by using $\rho = \mu/L^2 \simeq a^{-2}\rho_{i} /\sqrt{1-v^2}$, where
$\rho_{i}$ is some initial density. The decrease in $\rho$ due
velocity redshifting is

\beq \dot{\rho}|_{vel} = \frac{\rho_{i}}{a^2 \sqrt{1-v^2}} v \dot v
= -v^2 \frac{\dot{a}}{a} \rho \frac{1}{1-v^2} \approx -2v^2
\frac{\dot{a}}{a} \rho \eeq

On the far right we "cheated" by replacing the denominator
$1/(1-v^2)$ by "2". We did this because the denominator is
${\cal{O}}(1)$ since cosmic strings are not that relativistic and on
average $v^2  \le 1/2$ making $1/(1-v^2) \ \laq \ 1/(1-\frac{1}{2})
= 2$.  Thus instead of (\ref{dotrho}) we find

\beq \dot{\rho} = -2\frac{\dot{a}}{a} (1 +  v^2  )\rho - v f(P)
\frac{\rho}{L} \label{vos} \eeq

\n which a general relativistic calculation validates. Note, $v^2$
is really the average velocity squared defined by $v^2 = \int
\epsilon \dot{\x}^2 d\ell/ \int \epsilon d\ell$ where $\epsilon$ is
the energy per unit length. Also, instead of (\ref{gammaeq}) we find

\beq \frac{\dot{\gamma}}{\gamma} = - \frac{1}{2t} \left (2 -
2\beta(1 + v^2 ) - \frac{vf(P)}{\gamma} \right) \label{gammav} \eeq

\n To solve (\ref{gammav}) we need the evolution equation
(\ref{eveq}) for $v$

\beq \dot{v} = (1-v^2) \left (\frac{k}{R} - 2\frac{\dot{a}}{a} v
\right ) \label{velevolution}\eeq

\n where $R$ is the radius of curvature of an average string
segment. $k$ is the so-called momentum parameter which measures the
angle between the curvature vector of a segment of the string
$a^{-1} d^2 \bx/ds^2$, and the velocity of the string segment. A
fast moving string has lots of small scale structure/wiggles. Hence,
the curvature vector and velocity are uncorrelated implying $k \sim
0$. A slowly moving string  is smooth and the curvature vector
points in the same direction as the velocity leading to $k \sim 1$.
$k$ is defined as

\beq kv(1-v^2)/R = \langle \bxd \cdot \bu(1- \bxd) \rangle \eeq

\n where $\bu$ is defined as $a^{-1} d^2 \bx/ds^2  = \bu =
\hat{\bu}/R$.

Equation (\ref{gammav}) has the scaling solution

\begin{eqnarray}
\gamma^2 = \frac{k(k + f(P))}{4 \beta(1-\beta)} & {} & v^2 =
\frac{k(1-\beta)}{\beta(k + f(P))} \label{vossol}
\end{eqnarray}

\n for constant $P$.  Note that $\gamma$ depends less weakly on
$f(P)$ once velocity effects are taken into account than for the
one-scale-model (\ref{gammaeq}).   For example suppose  $f(P) \sim
P$.   If the string is moving fast $k \sim 0$, then $\Omega_{cs}
\sim P^{-1}$. If the string is moving slowly, $k \sim 1$, then
$\Omega_{cs} \sim (1+P)^{-1}$.  Note in later discussions we will
usually assume $f(P) \sim P$.

\subsection{Modeling $(p,q)$ strings using the 4D VOS model}
\label{sectionpqnetwork}

Scaling is a complex issue for superstrings.  They live in more than
4 dimensions and $(p,q)$ strings come in an infinite number of
flavors since  $p$ and $q$ can range over all the
integers.\footnote{We can however, restrict say $p$ to be positive
and let $q$ be any integer.} Furthermore, a $(p,q)$ string can decay
to a loop only if it self-intersects or collides with another
$(p,q)$ or $(-p,-q)$ string. A $(p,q)$ string which runs into a
$(p',q')$ string can either create  a heavier $(p+p',q+q')$ string
or a lighter $(p-p',q-q')$ string. As discussed in
$\S$\ref{sectionpq} kinematics dictates whether the heavier or
lighter product is formed.

\begin{figure}
\begin{center}
\includegraphics[width=2.7in]{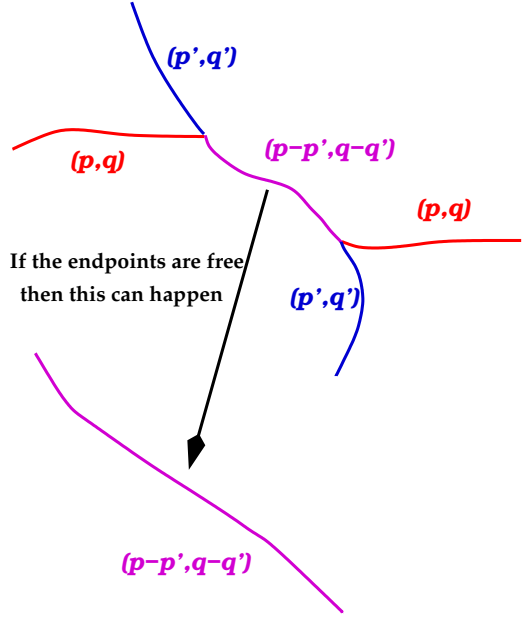}
\leavevmode
 \end{center}
  \caption{If the endpoints of the $(p,q)$ and $(p',q')$ strings
   are free then for example, a
  $(p-p',q-q')$ string can ``zip up'' the other strings and form a single $(p-p',q-q')$
   string.}
 \label{freestringendpoints}
\end{figure}

Colliding $(p,q)$  and $(p',q')$ strings do not generally annihilate
and produce a daughter. Instead a 3-string vertex is usually created
where the $(p,q), (p',q')$ and $(p \pm p',q \pm q')$ strings meet.
See figure \ref{pqstringfig}. If the endpoints of the $(p,q)$ and
$(p',q')$ strings are free (not attached to other 3-string vertices)
then it is energetically favorable for the two strings to move
toward each other, merge, dissolve into each other and form a $(p
\pm p', q \pm q')$ string. Thus $(p,q) + (p',q') \rightarrow (p \pm
p', q \pm q')$ only occurs if the initial two strings are not part
of a web such that their endpoints are free. See figure
\ref{freestringendpoints}.

If two strings can merge and coalesce into one string after a
collision quickly, and do so far before the next string collision,
then a web may not form - i.e. $\tau_{merge} \ll \tau_{collision}$.
The strings will on average then be non-intersecting.  What can
conceivably then happen is the following.  A gas of strings with
initially widely varying $p$ and $q$ can collide and self-intersect.
It will be energetically favorable for the collisions to lead to
lighter daughter strings. Thus, in the long term almost all of the
strings will be the lightest possible strings, either $(\pm 1,0)$ or
$(0,\pm 1)$, $\pm (1,1)$ or $\pm (1,-1)$ strings.  These remaining
strings may then self-intersect, form loops, and scale individually.
Then  $\Omega_{cs} = \Omega_{(\pm 1,  1)} + \Omega_{(1, \pm 1)}+
\Omega_{(\pm 1, 0)}  + \Omega_{(0, \pm 1)}$. If each of the
$\Omega_{(p,q)}$ is small then $\Omega_{cs}$ will be small and the
strings will not cause the universe to close.

A recent simulation by Tye, Wasserman and Wyman simulated such a
string gas  \cite{Tye05}. They assumed the endpoints of the strings
were free and the lengths of all the strings were the same. This
allowed them to assume that once a $(p,q)$ and a $(p',q')$ string
collide, they annihilate and form a $(p \pm p', q \pm q')$ string.
They used essentially the same equation as the Velocity One Scale
model in equation (\ref{vos}) for each $(p,q)$ species of string.
For convenience, we denote each $(p,q)$ species by $\alpha$ such
that $\alpha \equiv (p,q)$. They wrote a version of (\ref{vos})
using the number density $n_{\alpha}$.

\beq \dot{n}_{\alpha} = -2 \frac{\dot{a}}{a} n_{\alpha} -\frac{c_2
n_{\alpha} v}{L}  - \tilde{P} n_{\alpha}^2 v L + F v L \left (
\sum_{\beta, \gamma} P_{\alpha \beta \gamma} n_{\beta}
n_{\gamma}\frac{(1 + \delta_{\beta \gamma})}{2}  - \sum_{\beta,
\gamma} P_{\beta \gamma \alpha} n_{\gamma} n_{\alpha} (1+
\delta_{\gamma \alpha}) \right ) \label{ndot} \eeq

\n This can be rephrased in terms of the energy density by using
$\rho_{\alpha} = (1- v^2)^{-1/2} \mu_{\alpha} n_{\alpha}$ and the
velocity evolution equation (\ref{velevolution}) with $k/R$ replaced
by $c_2/L$. Thus $c_2$ represents the momentum parameter $k$ and $R$
is taken to be the correlation length $L$.

On the right hand side of (\ref{ndot}), the second term represents
string loss via loop production from string self-intersections. The
third term represents string loss from the collision of strings of
the same type.  $L$ appears to make the dimensions correct.
$\tilde{P}$ is the probability that a collision of strings of the
same type will create a loop of $\alpha$ string.

The fourth term represents the process $\beta + \gamma \rightarrow
\alpha$. $F$ is the ``overall interaction probability" of the
process. The $v L$ appear for the same reasons as previously
mentioned. $P_{\alpha \beta \gamma}$ is the branching ratio for a
$\beta$ and $\gamma$ string to become an $\alpha$ string.  The $1/2$
appears because of double-counting since the sum is symmetric in
$\beta$ and $\gamma$. However, the diagonal term $n_{\beta}^2
P_{\alpha \beta \beta} $ occurs only once.  To insure that it occurs
twice so that $\frac{1}{2}$ (two occurrences) = 1 occurrence, Tye
Wasserman and Wyman insert the factor $(1+ \delta_{\beta \gamma})$.

The fifth term represents the process  $\alpha + \gamma \rightarrow
\beta$ and appears with a minus sign because one $\alpha$ string is
destroyed.  The diagonal term in the sum $n_{\alpha}^2 P_{\beta
\alpha \alpha}$ represents the loss of two $\alpha$ strings.  That
is why the sum must be corrected with the factor $(1+ \delta_{\gamma
\alpha})$.

The branching ratio $P_{\alpha \beta \gamma}$ can be estimated as
follows. As shown in $\S\ref{sectionpq}$,  the critical angle
determining whether a $(p+p',q+q')$ string or a $(p-p',q-q')$ string
will form is given by (\ref{thetac}).


\beq \cos \theta_c = \frac{{\mbox{\boldmath $\pq \cdot \p'$}}}{
{\mbox{\boldmath $| \pq | | \p' |$ }}} = \frac{pp' +
g_s^{-2}qq'}{\sqrt{p^2 + g_s^{-2}q^2} \sqrt{p'^2 + g_s^{-2} q'^2}
}\eeq

\n where we have chosen the ``+" sign in (\ref{thetac}).  For
$\theta < \theta_c$ a $(p+p',q+q')$ string may form. For $\theta
> \theta_c$ only a $(p-p',q-q')$ string can form. The possible range
of $\cos \theta $ is $[-1,1]$.  If any value of $\theta$ is equally
likely then the probability $\IP(\cos \theta)$ is $1/2$. Thus the
probability $\IP(\cos \theta > \cos \theta_c) = (1- \cos
\theta_c)/2$ and  $\IP (\cos \theta < \cos\theta_c) = (1 + \cos
\theta_c)/2$.  Thus if $\beta = (p,q)$ and $\gamma = (p',q')$,

\begin{eqnarray}
P_{\beta + \gamma,\beta \gamma} & = & \frac{1}{2}\left (1 -
\frac{pp' + g_s^{-2}qq'}{\sqrt{p^2 + g_s^{-2}q^2} \sqrt{p'^2 +
g_s^{-2} q'^2}
} \right ) \label{plus} \\
P_{\beta - \gamma,\beta \gamma} & = & \frac{1}{2} \left (1 +
\frac{pp' + g_s^{-2}qq'}{\sqrt{p^2 + g_s^{-2}q^2} \sqrt{p'^2 +
g_s^{-2} q'^2} } \right ) \label{minus}
\end{eqnarray}

Some properties of (\ref{plus}) are: if $\beta  = (p,q)$ the
probability of creating a $\beta + \beta  = (2p,2q) $ string is zero
since $P_{2\beta,\beta \beta} = 0$. This is because the daughter
$(2p,2q) $ string is not actually a bound state since $2\sqrt{p^2 +
q^2/g_s^2} = \sqrt{(2p)^2 + (2q)^2/g_s^2}$. Hence $(2p,2q)$ strings
are not long-lived and can decay back to two $(p,q)$ strings. I.e.,
identical strings either collide and pass through each other and in
the long run remain as two identical strings or they form loops;
they do not create heavier strings. In a similar way the process
$(p,q) + (p',q') \rightarrow (Nk,Nl)$ doesn't happen because an
$(Nk,Nl)$ state is not a boundstate. There is nothing to prevent the
$(Nk,Nl)$ string from decaying to $N$ $(k,l)$ strings because
$\mu_{(Nk,Nl)} = N \mu_{(k,l)}$ (if $k$ and $l$ are relatively
prime). Hence, one must be careful how one inserts $P_{\alpha\beta
\gamma}$ into (\ref{ndot}). Tye Wasserman and Wyman therefore take $
P_{(Nk,Nl)(p,q)(p',q')} = N P_{(k,l)(p,q)(p',q')} $ if $(p \pm p',
q\pm q') = (Nk,Nl)$.

Also note that the process $\beta + \beta \rightarrow (0,0)$
corresponds to loop formation from same type collisions and
$P_{0\beta \beta}$ using (\ref{minus}). (A string with $(p,q) =
(0,0)$ is a closed string loop). However, this process is already
taken into account by the second term in (\ref{ndot}). Thus to
prevent this interaction from appearing twice in (\ref{ndot}) we set
$P_{0 \beta\beta} =0$ by hand.

 Using (\ref{plus}), (\ref{minus}) inserted into (\ref{ndot}),
 Tye Wasserman and Wyman simulated a gas of interacting $(p,q)$
strings in 3+1 dimensions. They chose widely varying initial
conditions for the number of each type of $(p,q)$ strings in the
string gas and found that initial conditions were largely
irrelevant.   Numerically, they found that for most reasonable
values of $P \lsim F \neq 0$ that the string network scaled, the
number of strings with large $p$ or $q$ was suppressed and that most
of the string energy was concentrated in the lowest tension $(\pm
1,0)$ or $(0,\pm 1)$, $\pm (1,1)$ or $\pm (1,-1)$ strings.

For $F \gg \tilde{P}$ they found that the number densities of higher
tension strings were highly suppressed. This is expected since
string collisions ($F$ interactions) tend to lower $p$ and $q$.

For $P \sim F$, they found that the number of higher tension
strings, $N_{(p,q)}$ was roughly power law suppressed as $N_{(p,q)}
\propto \mu_{(p,q)}^{-n}$ with $6< n  \; \laq \; 10$.  The strings
reach an approximate scaling solution where for example,
$\frac{\Omega_{(0,1)}}{ \frac{8}{3} \pi G \mu_{(0,1)} } =
\frac{46}{F + 0.53 P}$ for $g_s = 1/2$. The scaling is approximate
because the strings continue to evolve at late times. However, they
claim that this late evolution is not cosmologically problematic.

For $F \rightarrow 0$, they found that each $(p,q)$ species scaled
individually. Higher tension strings were not suppressed and
$\Omega_{(p,q)}/\mu_{(p,q)}$ for each $(p,q)$ species converged to
the same value leading to possible disaster since
$\sum_{p,q}^{\infty} \Omega_{(p,q)}$ may explode.


 Unfortunately, the authors do not present
results for realistic case of $g_s \ll 1$. The authors show that for
$g_s =1$, F and D-strings are interchangeable as expected. They also
show that for $g_s = 0.5$ that the number of F-strings is 2-3 times
the number of D-strings.  It would be nice to know for say $g_s \sim
1/10$ whether $D$-strings and $(p,q)$ strings with $q > 1$ are power
law suppressed like the heavier $(p,q)$ strings and/or whether they
scale individually. It would also be helpful to understand what
happens when both $F$ and $P$ are very small since $F, P \sim g_s^2
\ll 1$.

\subsection{Using the $D$ dimensional VOS model with F and D
strings}

Unfortunately, string physics is more complicated than the previous
section's approach.    Classically ($g_s \ll 1$) strings will evolve
only on a 3+1 dimensional slice of the higher dimensional spacetime
\cite{Sen04,Sen02}. However, quantum mechanically they will wander
off this slice and move in the extra dimensions. If the strings form
at the bottom of a throat with significant redshifting, they will be
localized at the bottom of the throat. Unless they are localized in
the compact directions by a confining potential they will be able to
move longitudinally and explore the extra-dimensional space at the
throat bottom  which for the special case of a deformed conifold
throat is an $S^3$ (assuming no $SU(2)\times SU(2)$ breaking).

Motion in the extra dimensions leads to the following interesting
consequences. (1) Moving strings generically intersect in 3+1
dimensions.  In higher dimensions, they miss each other as
non-interacting $p$ dimensional objects generically collide  in
$d$-dimensional spacetimes only if $d \le 2(p+1)$
\cite{Brandenberger88}. (2) If the extra dimensions are fixed, the
extra dimensional velocities do not redshift.  Cosmic strings in
expanding universes are constrained by $ v^2  \le 1/2$. Hence, if
extra dimensional velocities grow then 3+1 dimensional velocities
must decrease.  This may happen because while extra-dimensional and
4D velocities have similar sized source terms, the  4D velocities
are redshifted away while the extra-dimensional velocities build up.
All the velocity may end up in the extra dimensions. The strings
will then stop moving in the large dimensions.  They will not scale
because loop forming collisions will be infrequent and they will be
unable to  discharge their growing lengths/energies. Avgoustidis and
Shellard (AS) recently analyzed these consequences of string
propagation in extra dimensions \cite{Avgoustidis04}.

AS started with a metric of the form

\beq ds^2 = N(t)^2 dt^2 - a(t)^2 d \x^2 - b(t)^2 d {\mbox{\boldmath
$y$}}^2 \eeq

\n which is isotropic in 3 large spatial directions and separately
isotropic in $D-3$ small extra dimensions ($D$ is the number of
spatial dimensions). Note, no warping appears in the metric. The
inclusion of $N(t)$ allows one to switch between physical and
conformal time very easily. If $N = 1$ then $t$ is the physical
time. If $N = a =b$ then $t$ is very conveniently the conformal
time.

After deriving the equations of motion and expression for the energy
AS wrote down how the energy redshifts over time

\beq \dot{\rho} = - \frac{\dot{a}}{a} \left (2(1 + v_x^2) + v_y^2 +
W_1 \right )\rho - \frac{\dot{b}}{b} \left [(D-4 + v_y^2) + v_x^2 +
W_2 \right ] \rho \label{xtrarhodot}\eeq

This is a very similar to (\ref{vos}).  Compared to (\ref{vos}) the
first term on the right has an extra term $v_y^2 +W_1$.  This
represents the effects of extra dimensions and extra-dimensional
velocities.  Note, even if the extra-dimensional velocity vanishes,
$v_y =0$, and the extra dimensions are fixed, $\dot{b} =0$,
(\ref{xtrarhodot}) differs from the 4D VOS model.

In the second term $(D-4)/2$ instead of "2" appears because while
the codimension of  strings in  3 spatial dimensions is 2, the
codimension of a string in  $D-3$ extra dimensions is $(D-3) - 1$.
Otherwise the two terms are interchangeable because under the
exchange of $(a,\x) \rightarrow (b,\l)$, we have $W_1 \rightarrow
W_2$.

The velocities, $v_x, v_y$ have been defined to be

\begin{eqnarray}
v_x^2 = \left \langle \frac{a^2}{N^2} \dot{\bx}^2 \right \rangle &
{} & v_y^2 = \left \langle \frac{b^2}{N^2} \dot{\by}^2 \right
\rangle
\end{eqnarray}

\n where the averaging is done as

\beq \langle f \rangle  = \frac{ \int f \epsilon d \sigma^1}{ \int
\epsilon d\sigma^1 } \label{averaging} \eeq

\n and $\epsilon$ is the energy per unit string length. The $W$
functions are defined as

\begin{eqnarray}
W_1 \! \!& = & \! \! \left \langle \frac{b^2 \by'^2}{a^2 \bx'^2  \!
\!+ b^2 \by'^2} \frac{N^2 \! -  \! a^2 \dot{\bx}^2 \! \! - \! b^2
\dot{\by}^2}{N^2} \right \rangle \approx \left \langle \frac{b^2
\by'^2}{a^2 \bx'^2 + \by'^2} \right
\rangle (1 \!- \!v_x^2 \!- \! v_y^2) \\
W_2 \! \! & = & \! \! \left \langle \frac{a^2 \bx'^2}{a^2 \bx'^2 \!
\! + \! b^2 \by'^2} \frac{N^2 \! -  \! a^2 \dot{\bx}^2 \! \!- \! b^2
\dot{\by}^2}{N^2} \right \rangle \approx \left \langle \frac{a^2
\bx'^2}{a^2 \bx'^2 \! + \! \by'^2} \right \rangle (1 \!-\! v_x^2 \!
- \! v_y^2)
\end{eqnarray}

\n On the L.H.S we have used the approximation that worldsheet
spatial derivative terms like $f'$ are uncorrelated with worldsheet
time terms like $\dot{f}$.\footnote{The gauge conditions are $ \bxd
\cdot \bx' - \dot{\by} \cdot \by'/a^2 = 0$ and $\sigma^0 = t$.}
$W_1$ and $W_2$ are related to the amount of string that lies in the
extra and 3+1 dimensions respectively.

If the extra dimensions are fixed such that $b =1$ then we will call
$W_1(b=1) = w_{ls} (1 - v_x^2 - v_y^2)$. $w_{ls}$ measures the
proportion of the string in the extra dimensions. Equation
(\ref{xtrarhodot}) becomes

\beq \dot{\rho} = -\frac{\dot{a}}{a} \left [ (2+ w_{ls}) +
(2-w_{ls})v_x^2 + (1-w_{ls})v_y^2 \right ] \rho \label{extradimvos}
\eeq

\n Note, even if $v_y = 0$, (\ref{extradimvos}) differs from the 4D
VOS model of (\ref{vos}) because of $w_{ls}$.

If we assume that {\em on average} the partitioning of gradient
energy in $w_{ls}$ = \textsf{(gradient energy in
$\by'^2$)}/\textsf{(total gradient energy)}, is equal to the
partitioning of kinetic energy  in \textsf{(KE in
$\dot{\by}^2$)}/\textsf{(total KE)}, then we can model $w_{ls}$ as

\beq w_{ls} = \left \langle \frac{\by'^2}{a^2 \bx'^2 + \by'^2}
\right \rangle \simeq \left \langle \frac{\dot{\by}^2}{a^2
\dot{\bx}^2 + \dot{\by}^2} \right \rangle = \frac{v_y^2}{v^2} \eeq

\n Note, in this approximation if $v_y = 0$ then $w_{ls} =0$ and
(\ref{vos}) coincides with the $D+1$ dimensional VOS model
(\ref{extradimvos}).

AS argue that if the correlation length $L$ of the strings is much
larger than the size of the extra dimensions then string motion is
effectively 3-dimensional, and $\rho = \mu/L^2$ instead of say $\rho
= \mu L^{-2} L^{-(D-3)}$. Even if this is not initially true, if the
strings scale such that $L= \gamma t$, then  $L$ will eventually
grow larger than the size of the (fixed) extra dimensions. Inputting
$L=\gamma(t) t$ and $\rho = \mu/L^2$ and $w_{ls} = v_y^2/v^2$ into
(\ref{extradimvos}) yields

\beq \frac{\dot{\gamma}}{\gamma} = \frac{1}{2t}\left (\beta
(2+2v_x^2 + v_y^2/v^2) -2 + \frac{P c v_x}{\gamma} \right )
\label{vosxtra} \eeq

The evolution equations for $v_x$ and $v_y$ under the same
approximations are as copied out of AS

\begin{eqnarray}
\frac{\dot{v}_x}{v_x} & = & \frac{1}{t} \left ( \frac{k_x
v_x}{\gamma}(1-v^2) - \beta v_x^2 (2-2v_x^2 -v_y^2/v^2) \right )
\label{vxsource} \\
\frac{\dot{v}_y}{v_y} & = & \frac{1}{t} \left ( \frac{k_y
v_y}{\gamma}(1-v^2) - \beta v_y^2 (1-2v^2)(1 -v_y^2/v^2) \right )
\label{vlsource}
\end{eqnarray}

\n As before $k_x$ and $k_y$ are momentum parameters measuring how
correlated the string curvature is with the directions of
${\mbox{\boldmath{$v_x$}}}$ and ${\mbox{\boldmath{$v_y$}}}$
respectively. $k_x$ and $k_y$ are related to $k$ by $kv = k_x v_x +
k_y v_y$.  The formal definitions of $k_y$ and $k_x$ are

\begin{eqnarray}
\frac{k_x v_x (1 \! - \! v^2)}{R} \! = \!  \langle \bxd \cdot \bu(1
\!-  \! \dot{\bx}^2 \! -  \! \frac{\dot{\by}^2}{a^2} ) \rangle \! \!
& {} \! \!&
 \frac{k_y v_y
(1\! - \! v^2)}{R}  \!= \! \langle \frac{\dot{\by}}{a} \cdot \bu (1
\! - \! \dot{\bx}^2 \! - \! \frac{\dot{\by}^2}{a^2} )
\label{momentumpar} \rangle
\end{eqnarray}

If the string motion is effectively 3-dimensional, then the
curvature vector $\bu$ will mostly lie along the 3 large dimensions
making $\dot{\by} \cdot \bu \ll 1$ and $\bxd \cdot \bu \sim 1 $ in
(\ref{momentumpar}). Thus $k_x \gg \ k_y$. Since the source term for
$v_x$ is $\propto k_x$ from (\ref{vxsource}) and the source term for
$v_y$ is $\propto k_y$ from (\ref{vlsource}), the source term  for
3D velocities is much greater than for $v_y$.  However, note from
(\ref{vxsource}) and (\ref{vlsource}), that the damping term for
$v_x$ is significantly greater than the damping term of $v_y$
because the extra dimensions do not expand.

Equation (\ref{vosxtra}) was numerically solved for: (a) varying
amounts of "3Dness;" (b) varying initial $v_y$ and $v_x$; (c)
varying intercommuting probability $P$.

\begin{figure}
\includegraphics[width=1.9in]{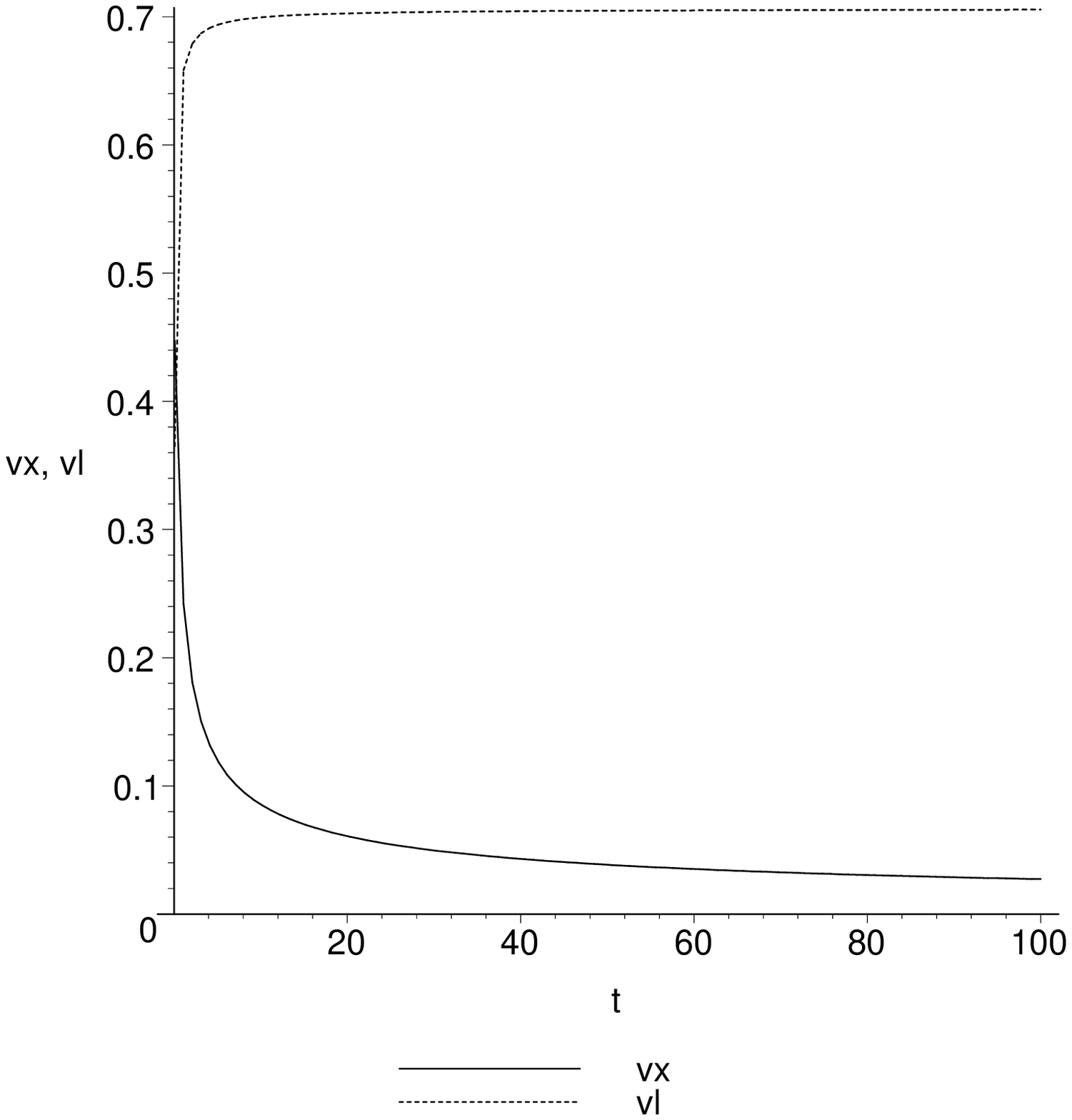}
\includegraphics[width=1.9in]{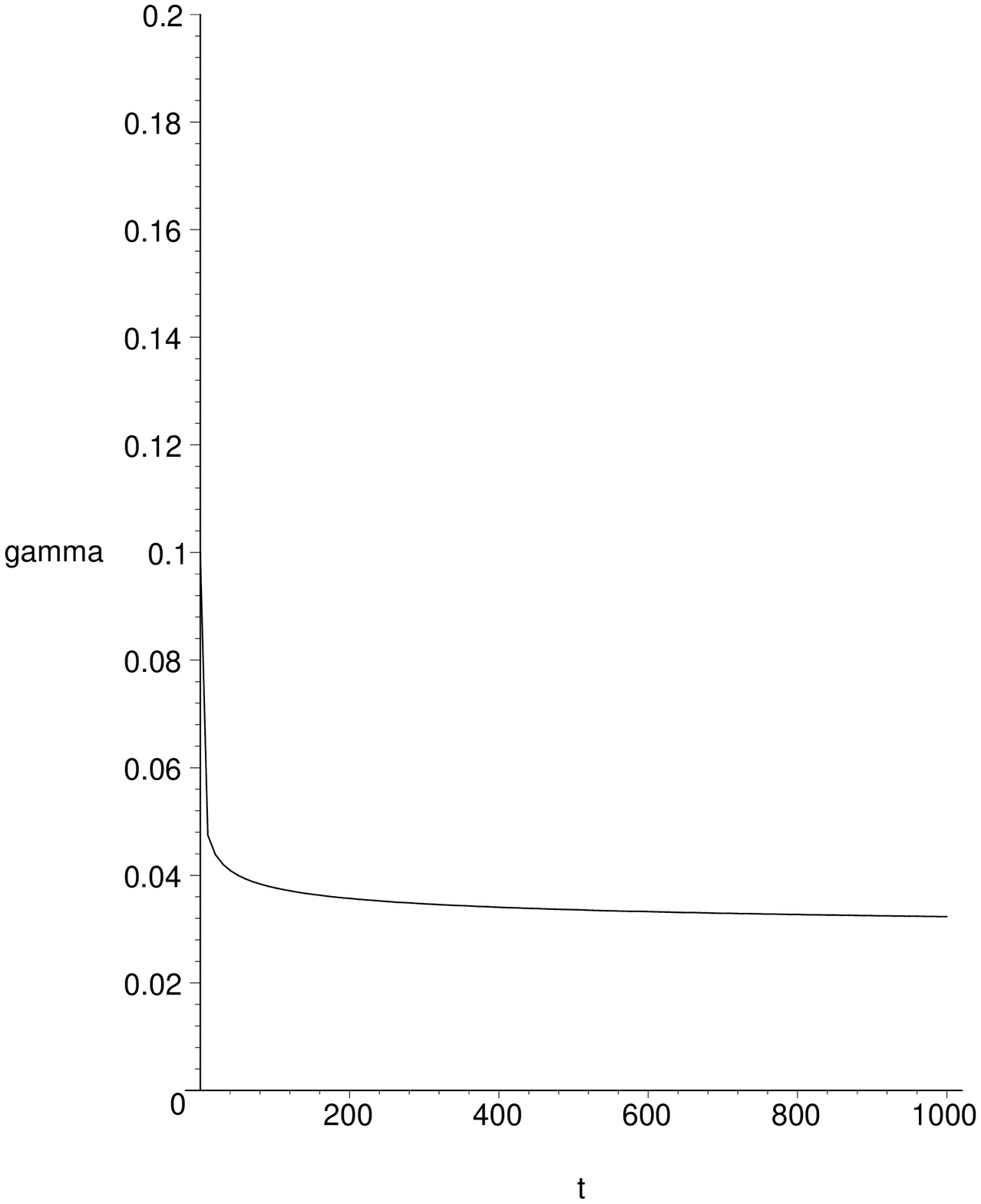}
\includegraphics[width=1.9in]{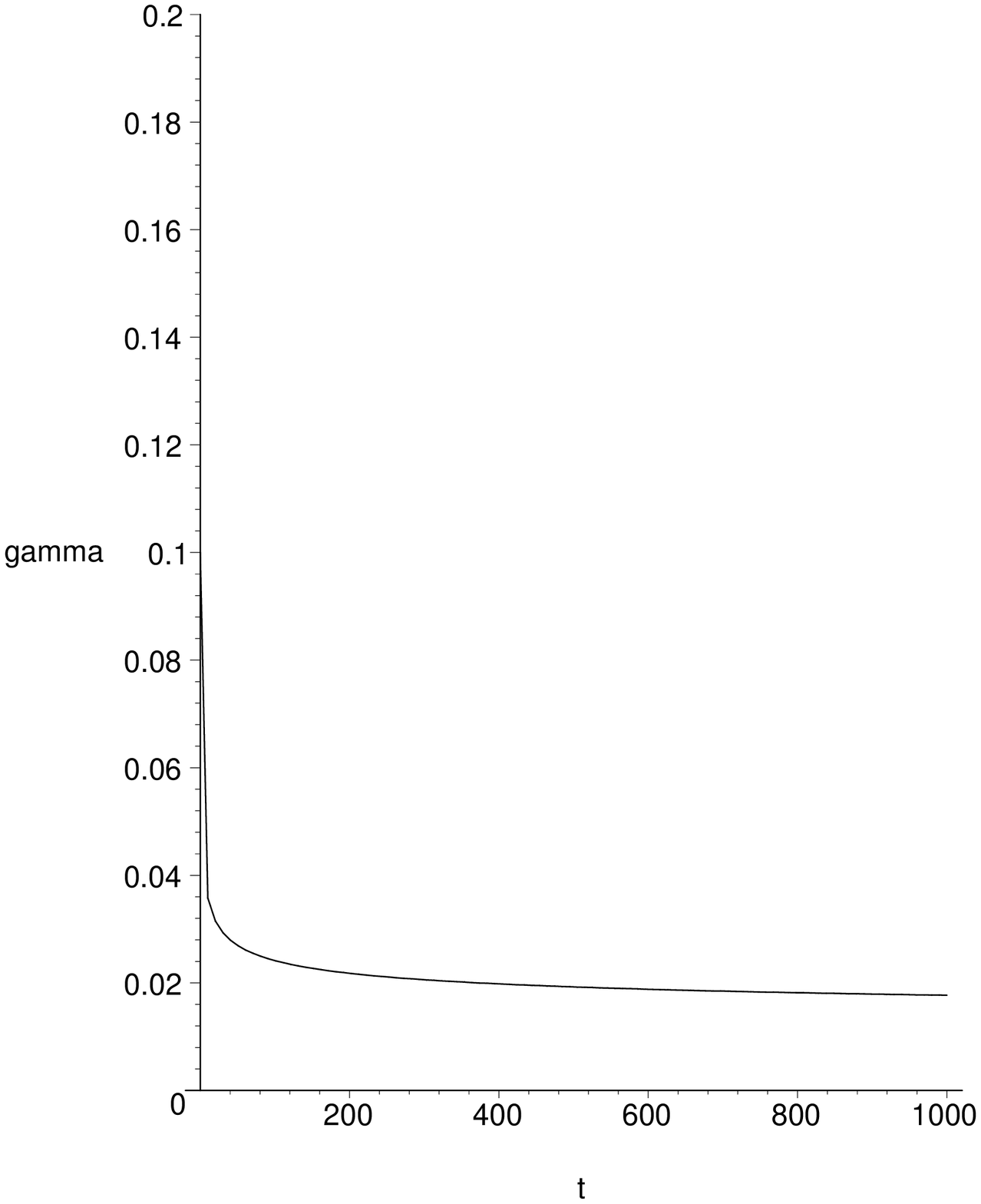}
\caption{From left to right: (1) Velocity evolution if the strings
evolve in more than 3 dimensions such that $k_x \sim k_y$. Note here
$v_y \equiv v_\ell$. For $D$ dimensional evolution  $v_y \equiv
v_{\ell}$ quickly dominates $v_x$ and the strings stop moving in the
extra dimensions. (2) Evolution of $\gamma$ for 3D motion if $v_y
\simeq 0.3$ initially.  (3) $\gamma(t)$ if $v_y \simeq
0.95/\sqrt{2}$ initially.  } \label{vydomination}
\end{figure}

If the motion of strings is not very 3-dimensional such that the
curvature vector $\bu$ explores the extra dimensions and $k_y$ is
not small then,  $v_x^2 \rightarrow 0$ and $v_y^2 \rightarrow 1/2$.
See the first graph in figure \ref{vydomination}.  This is because
in this case $v_x$ and $v_y$ have similar source terms, but $v_x$ is
more damped than $v_y$. Since $v_x^2 + v_y^2 \le 1/2$, this means
that $v_y$ grows relative to $v_x$ and drives $v_x \rightarrow 0$.
As the loop production term is $\propto v_x$, loop production then
ceases, no scaling solution is reached and strings presumably
dominate the energy density of the universe.

For 3D motion, even if $v_y \, \, \laq \, \, \frac{1}{\sqrt{2}}$
initially, a scaling solution is reached although the scaling value
of $\gamma$ is smaller by a factor of 2 then were say $v_y = 0.3$
initially, see figure \ref{vydomination}. Since $\rho \sim
\gamma^{-2}$,  the string density is a factor of $4$ larger in the
$v_y \simeq \frac{0.95}{\sqrt{2}}$ case compared to the  $v \simeq
0.3$ case. Nevertheless, unlike the $D$ dimensional case, a scaling
solution is reached in this 3D case because unlike the previous
case, here $v_x$ is not driven to zero. It reaches the asymptotic
value $v_x \approx 0.2$ for $v_y =0.95/\sqrt{2}$.  See figure
\ref{vydomination}. For smaller initial values of $v_y$, $v_x$
quickly dominates $v_y$ because $k_x \gg k_y$.

\begin{figure}
   \includegraphics[height=2in,width=1.9in]{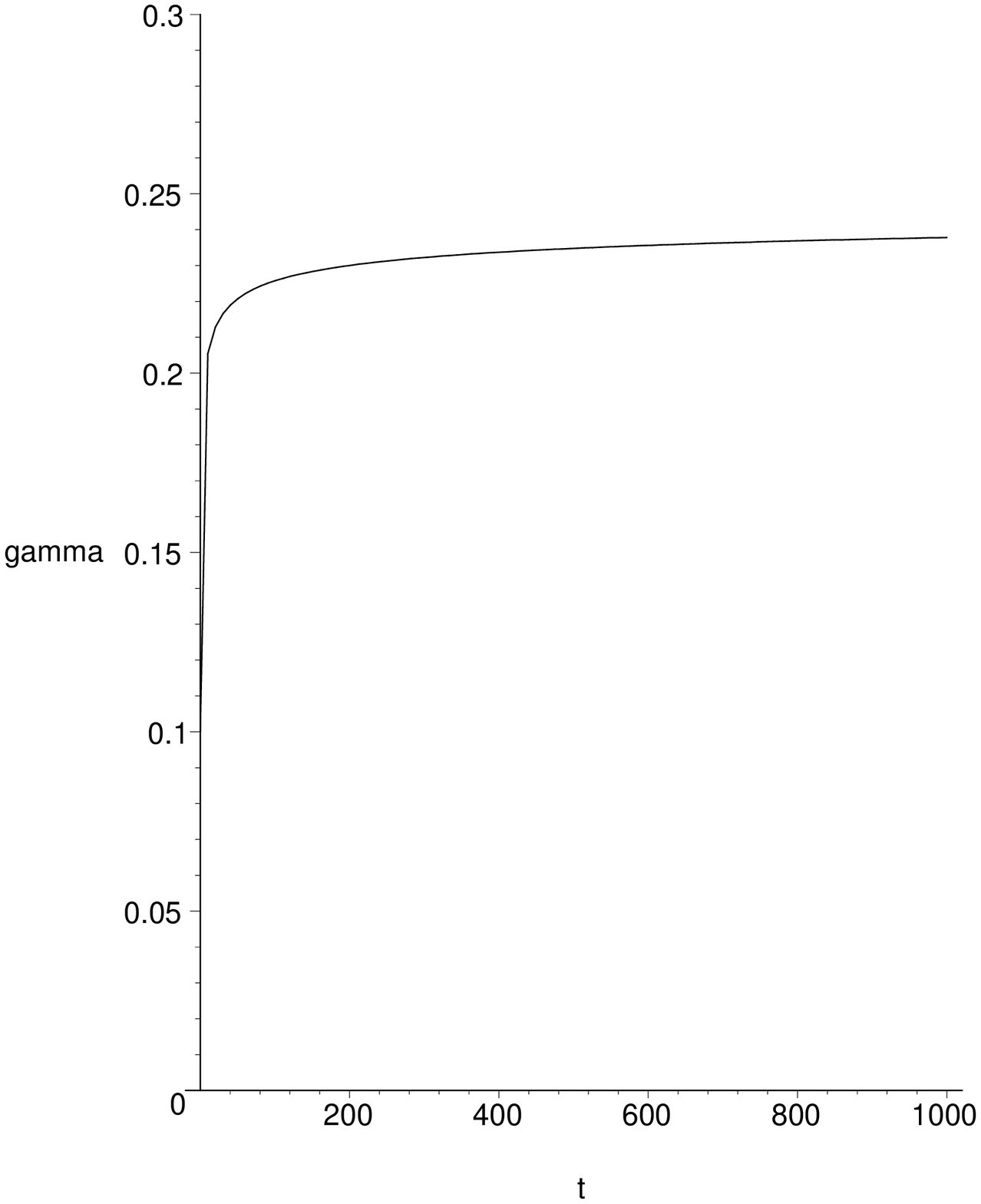}
   \includegraphics[height=2in,width=1.9in]{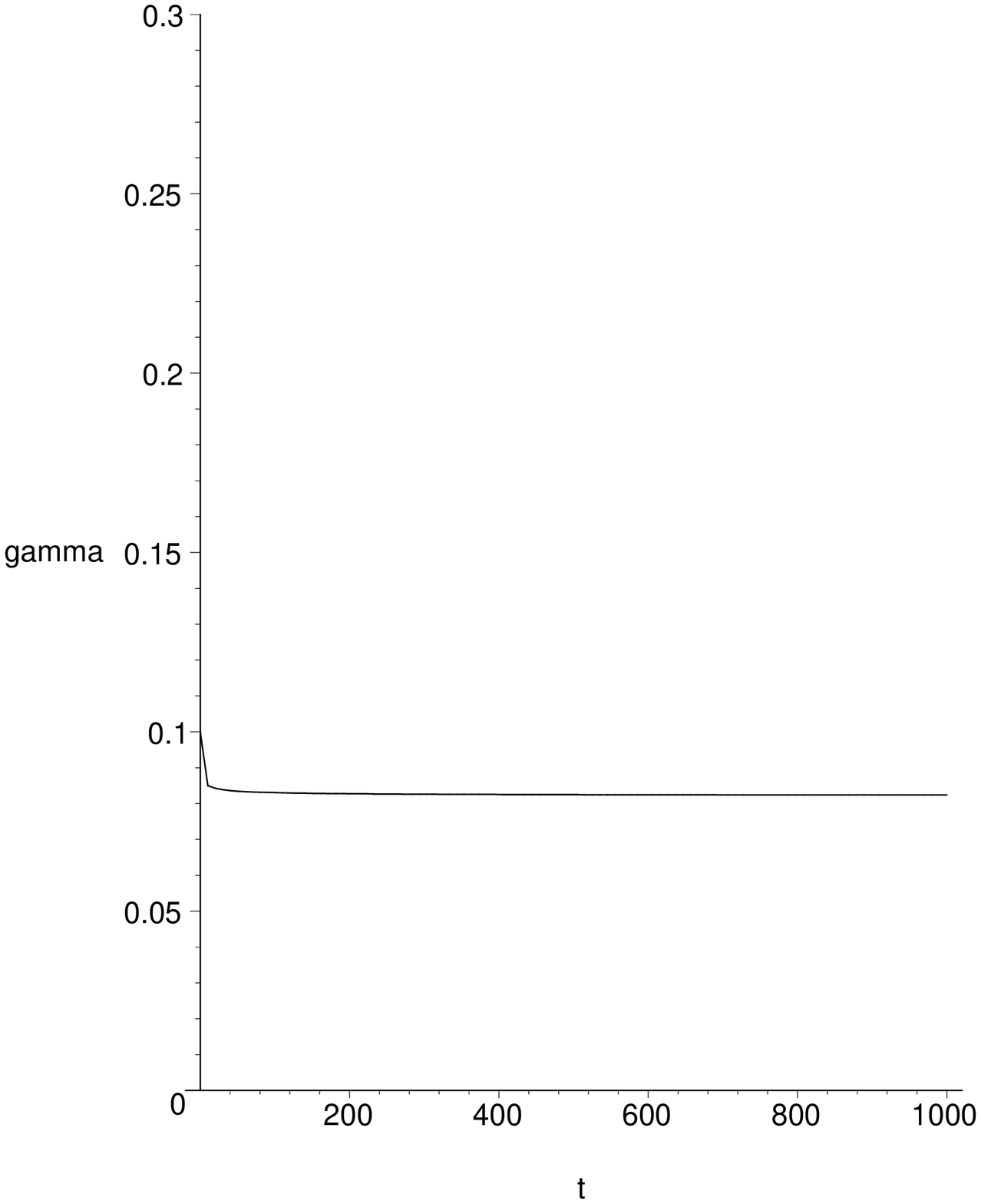}
   \includegraphics[height=2in,width=1.9in]{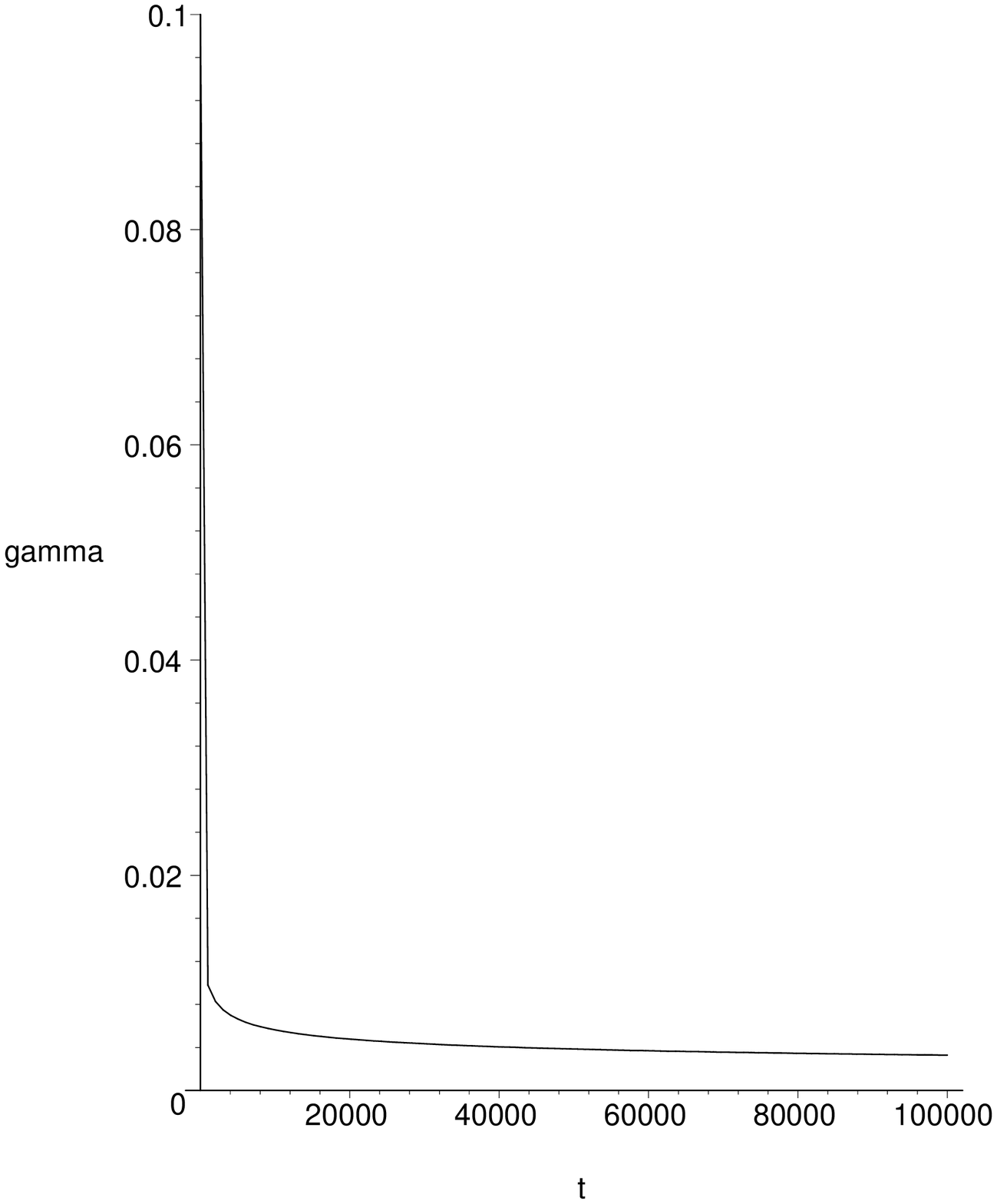}
   \caption{\label{effectofP} Time evolution of the scaling parameter $\gamma$ for
   varying intercommuting probability $P$. From left to right $P=1,
    \; 0.3, \; 10^{-3}$.
   Even for $P = 10^{-3}$ a scaling solution exists with $\gamma \simeq P$.
    However, it takes much longer to reach
   it.}
  \end{figure}

For 3D motion the effects of small $P$ are not overwhelming.  Even
if $P \sim 10^{-3}$  and $v_y =0.36$ strings apparently still scale.
However, the time needed to scale dramatically increases.  See
figure \ref{effectofP}.

\section{Gravitational Emission from Strings with Cusps}

Cosmic strings strongly emit gravitational waves. It was assumed
that the spectrum is Gaussian.  Hence, for $G\mu < 10^{-7}$
gravitational waves from strings were thought to be too weak to be
detected.  Recently Vilenkin and Damour showed that  gravitational
emission from cusps on cosmic strings is strongly non-Gaussian
\cite{damourvil1, damourvil2, damourvil3}. Hence even if $G\mu \sim
10^{-10}$, gravitational waves from strings are detectable by LIGO
and LISA. Cosmic superstrings are distinguished from field theory
strings because their intercommuting probability $P$ may be very
small, while field theory strings always intercommute.  The smaller
$P$ is the stronger the gravitational wave signal is.  It is
therefore remarkable that not only may galaxy-sized superstrings be
detectable, but they may be the  "brightest things in the sky"
\cite{reviewpol}.

\subsection{Cusp formation}

The worldsheet metric $\gamma_{ab}$  is not unique. It is invariant
under worldsheet reparameterizations, $(\sigma,\tau) \rightarrow
(f_{\sigma}(\sigma, \tau), f_{\tau}(\sigma,\tau))$.  The metric can
be fixed by imposing gauge conditions.  First, impose the gauge
condition $\gamma_{00} + \gamma_{11}=0$.  Since $\gamma_{ab} =
\partial_a X^{\mu} \partial_b X_{\mu}$ this leads to $\dot{X}^2 +
X'^2 = 0$.  Next align worldsheet and physical time $\tau = t =X^0$.
In this section we will ignore the extra dimensions and only work
with the strings' 3D properties.  Hence, we take $X^{\mu} = (X^0,
\bx)$ where $\bx$ is a 3-vector. We also impose $\gamma_{01} = 0$.
This gives the two conditions

\begin{eqnarray}
\bxd^2 + \bx'^2 & = & 1 \\
\bxd \cdot \bx' & = & 0 \label{gaugeconstraints}
\end{eqnarray}

We decompose the target space coordinates into left and right moving
components, $ \bx(\sigma,\tau) = \frac{1}{2}[\bx_+(\sigma_+) +
\bx_-(\sigma_-)]$ and $X^0_{\pm} = \sigma_{\pm}$ where $\sigma_{\pm}
= \tau \pm \sigma $. Since $\partial_{\pm} = \frac{1}{2}(
\partial_{\tau} \pm \partial_{\sigma})$, we have $\dot{ \bx} =
(\partial_+ +
\partial_-) \bx = \frac{1}{2}( \dot{\bx}_+ + \dot{\bx}_-)
$ and $\bx' =  \frac{1}{2} (\dot{\bx}_+  -\dot{\bx}_-)$ where we
defined $\dot{\bx}_{\pm} \equiv
\partial_{\pm} \bx_{\pm}$ and $\dot{\bx} = \partial_{\tau} \bx$.
The two gauge conditions then imply ${\dot{\bx}_+}^2(\sigma_+) =
{\dot{\bx}_-}^2(\sigma_-) = 1$ or in 4-vector notation
$(\dot{X}^{\mu}_{\pm})^2 = 0$.  Hence, $\pm \dot{\bx}_+(\sigma_+)$
and $\pm \dot{\bx}_-(\sigma_-)$  live on the unit sphere and trace
out closed curves since they are periodic with respect to $\sigma_+$
and $\sigma_-$.   $\dot{\bx}_+$ and $\dot{\bx}_-$ will generically
intersect on the sphere. In fact, because of the periodicity of
$\bx_{\pm}$, we find that $\int_0^L \dot{\bx}_{\pm} d\sigma_{\pm}
=0$ implying that neither curve can lie in a single hemisphere,
making it nearly impossible for $\dot{\bx}_{+}$  and $\dot{\bx}_-$
not to intersect at some worldsheet point
$(\sigma_+^{(c)},\sigma_-^{(c)})$. At the intersection point:
$\dot{\bx}_+(\sigma_+^{(c)}) = \dot{\bx}_-(\sigma_-^{(c)})$ and the
string velocity is $\bxd^2(\sigma_+^{(c)},\sigma_-^{(c)}) =
\frac{1}{4}[\dot{\bx}_+(\sigma_+^{(c)}) +
\dot{\bx}_-(\sigma_-^{(c)})]^2 =1$. Such a point where parts of a
massive string reach the speed of light are called cusps, and the
string shape near the cusp is of the form $y^3 = x^2$. The cusp is
at the singular point  $(x,y) = (0,0)$.

The importance of cusps is that they strongly emit radiation which
is not exponentially damped in the mode number.  String loops with a
non-singular shape emit their energy as radiation. However, their
spectral power, $P_n$ falls exponentially at large $n$, while for
cusps $P_n$ falls as some negative power of $n$.  Hence, high power
observational signals from cuspless strings are observationally very
difficult to observe. For strings with cusps however, such high
power radiation is much easier to detect. Any detection of such a
high power signal would be provocative circumstantial evidence for
the existence of cosmic strings and depending on the shape of the
observationally measured power spectrum -- evidence of cosmic
superstrings.

\subsection{Gravitational waves from cusps}

Below we give a detailed and hopefully transparent account of
Vilenkin and Damour's calculation of the gravitational wave
amplitude from cosmic string cusps.  More details can be found in
Vilenkin and Damour's original 3 papers,
\cite{damourvil1,damourvil2,damourvil3}.

A gravitational wave in the linearized approximation satisfies

\beq \label{eq2.7} \Box \, \bar{h}_{\mu \nu} = - \, 16 \pi \, G \,
T_{\mu \nu} \, , \eeq

\n which has a solution

\beq \label{eq2.8} \bar{h}_{\mu \nu} (\mbox{\boldmath$x$} , t) =
\frac{\kappa_{\mu \nu} (t-r , \mbox{\boldmath$n$})}{r} + {\cal O}
\left( \frac{1}{r^2} \right) \, . \eeq

\n where

\begin{equation}
\label{eq2.9}
 \kappa_{\mu \nu} (t-r , \mbox{\boldmath$n$}) =
4 \, G \ \sum_{\omega} \ e^{-i \omega (t-r)} \, T_{\mu \nu}
(\mbox{\boldmath$k$} , \omega)    \simeq   2 \, G \, \ell \int
\frac{d \omega}{2\pi} \ e^{-i \omega (t-r)} \ T_{\mu \nu}
(\mbox{\boldmath$k$} , \omega).
\end{equation}

\n On the right hand side we took a high frequency limit (where
cusps are most important) enabling the replacement of
$\sum_{\omega}$ by  $\int d \omega $. Note that $k^{\mu}_m =
(\omega_m, \mbox{\boldmath$k$}_m) \equiv \omega_m (1, \,
\mbox{\boldmath$n$})$, where $\mbox{\boldmath$n$}$ is the direction
of emission.

If we define the {\em logarithmic} Fourier component $\kappa (f)$ as

 \be
\label{eq2.21} \kappa (f) \equiv \vert f \vert \, \widetilde{\kappa}
(f) \equiv \vert f \vert \int dt \ e^{2\pi i f t} \, \kappa (t) \, .
\ee

\n then $\kappa_{\mu \nu}(f)$ can be  conveniently related to
$T_{\mu \nu}(\mbox{\boldmath$k$} , \omega)$  by

\be \label{eq2.22} \kappa_{\mu \nu} (f , \mbox{\boldmath$n$}) = 2 \,
G \, \ell \, \vert f \vert \, T_{\mu \nu} (\mbox{\boldmath$k$} ,
\omega) \, . \ee

As the wave $h(f)$ propagates over cosmological distances it will be
redshifted by the expansion of the universe. One can show that
(\ref{eq2.8}) still holds provided $r \rightarrow a(t) r$ and the
frequency an observer measures today is replaced as $ f \rightarrow
(1+z) f$. Thus

\beq \bar{h}_{\mu \nu}(f) = \frac{\kappa_{\mu \nu}[(1+z)f]}{a(t) r}
\approx \kappa_{\mu \nu}[(1+z)f] \frac{1+z}{t_0 z} \eeq

\n where we set $t=t_0 =\textsf{today}$ and $a_0 = a(t_0)$, and
employed the redshift relation for a flat matter dominated universe,

\beq a_0 r = 3 t_0 \left ( 1 - \frac{1}{\sqrt{1+z}} \right ) \;
\approx \; t_0 \frac{z}{1+z} \label{redshiftrel} \eeq

The  energy momentum tensor for a string is

\be \label{emten} T^{\mu \nu} (x) = \mu \int d \tau \, d \sigma
(\dot{X}^{\mu} \, \dot{X}^{\nu} - X'^{\mu} \, X'^{\nu}) \,
\delta^{(4)} (x - X(\tau , \sigma)) \, . \ee

\n which for periodic motion in time has the Fourier transform

\begin{eqnarray}
\label{eq2.10}
 T_{\mu \nu} (\mbox{\boldmath$k$} , \omega)
&  =  & \frac{1}{{\cal{T}}} \ \int_0^{{\cal{T}}} dt \int d^3
\mbox{\boldmath$x$} \, e^{i (\omega t - \mbox{\boldmath$k$} \cdot
\mbox{\boldmath$x$})} \, T_{\mu \nu} (\mbox{\boldmath$x$} , t)\\
 & = & \frac{\mu}{{\cal{T}}} \ \int_{\sum'} d \tau \, d \sigma
(\dot{X}_{\mu} \, \dot{X}_{\nu} - X'_{\mu} \, X'_{\nu}) \, e^{-i k
\cdot X} \\
& = & \frac{\mu}{2 {\cal{T}}} \ \int_{\sum'} d\sigma_+ \, d\sigma_-
\, \dot{X}_+^{(\mu} \dot{X}_-^{\nu)} \, e^{-\frac{i}{2} (k_m \cdot
X_+ + k_m \cdot X_-)} \nonumber\\
& = &  \frac{\mu}{2 {\cal{T}}} \ I_+^{(\mu} I_-^{\nu)} \, .
\end{eqnarray}

 \n Here ${\cal{T}}
= 2 \pi/\omega_1 = \ell/2$ is the "fundamental" oscillation period
of the string and $\ell$ is the length of the string.\footnote{See
Vilenkin and Shellard, p. 158 for an explanation why the period is
$\ell/2$.}  $\Sigma'$ is the worldsheet strip swept out in one
period. In the third  line we moved to $\sigma_{\pm}$ coordinates.
In the last line, $I^{\mu}_{\pm}$ is defined as

\be \label{eq2.18l} I_{\pm}^{\mu}(k_m) \equiv \int_0^{\ell}
d\sigma_{\pm}
 \dot{X}_{\pm}^{\mu} e^{-\frac{i}{2} k_m \cdot X_{\pm}} \, . \ee

A cusp will form if there is some $\ell^{\mu}$ such that  $
\ell^{\mu} = \dot{X}_+^{\mu} (\sigma_+^{(c)}) = \dot{X}_-^{\mu}
(\sigma_-^{(c)}) $. Recall that $\dot{X}_{\pm}^{\mu}
(\sigma_{\pm}^{(c)})$ is null and hence $\ell^{\mu}$ must be null.
$I_{\pm}^{\mu}$ will decrease exponentially unless the phase of
$I_{\pm}^{\mu}$ possesses a saddle point such that $\partial_{\pm}
(k_m \cdot X_{\pm}) = 0$.   Since both $k_m$ and $X_{\pm}^{\mu}$ are
null, a saddle point at the worldsheet point $(\sigma_{+}^{(c)},
\sigma_-^{(c)})$, will exist if $k_m^{\mu} \propto
\dot{X}_{\pm}^{\mu} (\sigma_{\pm}^{(c)}) = \ell^{\mu}$. Since
$k^{\mu}_m = \omega_m (1, \, \mbox{\boldmath$n$})$, and
$\dot{X}_{\pm}^{\mu} = (1, \bxd_{\pm})$, and  $\bxd^2_{\pm} =1 =
\mbox{\boldmath$n$}^2$, we find that $k_m^{\mu} = \omega_m
\ell^{\mu}$. Note: motion of the cusp is parallel to the direction
of emission (assuming a saddle point) since
$\dot{X}^{\mu}(\sigma_+^{(c)}, \sigma_-^{(c)}) =
\frac{1}{2}[\dot{X}_+^{\mu}(\sigma_+^{(c)}) +
\dot{X}_-^{\mu}(\sigma^{(c)}_-)] = \frac{1}{2} [ \ell^{\mu} +
\ell^{\mu}] = \ell^{\mu} $.

To evaluate $I_{\pm}$ for a cuspy string we shift the origin of
$\sigma_{\pm}$ so that $\sigma_{\pm}^{(c)} = 0$, and the origin of
$X^{\mu}$ so that $X^{\mu}(\sigma_+^{(c)}, \sigma_-^{(c)}) = 0$. The
Taylor expansions around $X^{\mu}_{\pm}(\sigma_+^{(c)},
\sigma_-^{(c)})$ are

\be \label{eq3.2} X_{\pm}^{\mu} (\sigma_{\pm}) = \ell^{\mu} \,
\sigma_{\pm} + \frac{1}{2} \ \ddot{X}_{\pm}^{\mu} \, \sigma_{\pm}^2
+ \frac{1}{6} \ X_{\pm}^{(3)\mu} \, \sigma_{\pm}^3 \, , \ee \be
\label{eq3.3} \dot{X}_{\pm}^{\mu} (\sigma_{\pm}) = \ell^{\mu} +
\ddot{X}_{\pm}^{\mu} \, \sigma_{\pm} + \frac{1}{2} \
X_{\pm}^{(3)\mu} \, \sigma_{\pm}^2 \, , \ee

Differentiating the  constraint $\dot{X}_{\pm}^2 = 0$ mentioned in
the paragraph below (\ref{gaugeconstraints}) yields the relations
$\dot{X}_{\pm} \cdot \ddot{X}_{\pm} = 0$ and $\dot{X}_{\pm} \cdot
{X}_{\pm}^{(3)} + \ddot{X}_{\pm}^2 = 0$. Therefore, at the cusp, one
has $\ell \cdot \ddot{X}_{\pm} = 0$ and $\ell \cdot X_{\pm}^{(3)} =
- (\ddot{X}_{\pm})^2$. Thus the phase factor in (\ref{eq2.18l}) is

\be \label{eq3.4} k_m \cdot X_{\pm}  = \omega_m \ell_{\mu} \,
X_{\pm}^{\mu} (\sigma_{\pm}) = - \frac{\omega_m}{6} \
(\ddot{X}_{\pm}^{\mu})^2 \, \sigma_{\pm}^3 \, . \ee

Inserting these results in .~(\ref{eq2.18l}) leads to an expression
of the form

\be \label{eq3.5} I_{\pm}^{\mu} = \int  d \, \sigma_{\pm}
(\ell^{\mu} + \ddot{X}_{\pm}^{\mu} \, \sigma_{\pm} + \cdots) \, \exp
\left (\frac{i\omega_m}{12} \ddot{X}_{\pm}^2 \, \sigma_{\pm}^3
\right )\, . \ee

Around the saddle point,$(\sigma_+^{(c)},\sigma_-^{(c)}) = (0,0)$ we
have $I_{\pm}^{\mu} \simeq a_{\pm} \, \ell^{\mu} + b_{\pm}^{\mu}$.
The leading order term  is  $a_{\pm} \, \ell^{\mu}$ and
$b_{\pm}^{\mu}$ is the subleading term. Since $\kappa_{\mu \nu} \sim
I_+^{(\mu} I_-^{\nu)}$, the leading-leading order term is $a_+ a_-
\ell^{\mu} \ell^{\nu}$ and the leading-subleading terms are
$a_+\ell^{\mu} b^{\nu}$ and $a_-\ell^{\nu} b^{\mu}$.  These 3 terms
can be gauged away by a coordinate transformation under which $
\kappa_{\mu \nu} \rightarrow \kappa_{\mu \nu} + k_{\mu} \, \xi_{\nu}
+ k_{\nu} \, \xi_{\mu}$ where $k^{\mu} = \omega \ell^{\mu}$.  Thus,
the leading order physical part of $\kappa_{\mu \nu}$ is
$\kappa_{\mu \nu} \sim I_+^{(\mu} I_-^{\nu)} = b_+^{(\mu}
b_-^{\nu)}$ with

\be \label{eq3.7} b_{\pm}^{\mu} \! \simeq \! \ddot{X}_{\pm}^{\mu}
\int d \, \sigma_{\pm} \, \sigma_{\pm} \, \exp \left( \frac{i}{12}
\, \omega_{m} \, \ddot{X}_{\pm}^2 \, \sigma_{\pm}^3 \right) \!  = \!
\frac{\ddot{X}_{\pm}^{\mu}}{N_{\pm}^2} \int_{-\infty}^{\infty} \! \!
\! \! d \, u_{\pm} \, u_{\pm} \, e^{{ i \, {\rm sign}(m) u_{\pm}^3}}
\! = \! i \, {\rm sign}(m) \frac{\ddot{X}_{\pm}^{\mu}}{N_{\pm}^2}
\frac{2\pi}{3 \Gamma(\frac{1}{3})} \ee

Recall that $\omega_m = m \omega_1$.  The integral depends on the
sign of $m$.  Here, $u_{\pm} = N_{\pm} \sigma_{\pm}$ where
$N_{\pm}^3 = |\omega_m| \ddot{X}_{\pm}^2/12$. Since most of the
integral over $d \sigma_{\pm}$ comes from around the saddle point,
$\sigma_{\pm} = 0$, we extended the integration limits of $u_{\pm}$
to $\pm \infty$.

The logarithmic Fourier transform is real, independent of the sign
of $m$ and is

\beq  \kappa^{\mu \nu} (f) \   \simeq \ - C \frac{G \mu}{(2 \pi
|f|)^{1/3}}\frac{ \ddot{X}_+^{(\mu} \ddot{X}_-^{\nu)}}{( |
\ddot{X}_{+}| | \ddot{X}_{-}|)^{4/3}} \  \sim  \ \frac{G\mu
\ell}{\sqrt[3]{|f| \ell}} \ \Theta(\, \theta_0(f) - \theta \, )
\label{kmunu} \eeq

\n where $C = \frac{4 \pi (12)^{\frac{4}{3}}}{\left( 3 \Gamma \left(
\frac{1}{3} \right) \right)^2  }$ and we have estimated
$|\ddot{X}_{\pm}| \sim 2 \pi/\ell$. We could do this because
$X^{\mu}_{\pm}$ has the Fourier expansion $X^{\mu}_{\pm} \sim
\frac{\ell}{2\pi} \sum \frac{\alpha_n^{\mu}}{n} e^{2 \pi i n
\sigma_{\pm}/\ell}$ which gives $|\ddot{\bx}_{\pm}|^2 =
(\frac{2\pi}{\ell})^2\sum n^2|{\mbox{\boldmath $
\alpha$}}_{\pm}|^2$.  Thus for not very large $n$ we find
$|\ddot{\bx}_{\pm}|\sim \frac{2\pi}{\ell}$ since $1 =|\bxd_{\pm}|^2
= \sum |{\mbox{\boldmath $ \alpha$}}_{\pm}|^2$. The Heaviside
$\Theta(\theta_0 - \theta)$ function appears because as Damour and
Vilenkin show: if the direction of emission, $\bk$, (the direction
to the observer) is misaligned from the direction of motion of the
cusp ${\mbox{\boldmath$\ell$}}$, by more than an angle $\theta_0$,
the emission amplitude is exponentially suppressed. For example,
Damour and Vilenkin show that if  $\bk \nparallel
{\mbox{\boldmath$\ell$}}$ by more than an angle $\theta_0 \simeq
(\frac{2}{|f| \ell})^{1/3}$ then $I_{\pm}^{\mu}$ doesn't possess a
saddle point and decays exponentially.

Note (\ref{kmunu}) crucially depends of the absolute value $|f|
\propto |m|$ because the integrals in (\ref{eq3.7}) depend on the
sign of $m$. This absolute value dependence is what causes
distinctive ``spikiness" of gravitational emission from cusps. Also
note that because the vectors
$\ddot{X}_{\pm}/|\ddot{X}_{\pm}|^{4/3}$ are spacelike and orthogonal
to $\ell^{\mu}$, that gravitational wave in (\ref{kmunu}) is
linearly polarized.

From now on we will usually assume that the observed frequency $f$
is positive and will thus drop the absolute value signs.  Using
(\ref{kmunu}) and (\ref{redshiftrel}), the wave amplitude $h(f)$ for
a flat matter-dominated universe is

\beq h(f) \sim \frac{G \mu \ell}{\sqrt[3]{(1+z)f \ell}}\;
\frac{1+z}{t_0 z} \;\Theta \left (\theta_0(f,z) - \;\;\theta \right
) \label{logamp} \eeq

Cosmic string loops will have a size of order the correlation length
of the string network which is $L$.  If the strings scale such $L
\sim t$ is a fixed proportion of the horizon size $t$, then the loop
size $\ell$ will be $\ell = \alpha t$. We define $\alpha = \epsilon
\Gamma G\mu$ where $\Gamma \sim 50$ and $\epsilon$ are dimensionless
numbers.  $\epsilon$ measures how close $\alpha$ is to $\Gamma G
\mu$.  A very wiggly string with lots of small-scale structure may
have a very small $\epsilon \sim 10^{-10}$.  Smooth strings
correspond to $\epsilon \sim 1$. Since $\ell \sim t$, and we would
like to express (\ref{logamp}) in terms of the redshift $z$, we need
$t$ as a function of $z$. This is a complicated function because
$a(t)$ takes different functional forms during the matter and
radiation epochs. An approximate interpolating function relating $z$
to $t$ in both epochs is $t= t_0 \varphi_{\ell}$ where
$\varphi_{\ell}(z) = (1+z)^{-3/2} (1 + z/z_{eq})^{-1/2} $.  If for
convenience, we define the new function $\varphi_h(z) =
[\varphi_{\ell}(z)]^{2/3} z^{-1} (1+ z)^{2/3} $ then we can write

\beq  h(f) \sim \frac{G \mu \alpha^{2/3} }{\sqrt[3]{f t_0}}\;
\varphi_h(z)  \;\Theta \left ( \theta_0(f,z)  \;- \; \theta \right )
\label{idiot} \eeq

\n where $f\rightarrow f(1+z)$ and $t = t_0 \varphi_\ell$.

\beq \theta_0 = \frac{1}{\sqrt[3]{\alpha f t_0 (1+z)
\varphi_{\ell}}} = \frac{(1+z)^{1/6}
(1+z/z_{eq})^{1/6}}{\sqrt[3]{\alpha f t_0}} \label{anglered} \eeq

We would like to understand how $h(f)$ depends on the  burst rate
$\dot{N}$ for gravitational waves from cosmic strings. The burst
rate changes with redshift as

\beq  \frac{d \dot{N}}{dz} \sim \frac{\theta_0^2}{4} (1+z)^{-1}
\frac{c n_{\ell}}{{\cal{T}}} \frac{dV}{dz} \label{bursty} \eeq

We understand (\ref{bursty}) as follows.  The number of bursts per
string oscillation period $c$, divided by the string period
${\cal{T}}$, gives the number of cusps produced per unit time,
$c/{\cal{T}}$. Multiplying that by the density of loops $n_{\ell}$,
gives the (number of cusps)/(spacetime volume). Multiplying that by
$dV(z)$, the spacetime volume in the interval $(z+dz, dz)$, gives
the number of cusps produced between redshifts $z+dz$ and $z$ --
this is $c n_{\ell}{\cal{T}}^{-1} dV(z)$. We now multiply by the
fraction of cusps whose motion is in the same direction as an
earthbound observer, ($\theta < \theta_0$).  This is the amount of
solid angle subtended by a cone with opening angle $\theta_0$. Since
$\theta_0$ is small, this is approximately $ \pi \theta_0^2/(4 \pi)
= \theta_0^2/4$. Finally, since $\dot{N}$ is a rate, and observed
time redshifts as, $dt_{obs} = (1+z) dt$, we must multiply by
$(1+z)^{-1}$.

For a flat universe,

 \beq \frac{dV}{dz} =
 \left \{\begin{array}{cc}
  54 \pi t_0^3 [(1+z)^{1/2} -1]^2(1+z)^{-11/2}  &  z <z_{eq} \\
  72 \pi t_0^3 (1+z_{eq})^{1/2} (1+z)^{-5} &  z > z_{eq}  \\
\end{array} \right \} \label{volred}
\eeq

The loop density $n_{\ell}$ is the number of strings in a
correlation length volume, $L^3 \sim t^3 = (t_0 \varphi_{\ell})^3$.
In order for strings to scale, long strings must release their
energy in the form of loops.  If all the length in a correlation
volume were to go to loops, then the number of loops formed per time
$t\sim L$ and per volume $L^3$ would be $\frac{L(t)}{\ell(t)}$. This
implies that at least $\frac{L(t)}{\ell(t)}$
collisions/self-intersections are required. However, if probability
of a collision leading to the formation of a loop is $P$, then more
collisions are needed for the strings to scale. In particular, $1/P$
times more collisions are required. This can only happen if there
are at least $1/P$ strings per $L^3$, instead of just 1 string per
$L^3$.  Then the number of loops needed  for scaling is $N_s \sim
\frac{1}{P} \frac{L(t)}{\ell(t)} \sim \frac{1}{P \alpha}$ using
$L\sim t, \ell = \alpha t$.  The lifetime of a loop is $\tau \sim
\epsilon t$.\footnote{Very approximately: using the quadrapole
formula a string's rate of energy loss by radiation is $\dot{E} =
\Gamma G \mu^2$. Thus $\tau \sim \ell/\dot{E}
 = \mu \alpha t/\Gamma G \mu^2 = \epsilon t$.} Thus, the number of strings
which survive till time $t$ is $\frac{\tau}{t} N_s$. Hence the loop
density at time $t$ is then $n_{\ell} \sim
\frac{\epsilon}{P\alpha}t^{-3}$. (Note the energy density goes as
$n_{\ell} \ell \sim \mu/Pt^2  \equiv \mu/L^2$ if $\epsilon =1$
implying that the scaling solution is $L \sim \sqrt{P} t$.)

If we plug in $n_{\ell}$, the period ${\cal{T}} = \frac{\ell}{2} =
\frac{\alpha t_0 \varphi_{\ell}}{2}$, and equations (\ref{anglered})
and (\ref{volred}) into (\ref{bursty}), we get a piecewise very
complicated function. But, by introducing a new interpolating
function $ \varphi_n(z) = z^3 (1+z)^{-7/6}(1+ z/z_{eq})^{11/6}$ we
can write a simpler expression for $d\dot{N}/dz$:

\beq \frac{d \dot{N}}{d \ln z} \sim \frac{100 \epsilon c}{ P t_0
\alpha^{8/3}} \frac{ \varphi_n(z) }{ (f t_0)^{2/3}}
\label{dNdlnz}\eeq

\n $\varphi_n(z)$ is a strongly increasing function and varies as
$z^{11/3}$ for large $z$.  Thus $d \dot{N}/d \ln z$ is strongly
peaked at large $z$. If $z_m$ is the largest allowed redshift, then
we can approximate

\beq \dot{N} = \int^{z_m} \left (\frac{d\dot{N}}{d \ln z} \right ) d
\ln z \approx \frac{d \dot{N}}{d \ln z}(z_m)  \label{lnz} \eeq

Thus using (\ref{dNdlnz}) and (\ref{lnz}), we can write

\beq \varphi_n (z_m) =  10^{-2} \frac{P\dot{N}}{\epsilon c} t_0
\alpha^{8/3} (ft_0)^{2/3} \equiv y(\dot{N}, f) \label{yfunc} \eeq

Note, the L.H.S of (\ref{yfunc}) depends on only $z_m$, while the
R.H.S is a function of only  $\dot{N}$ and $f$.  We call this
function $y(\dot{N},f)$.   Equation (\ref{yfunc}) can thus be
inverted. We can then write the maximum redshift $z_m$ as a function
of the observed $\dot{N},f$ as

\beq z_m(y) = y^{1/3} (1+y)^{7/33}(1+y/y_{eq})^{-3/11} \eeq

\n with $y$ given by (\ref{yfunc}).   Note, if $y < 1 \Rightarrow
z_m<1$, if $1<y< y_{eq} \Rightarrow 1 < z_m < z_{eq}$, and if $y >
y_{eq} \Rightarrow z_m > z_{eq}$ where $y_{eq} = z_{eq}^{11/6}$.  We
can then stick $z_m$ into (\ref{idiot}), and $\theta_0(z,f)$ to get
an expression for the wave amplitude for given values of $\dot{N}$
and $f$. The dependence on $z$ has thus been replaced by the
dependence on $\dot{N}$ and $f$. Thus,

\beq  h(f,\dot{N},y) = G \mu \alpha^{2/3}
\frac{g(y)}{\sqrt[3]{ft_0}} \Theta ( \theta_0 - \theta) =
\Gamma^{2/3} (G \mu)^{5/3}\epsilon^{2/3} \frac{g(y)}{\sqrt[3]{ft_0}}
\Theta ( \theta_0 - \theta) \eeq

\n where on R.H.S, we have replaced $\alpha$ by $\alpha = \epsilon
\Gamma G\mu$ and where

\beq g[y(\dot{N}, f)] = \varphi_h(z_m(y))  = y^{-1/3} (1+y)^{-13/33}
(1+ y/y_{eq})^{3/11} \eeq

\begin{figure}
\includegraphics[height=2.7in,width=3in]{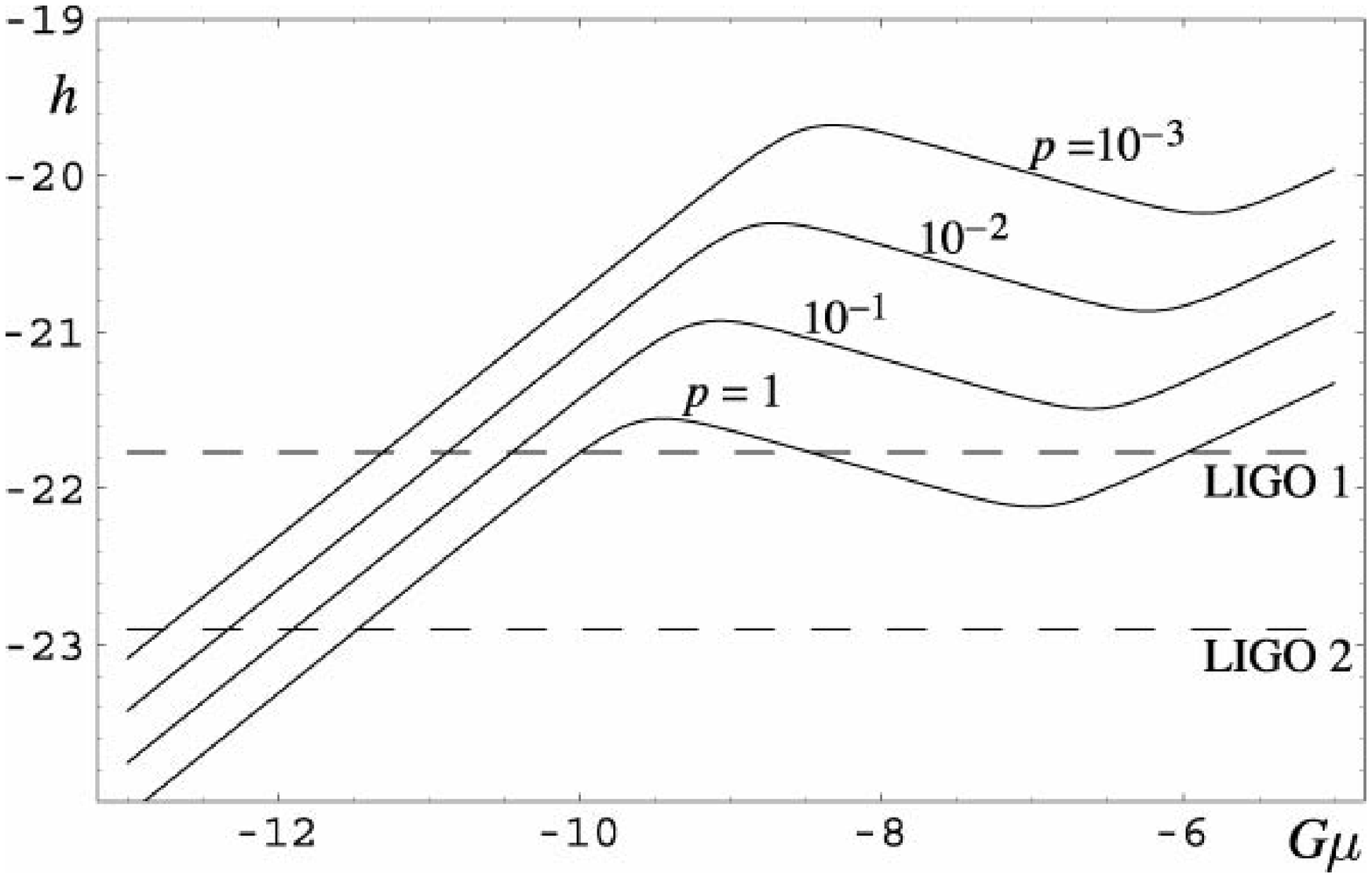}
\includegraphics[height=2.7in,width=3in]{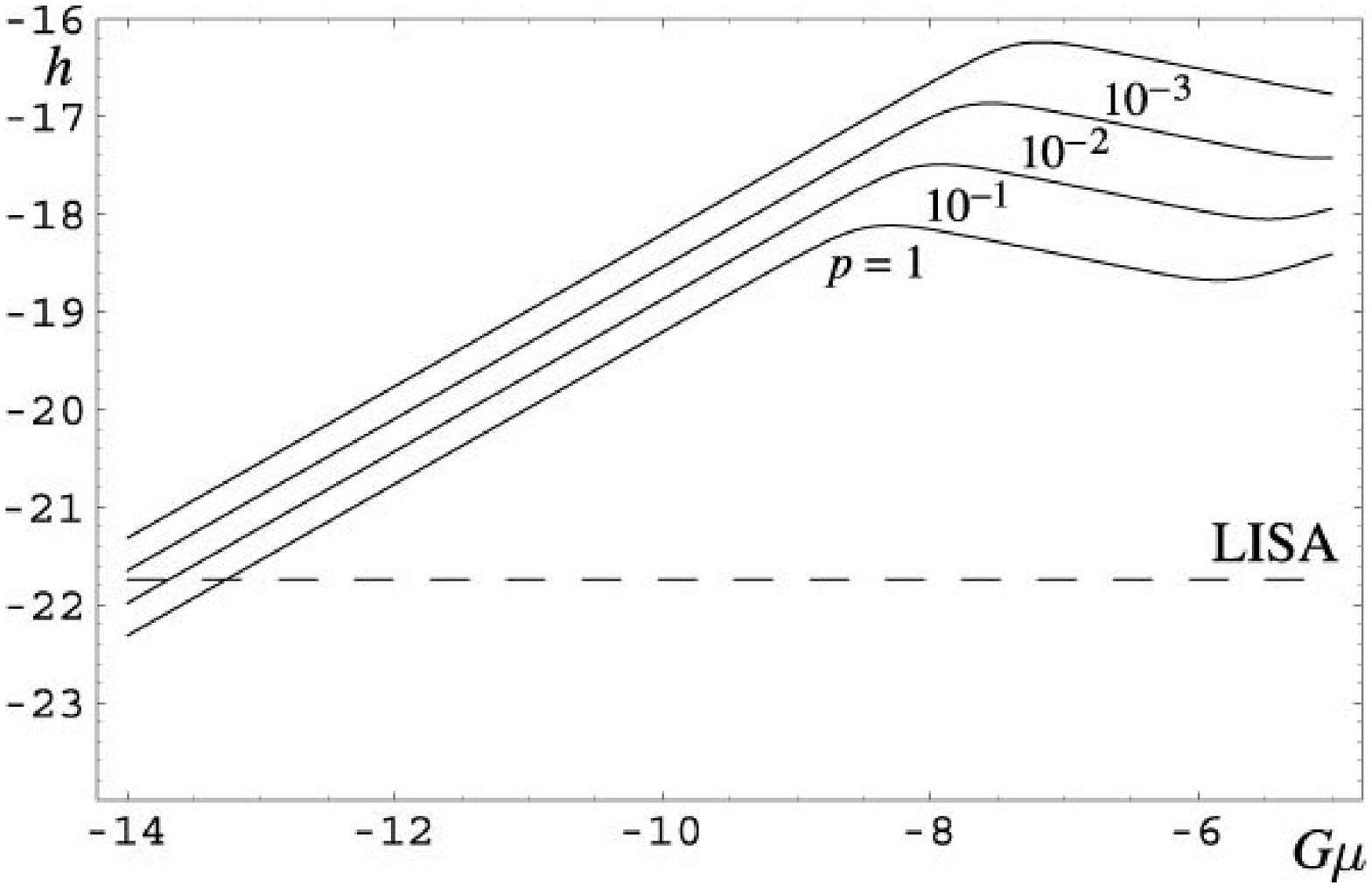}
\includegraphics[height=2.7in,width=3in]{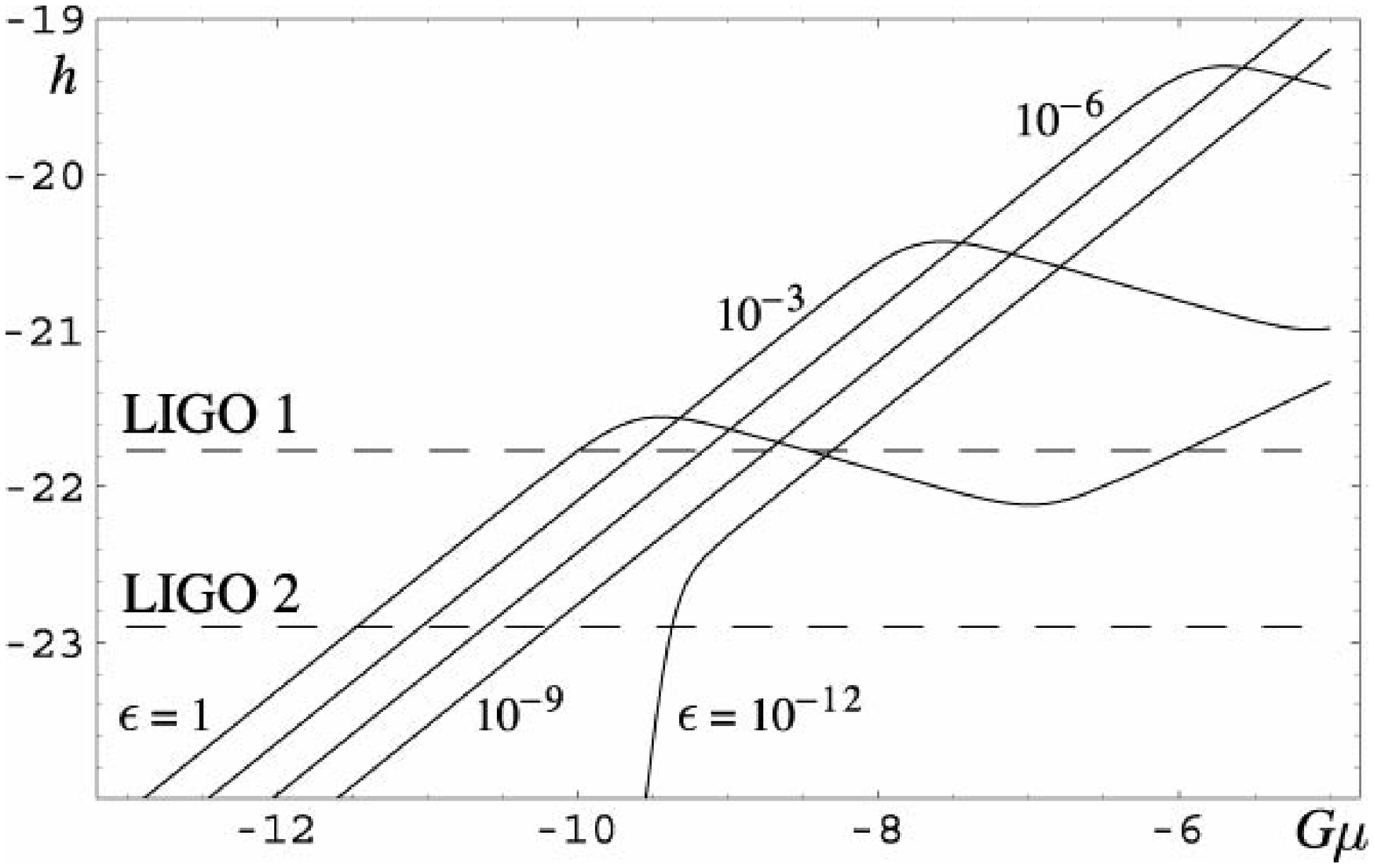}
\includegraphics[height=2.7in,width=3in]{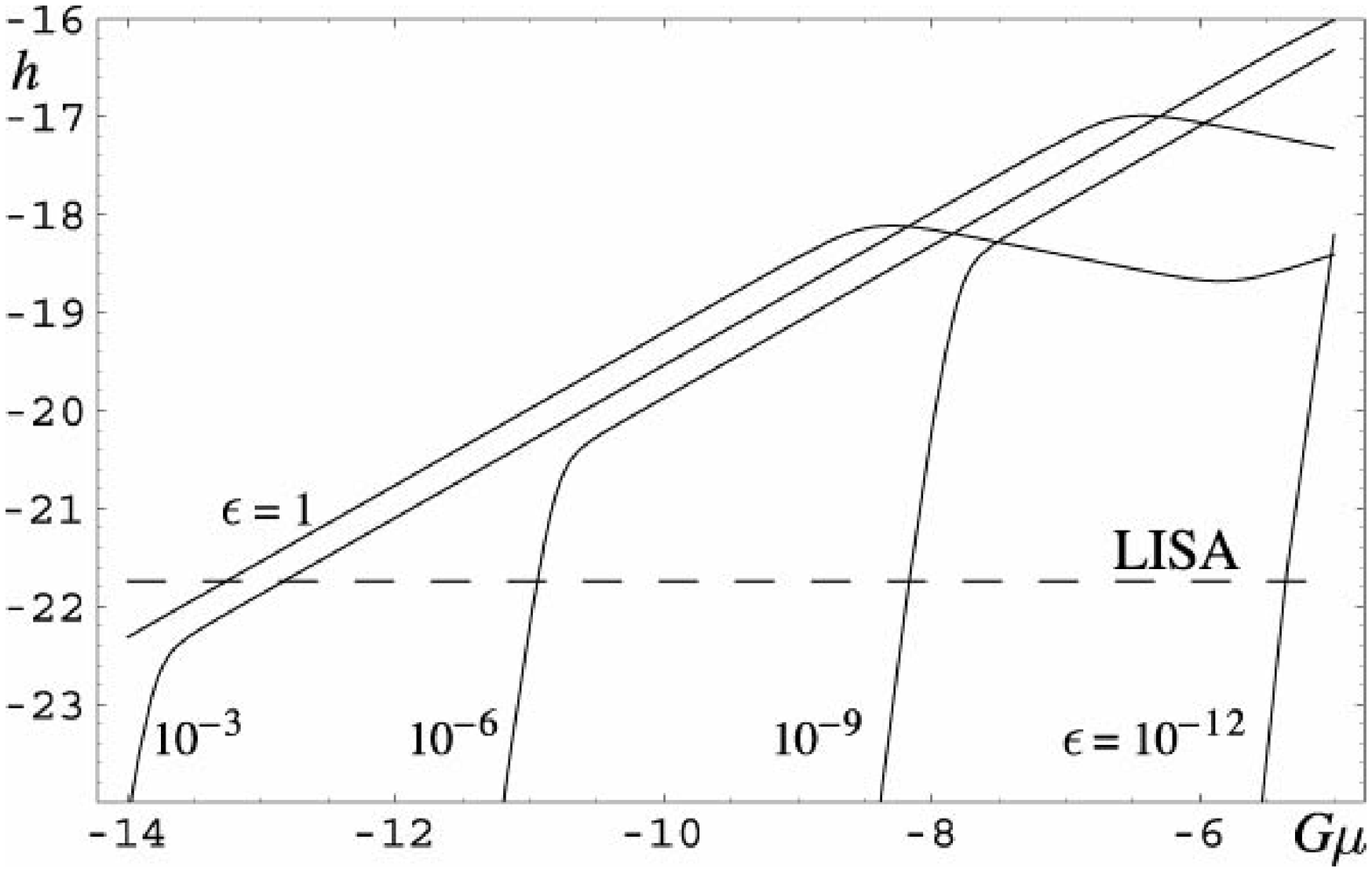}
\caption{Log-log plots of $h(f)$ versus $G\mu$.  Starting from top
left: (1) at $f_{LIGO} = 150$ Hz for $10^{-3} \le P \le 1$; (2) at
$f_{LISA} = 3.88 \cdot 10^{-3}$ Hz for $10^{-3} \le P \le 1 $; (3)
at $f_{LIGO}$ for $10^{-12} \le \epsilon \le 1$; (4) at $f_{LISA}$
for $10^{-12} \le \epsilon \le 1$. } \label{cuspfigures}
\end{figure}

Figure \ref{cuspfigures} shows log-log plots of $h(f)$ versus $G
\mu$ for: (1) $f = f_{LIGO} = 150$ Hz for $1 \le P \le 10^{-3}$;
  (2) $f = f_{LISA} = 3.88 \cdot 10^{-3}$ Hz for $1 \le P \le 10^{-3}$;
  (3) $f = f_{LIGO} = 150$ Hz
for $10^{-12} \le \epsilon \le 1$; and (4)  $f = f_{LISA} = 3.88
\cdot 10^{-3}$ Hz for $10^{-12} \le \epsilon \le 1$.  All the plots
assumed $c =1$, i.e. that one cusp per period was produced and that
$\dot{N} = 1/year$. The burst amplitude rises as $G\mu$ becomes
larger than $10^{-12}$. Then it falls and then rises again. The
local maximum and minimum depend on $f, p$ and $\epsilon$.

The dependence of $h(f)$ on $G \mu$ is understood as follows. Now
from (\ref{yfunc}), $y \sim (G\mu)^{8/3}$.  Also, $h(f,y) \sim (G
\mu)^{5/3}g(y)$ has the power law behavior $h\sim (G \mu)^{5/3}g
\sim (G \mu)^{5/3}y^n$ where $n= -1/3, -8/11, -5/11$ for $y \lsim
1$, $1 \lsim y \lsim y_{eq}$ and $y \gsim y_{eq}$ respectively. Thus
$h \sim (G\mu)^k$ where $k = 7/9,-3/11,5/11$ for $0< z_m <1$, $1 <
z_m < z_{eq}$, and $z > z_{eq}$.

$G\mu$ measures the radiation power of the string. So it is not
surprising that apart from the middle regime $ 1 <z_m < z_{eq}$ that
the burst amplitude increases with $G \mu$.  However, the loop
density varies as $(G\mu)^{-1}$; hence the number of radiation
emitting loops decreases with increasing $G\mu$.  For $z_m<1$ and
$z_m
>z_{eq}$, the radiation power effect dominates the loop density
effect. However, in the middle regime, the two effects are
comparable and the fact that fewer loops are produced is actually
more important than the increase in radiation power. Hence, for $1
<z_m < z_{eq}$ the burst amplitude decreases with increasing string
gravity, $G \mu$.

The dependence on $P$ and $\epsilon$ is explained as follows. Since
$y \sim P \epsilon^{5/3}$, we find $h(f) \sim \epsilon^{2/3} (P
\epsilon^{5/3})^n \sim P^n \epsilon^{(2+5n)/3}$. Therefore the
$\epsilon,P$ dependence is $h \sim P^{-1/3} \epsilon^{1/9}$ and
$P^{-8/11} \epsilon^{-6/11}$ and $P^{-5/11} \epsilon^{-1/11}$ for
the three regimes respectively.

Thus decreasing $P$ increases $h$ slightly. If $P$ decreases to $P
=10^{-3}$, $h$ increases by an order of magnitude.  The dependence
on $\epsilon$ is very weak, and unless $\epsilon$ decreases to
$10^{-10}$, $h(f)$ hardly changes.  However, if $G\mu \sim 10^{-10}$
as in the KKLMMT scenario, then taking $\epsilon \rightarrow
10^{-10}$ takes $\theta_0 \rightarrow 0$.  LISA and LIGO can't
resolve such angles $\theta < \theta_0 \sim 0$.  Thus strings with
$\epsilon \sim 10^{-10}$ will affect $h$, but LISA not LIGO will be
able to detect them.  In general, for $p < 1, \epsilon < 1$,
decreasing $P$ or $\epsilon$ moves  $h$ to the right and upwards and
therefore increases the local minimum and maximum.

\section{Conclusions}

The possibility that galaxy sized strings -- the same objects which
ordinarily live at $10^{19}$ GeV  -- may exist and be observed is
extraordinarily tantalizing.   Therefore, the recent claim that the
double galaxy known as CSL-1 is a double image produced by a comic
string of a single galaxy is remarkable \cite{csl}. It may
experimentally vindicate string theory. However, getting to the
point where this or another candidate object may be conclusively
identified as a cosmic superstring and distinguished from a gauge
theory string is a long and winding road.

Below we try to expose the reader to some the debates, controversies
and open questions regarding cosmic superstrings.


{\sc $\bullet$ String Production:}  Barnaby et al \cite{barnaby}
have claimed that the unusual nature of tachyon condensation allows
the Kibble mechanism to operate in the extra dimensions and allows
them to be populated with topological defects.  Their analysis is
based on the DBI effective action for a tachyon. In this
approximation the tachyon never fully condenses and interactions are
suppressed because $g_s$ is taken to be virtually zero. The DBI
action is very much an approximation to the full tachyon action
\cite{kutasov} and so it would be very interesting to know what
happens in a model of tachyon condensation where $g_s$ is not very
small. Another related issue is: are infinite F-strings created by
tachyon condensation? There is no phase transition language to
describe F-string creation using tachyon condensation (unless $g_s
\gg 1$) and hence it is difficult to understand whether the F-string
creation process percolates.

{\sc $\bullet$ Inflation and Reheating:}  We have completely ignored
issues of inflation and reheating.  However, since most of the
community tends to view string creation as a by-product of inflation
it is important to know which  brane inflation models produce cosmic
strings. It turns out that the tachyon is real in most brane
inflation scenarios not involving brane-anti-brane interactions.
Hence, codimension 2 defects (i.e. strings) are {\em not} formed
\cite{racetrack}. Hence, cosmic string creation scenarios tend to
need branes and anti-branes.

{\sc $\bullet$ String scattering:} An effective theory able to
accurately calculate intercommutation probabilities would be useful.
However, because string intercommutation, particularly  D-D
reconnection is a very stringy process  involving a tachyon, tower
of states, etc., it is unclear how an effective supergravity
approach can capture the essence of string reconnection. For
example, the supergravity D-term strings approach becomes invalid
once strings come very close to each other
\cite{fayetilio1,fayetilio2}.  However, the 2D SYM toy model of
Hanany and Hashimoto does seem to verify the non-perturbative nature
of D-D reconnection \cite{khashimoto05}.

{\sc $\bullet$ String scaling:} As we discussed in the paragraph
below (\ref{volred}) when we determine the scaling solution taking
into account only the inter-string distance and correlation length
we expect $\gamma \sim \sqrt{P}$, where $L = \gamma t$. However,
despite the simulation \cite{saka}, other simulations seem to
instead suggest that $\gamma \sim P^{1/3}$ or $P$ to some other
fractional power \cite{shellardtalk}. This is probably because the
small scale structure of the colliding strings is important in
determining $\gamma$ and cause $\gamma$ to deviate from $\sim \! \!
\sqrt{P}$. Also, despite numerical evidence that $(p,q)$ string
networks scale the issue is still unsettled. Even if they do scale
it is important to know how long it takes the strings to latch onto
the scaling solution. If it takes a very long time, then for all
practical purposes strings will not scale.

{\sc $\bullet$ String stability:} Scaling assumes that the string
loops which break off the long strings decay quickly.  However, if
fermion zero modes on the strings exist these zero modes may set up
a current stabilizing the strings.  So far though it seems that
there are no such fermionic zero modes on F-strings or D-string
loops to prevent collapse \cite{davisdavis}.  Also, it is now
suspected that (in the $SO(32)$ case) despite the different boundary
conditions on the right and left moving parts of the heterotic
string that heterotic strings can break (although their endpoints
may be very heavy) \cite{polchinskiheterotic}. Heterotic string
theory is attractive from a phenomenological point of view and so it
would be interesting to understand what kinds of networks heterotic
strings can form. Recent ideas about cosmic strings in strongly
coupled heterotic M-theory can be found in \cite{beckerbecker}.

{\sc $\bullet$ Gravitational radiation:} The non-Gaussianity of cusp
emission comes from the singular nature of the cusp.  However,
strings possess small scale structure which may potentially smooth
out the cusp.  A simulation by Siemens and Olum claims that the cusp
remains despite small scale structure \cite{olum}. Nevertheless, the
issue is very important and needs to be further investigated.
Furthermore, the calculation of the gravitational wave amplitude in
the text assumed no backreaction -- the initial and final state of
the emitting string are the same. It has been claimed that when a
fully quantum calculation is made taking into account some
backreaction effects that the power law decay of $h(f)$ disappears
and turns into an exponential decay \cite{chialvatalk}.  Clearly
this issue of considerable importance.

{\sc $\bullet$ Philosphical/Our Universe:} These are foundational
issues which at some point need to be settled in order a build a
theory of cosmic superstrings. For example, did a tachyonic phase
transition ever occur?  Where is our universe among the huge
multitude and perhaps infinite number of string vacua
\cite{kachruinfinite}? How does the standard model appear in this
vacuum and how do cosmic superstrings couple to this standard model?


\section*{Acknowledgements}

We thank the organizers of COSLAB and Ana Achucarro in Leiden for
the opportunity to speak about cosmic superstrings.  We also thank
various participants at the Cosmic Superstrings Conference in Paris
at the Institut Henri Poincare for discussions. We also thank Rob
Meyers, Joe Polchinski and especially Nick Jones for help in
understanding various issues. Without Anne Davis' request for
lecture notes this review would have never been written. Finally we
thank MSRS for their patience.




\bibliography{bibliography}
\bibliographystyle{unsrt}

\end{document}

%% file: definitions.tex
\def\IP{{\mathbb P}}

\def\IZ{{\mathbb Z}}

\def\IC{{\mathbb C}}

\def\I1{{\mathbb 1}}
\def\gaq{\raise 0.4ex\hbox{$>$}\kern -0.7em\lower 0.62ex\hbox{$\sim$}}
\def\laq{\raise 0.4ex\hbox{$<$}\kern -0.8em\lower 0.62ex\hbox{$\sim$}}
\def\lsim{\mathrel{\rlap{\lower 4pt \hbox{\hskip 1pt $\sim$}}\raise 1pt \hbox
    {$<$}}}
\def\gsim{\mathrel{\rlap{\lower 4pt \hbox{\hskip 1pt $\sim$}}\raise 1pt \hbox
    {$>$}}}
\def \eqn#1#2{\begin{equation}#2\label{#1}\end{equation}}
\def \tr{{\rm Tr}}
\def \pq{\mu_{(p,q)}}
\def \p'{\mu_{(p',q')}}
\def \ppm{\mu_{(p\pm p',q \pm q')}}
\def \da9{D9-\bar{D}9}
\def \cf{{\cal{F}}}
\def \ca{{\cal{A}}}
\def \n{\noindent}

\def \beq{\begin{equation}}
\def \eeq{\end{equation}}
\def \be{\begin{equation}}
\def \ee{\end{equation}}
\def \da9{D9-\bar{D}9}

\def \cf{{\cal{F}}}

\def \ca{{\cal{A}}}
\def \cm{{\cal{M}}}